\documentclass[
    aps,
    superscriptaddress,
    10pt,
    prd,
    notitlepage,
    twocolumn,s
    tightenlines,
    nofootinbib,
    floatfix]{revtex4-2}

\usepackage{anyfontsize}

\usepackage[linktocpage,breaklinks]{hyperref}
\usepackage[usenames,dvipsnames]{xcolor}

\usepackage{aas_macros}
\usepackage{amsmath}
\usepackage{amssymb}
\usepackage{amsfonts}
\usepackage{txfonts}
\usepackage{bm}
\usepackage{stmaryrd}
\usepackage{tensor}
\usepackage{mathrsfs}
\usepackage[utf8]{inputenc}
\usepackage{url}
\usepackage{makecell}
\usepackage{longtable}

\usepackage{float}

\usepackage{xcolor}
\usepackage[most]{tcolorbox}

\usepackage{graphicx}
\usepackage{epsfig}
\usepackage{epstopdf}
\usepackage[normalem]{ulem}
\usepackage{lineno}

\usepackage{natbib}
\usepackage{hyperref}
\usepackage{cleveref}
\hypersetup{colorlinks=true,
            citecolor=Black,
            linkcolor=Black,
            urlcolor=Black}

\makeatletter
\newcommand{\pseudoeqref}[1]{\textup{\tagform@{#1}}}
\makeatother

\graphicspath{{figures/}}
\newcommand{\IMRELEASENO}{LLNL-JRNL-2007804}

\newcommand{\icasu}{Illinois Center for Advanced Study of the Universe \&  Department of Physics, University of Illinois at Urbana-Champaign, Champaign, Illinois 61820 USA}
\newcommand{\icas}{International Center for Advanced Studies, Universidad de San Martin and CONICET, San Martin, Buenos Aires, Argentina}
\newcommand{\llnl}{Lawrence Livermore National Laboratory, 7000 East Ave., Livermore, CA 94550, USA}
\newcommand{\caps}{Center for AstroPhysical Surveys, National Center for Supercomputing Applications, Urbana, Illinois 61801 USA}

\begin{document}

\title{Dark Matter Clumps as Sources of Gravitational-Wave Glitches in LIGO/Virgo/KAGRA data}
\begin{abstract}

We consider the hypothetical possibility that non-stationary glitch features in the noise of ground-based gravitational-wave detectors could be produced by small dark matter clumps that  pass through the earth in the vicinity of gravitational-wave detectors. We first derive the gravitational-wave strain that would be generated by the passage of such a dark matter clump. We find that the strain is primarily sourced by the Newtonian gravitational acceleration of the mirrors toward the clump and by the Shapiro time delay of the photons in the laser beams as they pass through the gravitational potential created by the dark matter clump. We also find that the Newtonian acceleration effect dominates the gravitational-wave strain for both ground and space-based interferometers. We then compare our dark matter clump, gravitational-wave strain model to 84 Koi-Fish glitches detected during the second observing run of the LIGO/Virgo/KAGRA collaboration through a Markov Chain Monte Carlo Bayesian analysis. We find that all glitches but 9 can be confidently rejected as having originated from dark matter clumps. For the remaining glitches, the dark matter hypothesis cannot be excluded, and the maximum \textit{a posteriori} parameters yield minimum densities of about $10^{-7} {\rm{g}}/{\rm{cm}}^3$, within the model. These results allow us to place the first direct upper limits with gravitational-wave detectors on the local over-density of dark matter in the form of clumps in the local neighborhood of Earth, namely $\rho_{{\rm DM} \, {\rm clumps}} \lesssim 10^{-15} {\rm{g}}/{\rm{cm}}^{-3}$. 

\end{abstract}
\author{Ezequiel Alvarez}
\email{sequi@unsam.edu.ar}
\affiliation{\icas}
\author{Scott Perkins}
\email{perkins35@llnl.gov}
\affiliation{\icasu}
\affiliation{\llnl}
\author{Federico Ravanedo}
\email{fravanedo@unsam.edu.are}
\affiliation{\icas}
\author{Nicolas Yunes}
\email{nyunes@illinois.edu}
\affiliation{\icasu}
\affiliation{\caps}

\date{\today}
\maketitle

\section{Introduction}\label{sec:intro}

The mystery of dark matter (DM) continues to stand as one of the most profound and enduring puzzles in contemporary astrophysics, cosmology and particle physics. Despite accounting for approximately 27\% of the Universe's mass-energy content, dark matter remains elusive, characterized solely by its gravitational interactions. Unraveling the nature of dark matter is a major objective in today’s physics research, not only for our understanding of galaxies and large-scale structures but also for formulating a complete picture of the Universe itself.  While numerous indirect detection methods have focused on observing the gravitational effects of dark matter on galactic \cite{Simon_2019, Salucci_2019, Allen_2011} and cosmic scales \cite{planck18}, direct detection aims to capture the influence of dark matter as it interacts with baryonic matter at the Earth scale \cite{xenon,agnes,pandax,damic,Edelweiss}.   Although direct detection usually consists of dark matter interacting with Standard Model particles through Quantum Field Theory, in this work we pose a different paradigm relying on gravitational direct detection.

The extreme sensitivity of gravitational-wave detectors in measuring the strain makes these detectors, in conjunction, a perfect instrument to detect gravitational waves. Moreover, the detectors are such sensitive instruments that they are also activated through tiny signals, such as Earth tides \cite{tide}, nearby human activity \cite{human}, and microseisms from ocean waves \cite{micro}, among others. However, in these latter cases, the activation is either well differentiated between the detectors, or it occurs in only one of the detectors in the gravitational-wave network.  This allows us to discard the hypothesis that it was generated by a gravitational-wave source.  Furthermore, there are many cases where one detector in the gravitational-wave network reads out a transient signal in its strain, but there is no conclusive understanding of what produced the transient. These are usually referred to as glitches in the gravitational-wave literature \cite{glitches, tSNE, glitches2}. 

In this paper, we aim to study whether glitches detected in only one gravitational-wave interferometer could be produced by relative changes in the lengths of the optical paths of the instrument due to the passage of a Dark Matter (DM) clump.  This line of inquiry has been previously explored  in \cite{  vincentlee, Jaeckel:2020mqa, Hall:2016usm, Lee:2022tsw, Baum:2022duc, Kawasaki:2018xak}, where given different scenarios of DM density as clumps, primordial black holes, cosmic strings or domain walls, the authors compute the sensitivity of measuring them in a variety of experiments.  In this paper, in contrast, we use real data to estimate the density of DM clumps that could be producing them. Our approach has the complexity that the sought signal, if it exists, would be hidden in a background of glitches that are still not well understood.

The passage of a DM clump on a gravitational-wave interferometer has two main effects: a Newtonian force (due to the gravitational attraction of the mirrors to the passing clump) and a Shapiro effect (due to the time delay experienced by photons that cross the potential sourced by the passing clump). Given the unparalleled ability to measure minute variations of the distance between mirrors of about one part in $10^{20}$, one cannot immediately conclude whether the Shapiro effect or the Newtonian force over the mirrors would provide the main contribution in modifying the optical path length. One of the first conclusions in this work is that for the studied scales, the latter would dominate over the former.   

Once the model for DM clump strain is well understood,  we analyze 84 glitches found in LIGO Hanford data that have been previously classified as {\it Koi-Fish} \cite{Glanzer_2023,gravityspy} because of their spectrogram shape.  We study how a DM clump of a given mass, velocity, and trajectory would generate a signal at ground-based interferometers that would be registered as a glitch, and then infer the characteristics of a possible DM clump, given the glitch signals. In particular, we introduce, explore and apply two data analysis measures (a modified fitting factor, and a $\chi^2$ metric based on the expectation value of the likelihood at its peak) to our set of Koi-Fish glitches to determine whether a given glitch could have originated from the passage of a DM clump. We find that for all but 9 glitches, the DM clump hypothesis can be confidently rejected. For the remaining 9 glitches, a DM clump hypothesis cannot be excluded.

Given the above results, we then investigate the implications on the DM clump properties and population if we assume that all of the above 9 glitches were produced by the passage of DM clumps. First, we note that gravitational-wave detectors are sensitive to DM clumps whose size is not much larger than each observatory (${\sim}$4 km, since otherwise the laser interference is not modified). 
Using this, the assumption that all 9 glitches were produced by DM clumps and that DM is composed of DM clumps of this size and nature, and some assumptions on the velocity of small DM clumps that pass through Earth, we find an upper bound on the DM density of approximately $10^{-15} \; {\rm{g}}/{\rm{cm}}^3$. 
If we relax the assumption that these 9 glitches were all produced by a DM clump and consider that none of them comes from a DM clump passage, then the bound changes by at most one order of magnitude. The general idea here is that if the density of small DM clumps in the Solar System were high enough, then this would greatly increase the number of glitches detected by interferometers; thus, given the finite number of glitches with DM features detected, one can place an approximate bound on the density of DM.

The above inferences ought to be compared to existing estimates of the dark‑matter density near the Sun. Most current methods are indirect: fits to the Galactic rotation curve and the vertical kinematics of tracer stars in the Solar neighborhood typically yield $(1.0 \pm 0.3)\times 10^{-24}$ g/cm$^3$ \cite{localdensity, localdensity2, pdg}, which are clearly more stringent than the bounds placed in this paper. Other, more direct Solar‑System tests assume a heliocentric DM density distribution \cite{KHRIPLOVICH_2006, Frere:2007pi, Pitjev_2013} and, using precise planetary ephemerides, set limits of order $10^{-19}$ g/cm$^3$ at Earth’s orbit. In contrast, our analysis does not assume any particular global density profile: it is sensitive to compact, transiting dark‑matter clumps in the near‑Earth environment that can generate detectable, short‑duration tidal signals in a terrestrial interferometer, and to non‑spherically symmetric substructure intersecting Earth’s vicinity. This is distinct from—and not a re‑interpretation of—the smooth, Sun‑centered, spherically-symmetric component constrained by planetary ephemerides. Ephemeris‑based limits target a steady heliocentric distribution and do not directly apply to sparse, anisotropic, or transient clump populations (e.g., Earth‑bound substructure, streams, or ring‑like features); conversely, our constraint is effectively insensitive, on interferometer scales, to any spatially smooth component of the DM density. Our work highlights a novel use of ground‑based gravitational‑wave detectors, leveraging their sensitivity to local tidal fields to explore direct DM detection possibilities. Moreover, our inference is local and sensitive to variations in the DM environment along Earth’s orbit.

This paper is organized as follows.  In Sec.~\ref{sec:Glitches at LIGO}, we briefly describe the glitches in gravitational-wave detectors, their effect in the apparatus, and existing classifications.  In Sec.~\ref{sec:Newton_and_Shapiro}, we describe the physics of a DM clump passing through a gravitational-wave detector. We study the Shapiro effect on spacetime and the Newton effect on the mirrors attraction, concluding that the latter is more important.  In Sec.~\ref{sec:Dark Matter Clump Model}, we describe in detail the model and its magnitudes for a DM clump passing through gravitational-wave detectors.  In Sec.~\ref{sec:improved_methods}, we discuss how to infer the posterior distribution on the DM clump parameters. To this end, we describe the relevant features of the noise within our framework, and then we discuss how we use this and the DM clump model to infer its parameters.  In Sec.~\ref{sec:model-exploration}, we explore expected behaviors for the inference process. We discuss degeneracies in the DM clump model that yield similar signals and introduce two metrics that we use to assess how likely a dataset is to be originated by a DM clump passing in the vicinity of a detector. We test these metrics using an injected DM clump signal and our artificial parameterization of a glitch described by a Cos-Gaussian model.  In Sec.~\ref{sec:Results on a sample of glitches} we are ready to apply the above tools on a real dataset of glitches.  We use a set of 84 Koi-Fish glitches taken from LIGO Hanford data, and we infer the DM clump parameters. We then compute the assessment metrics on each one of these datasets and estimate rough, order of magnitude constraints on the DM density that could be attributed to this species of DM clump. Finally, Sec.~\ref{sec:conclusions} contains the conclusions and outlook of the work.  We collect in the Appendices some calculations and plots that are supplementary to the presented text.

\section{A Very Brief Introduction to Glitches}\label{sec:Glitches at LIGO}

\allowdisplaybreaks[4]

Since the beginning of data acquisition by the advanced LIGO/Virgo/Kagra (LVK) detectors, transient signals corresponding to non-Gaussian noise have been observed in the primary strain channel. These transient non-Gaussian signals are known as ``glitches.'' Glitches generally have instrumental or environmental origins, making them uncorrelated between detectors. However, there is a nonzero probability that two glitches could accidentally coincide in time and frequency across different detectors. Additionally, some glitches can resemble gravitational-wave signals emitted by certain astrophysical systems, leading to potential false alarms \cite{Nitz:2017lco, DalCanton:2014hxh, Merritt:2021xwh}.

Therefore, one of the most important goals of glitch experts is to minimize the occurrence rate of glitches. A first step toward this goal is to classify and characterize glitches and study how each resulting class correlates with the state of different components of the detector. This is why thousands of different channels are used to monitor the behavior of various parts of the detector and relate these observations to what is seen in the primary strain channel \cite{Colgan:2019lyo, Nuttall:2018xhi}. 

\begin{figure*}[th]
\centering
\includegraphics[height=0.32\textheight]{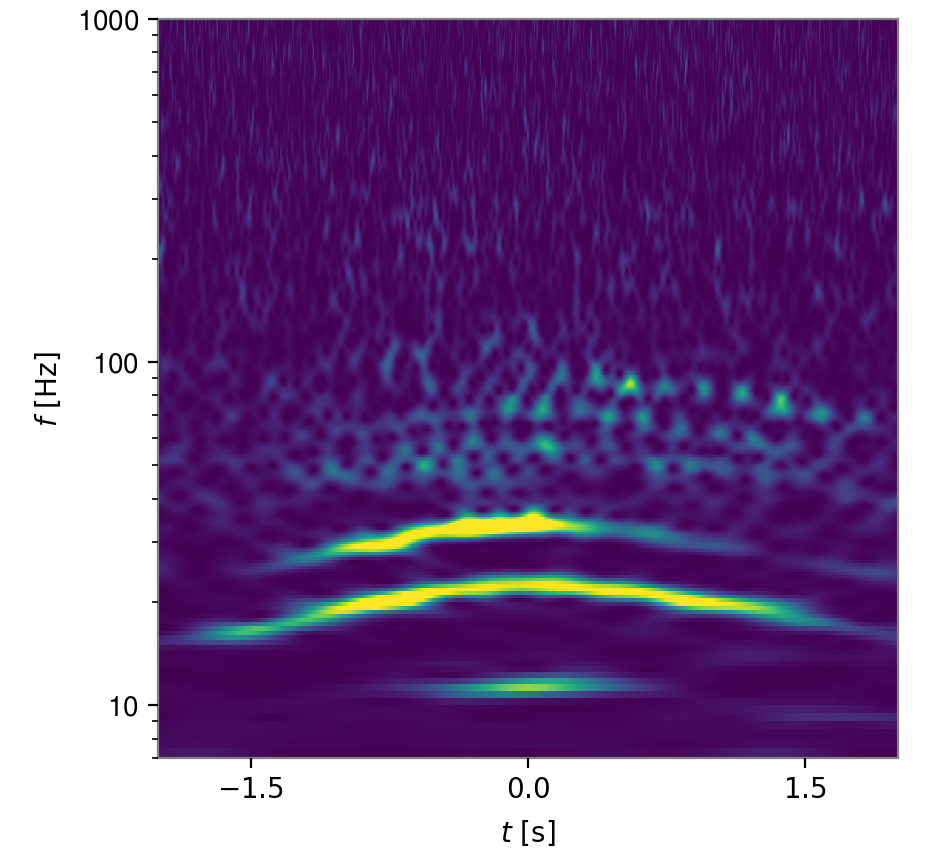}
\includegraphics[height=0.32\textheight]{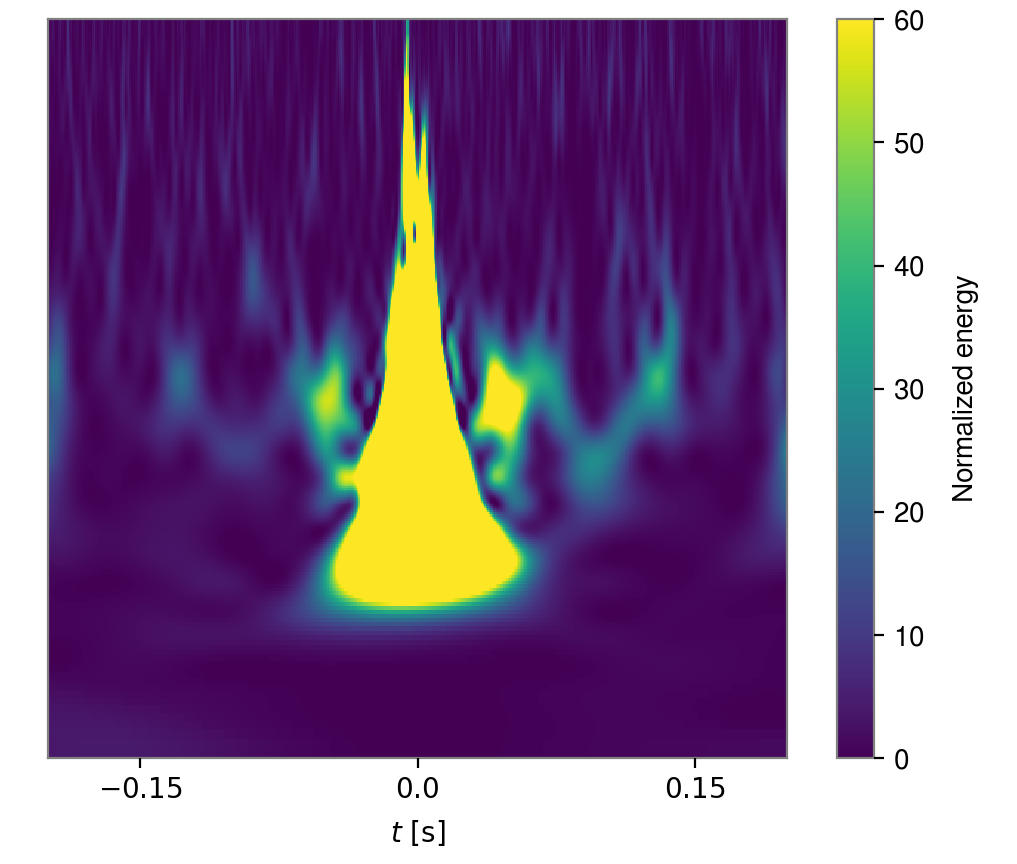}
\caption{Time-frequency spectrogram examples for Scattered Light (left) and Koi-Fish (right) glitches. Observe how the Scattered Light event is spread over long times, while the Koi-Fish event is of much shorter duration.}
\label{fig:glitches}
\end{figure*}

In their efforts to classify and search for correlations, researchers have identified about twenty different categories of glitches and discovered how some correlate with specific components of the detector, thus identifying their origins \cite{Glanzer:2022avx}. For example, one of the most common categories, known as ``Scattered Light,'' is caused by light scattering off optical components. These optical components move due to seismic activity, resulting in a phase shift in the scattered light \cite{Tolley:2023umc}. However, there are some categories of glitches for which researchers have not yet found a clear correlation with different components of the detector and have been unable to determine their origin; one such category is called ``Koi-Fish.'' Figure~\ref{fig:glitches} shows time-frequency spectrograms of a Scattered Light glitch in the left panel and a Koi-Fish glitch in the right panel. From the spectrograms, one can recognize that each category has different patterns. The Scattered Light event corresponds to long-duration glitches with peaks at specific frequencies, whereas the Koi-Fish event corresponds to short-duration glitches with a continuum of frequencies with preference for lower frequencies.

In this work, we investigate whether some glitches could be produced by the passage of Dark Matter (DM) clumps through the gravitational-wave detectors \cite{Jaeckel:2020mqa}. The question we wish to address is whether glitch categories that are not well correlated with instrumental effects could correspond to the passage of a DM clump through the detector. Only glitch categories with unclear origins and signal shapes similar to those expected from a DM clump are relevant to this hypothesis. In particular, this work focuses on the Koi-Fish category, whose origin remains unclear and whose signal shape could resemble that expected from a DM clump passage.

The data used in this work are available through Ref.~\cite{gwpy}. We use the glitch list shown in Table~\ref{tab:fullwidth}, which is a subset of the catalog presented in Ref.~\cite{gravityspy}. This list consists of a randomly selected sample of 84 Koi-Fish glitches, which we investigate in detail within our paradigm\footnote{We analyzed 84 Koi-Fish glitches because we initially selected 100 at random from a Gravity Spy catalog of 564 events, but 16 had data segments that were not retrievable. The 564 events are from the LIGO Hanford O2 run and were classified by Gravity Spy as Koi-Fish with classification confidence $>95\%$ and signal-to-noise ratios between $100$ and $200$.}.


\section{The Effect of the Passage of a Dark Matter Clump through a Ground-based Gravitational-Wave Interferometer}\label{sec:Newton_and_Shapiro}

\begin{figure*}[th]
\includegraphics[width=.45\linewidth]{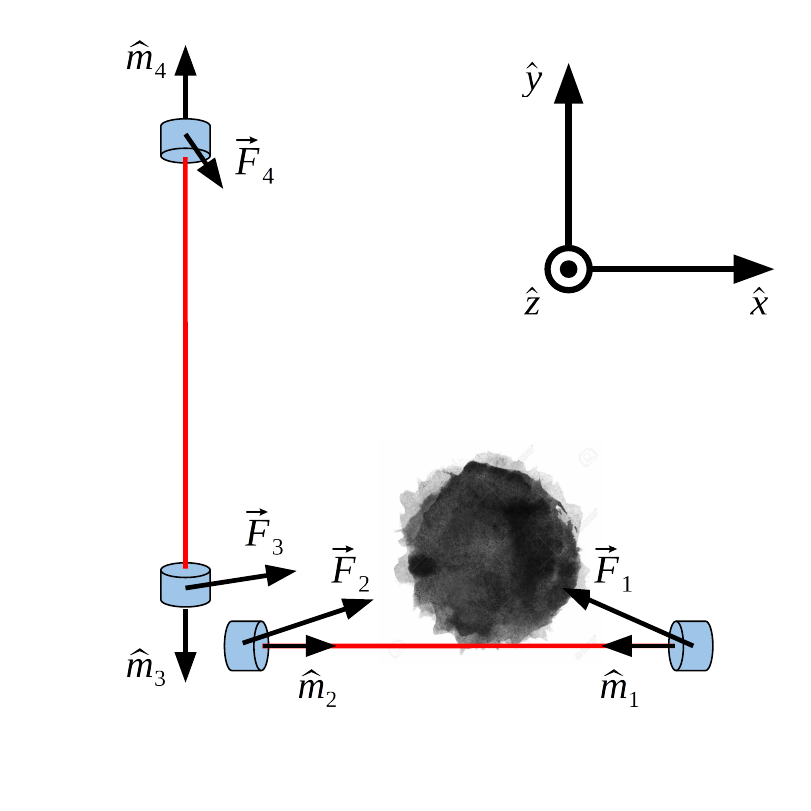}
\includegraphics[width=.45\linewidth]{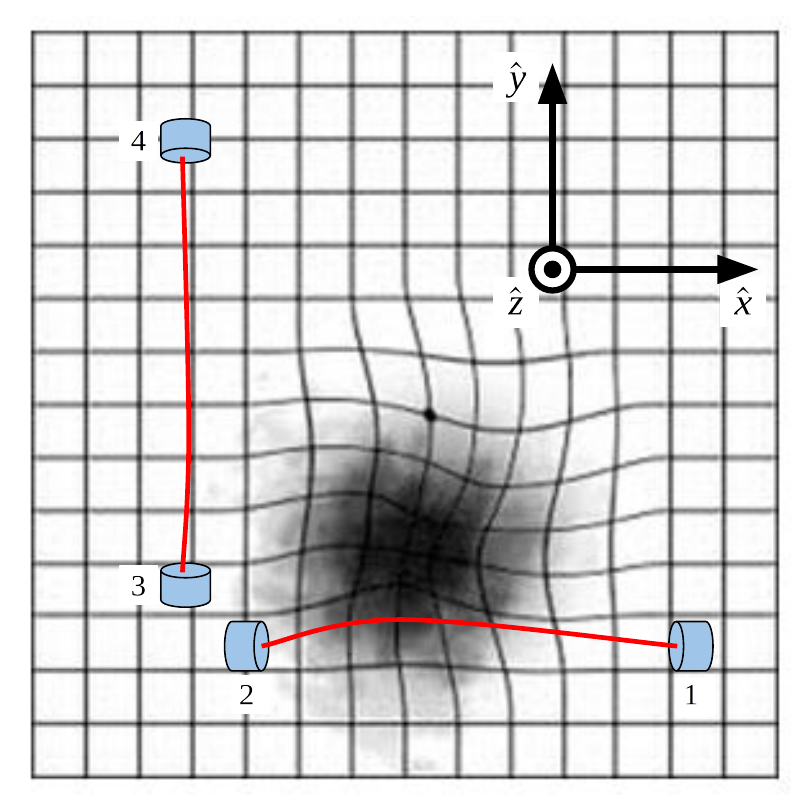}
\caption{Left: Illustration of the Newtonian effect, wherein each mirror experiences a gravitational force toward the DM clump.  
Right: Illustration of the Shapiro effect, wherein a DM clump time-delays the photons in the arms.
Additionally, this figure shows the system of unit vectors along the $x$, $y$, and $z$ axes that we use throughout this work.}\label{fig:Newton_Shapiro}
\end{figure*}

The variation in optical length in each detector is related to the strain observed at each detector, and denoted by $h(t)$, through
\begin{equation}
 h(t) = \frac{1}{L}\Big(-\Delta{L}_{x}(t) + \Delta{L}_{y}(t)\Big) \,,
\label{eq:strain} 
\end{equation}
where $L$ is the optical path length of a light ray in the detector's arms, and $\Delta{L}_{x}(t)$ and $\Delta{L}_{y}(t)$ are the changes in the length of the optical path in each arm, which are both functions of time. This expression can be straightforwardly derived from the solution to the geodesic deviation equation on a background that is perturbed away from Minkowski by an impinging gravitational wave. Throughout this work, we utilize the coordinate system illustrated in Fig.~\ref{fig:Newton_Shapiro}. We refer to the arm extending along the $x$ and $y$-axis directions as the ``$x$-arm'' and ``$y$-arm'', respectively.  Additionally, we set the position of the detector beam splitter at the origin of the coordinate system, and define the plane of the detector as the $z = 0$ plane.

We wish to understand how the presence of a DM clump affects the lengths of the optical paths of
the detector. In this section,  we examine two possible effects: the Newtonian gravitational interaction between the DM clump and each of the detector's mirrors, and the Shapiro effect that the DM clump would produce on the paths traveled by the light. The Newtonian effect is simply the Newtonian gravitational pull or force that a DM clump would exert on the mirrors. The left panel of Fig.~\ref{fig:Newton_Shapiro} shows an illustration of this effect, where each mirror experiences a gravitational pull towards the DM clump. The Shapiro effect is the gravitational time delay that the photons in the interferometer arms would experience if a DM clump gravitationally interacts with the detector. The right panel of Fig.~\ref{fig:Newton_Shapiro} shows a pictorial illustration of this effect, where the photons in the $x$-arm are the most Shapiro time-delayed.

A first question that arises is whether either of these effects dominates over the other. Since much of the statistical analysis of the noise in gravitational-wave detectors is in frequency space, we compare both effects in the Fourier domain through the Fourier transforms of the strain, $\widetilde{h}(f)$. In other words, we need to calculate the modifications to $\widetilde{h}(f)$ due to the Newtonian effect ($\widetilde{h}_{N}(f)$) and due to the Shapiro effect ($\widetilde{h}_{S}(f)$), and compare their strength as a function of the DM clump parameters. A work of this kind was previously carried out in \cite{vincentlee}, where the strengths of both effects were computed and compared. In the next section, we present a derivation and a comparison of these effects within our framework. For other similar derivations, particularly of the Newtonian effect, see \cite{Jaeckel:2020mqa, Hall:2016usm, Lee:2022tsw, Baum:2022duc, Kawasaki:2018xak}. 

In what follows we adopt the following Fourier transform convention:
\begin{equation}
\begin{split} 
& \widetilde{h}(f) = \int dt \, h(t) \, e^{-2 \pi i f t} \ ,
\label{eq:Fourier_Transform_conventions}
\end{split}
\end{equation}
which implies that, since $h(t)$ is dimensionless, $\widetilde{h}(f)$ has units of inverse frequency, $\text{Hz}^{-1}$.

\subsection{Newtonian effect}\label{sec:Newtonian_effect}

The Newtonian effect corresponds to an acceleration in the mirrors, so it produces a second derivative of the strain that can be expressed as \cite{Thorne:1998hq}
\begin{equation}
\displaystyle\frac{d^2{h_{N}}}{dt^2}= \frac{1}{L}\displaystyle\sum_{j=1}^4 \vec{g_j}\cdot{\widehat{m}_j} \ .
\label{eq:Newton1}
\end{equation}
Here $\widehat{m}_j$ is the unit vector of mirror $j$ along its corresponding arm direction (see Fig.~\ref{fig:Newton_Shapiro}), and $\vec{g}_j$ is the Newtonian acceleration of mirror $j$ due to the presence of the DM clump; arrows over symbols denote three-dimensional Euclidean vectors. Let $\vec{r}_{DM}(t)$ and $M_{DM}$ denote the position vector and the mass of the DM clump, and let $\vec{r}_j$ be the position vector of mirror $j$. Then, we have that 
\begin{equation}
\frac{d^2{h_{N}}}{dt^2} = \frac{M_{DM}\,G}{L}\sum_{j=1}^4 \frac{\vec{r}_{DM}(t)-\vec{r}_j}{\left |{\vec{r}_{DM}(t)-\vec{r}_j}\right |^3} \cdot{\widehat{m}_j} \ , 
\label{eq:Newton2}
\end{equation}
where for simplicity we assume that the DM clump has spherical symmetry around its center and that it does not collide with the mirrors. The second derivative of the strain, $d^2{h_{N}}/dt^2$, is a function of time because $\vec{r}_{DM}$ is. 

Given $\vec{r}_{DM}(t)$, we can compute the Fourier transform of $d^2{h_{N}}/dt^2$ and then obtain $\widetilde{h}_{N}(f)$ using the rules of Fourier transform for derivatives, namely
\begin{equation}
\widetilde{h}_{N}(f) = -\frac{1}{(2 \pi f)^2}
\widetilde{\frac{d^2h_{N}}{dt^2}}(f) \ . 
\label{eq:rule_FT}
\end{equation}
However, computing the Fourier transform of $d^2 h_{N} / dt^2$ is a challenging task that can be performed analytically only for a few special cases, corresponding to particular incoming directions of the DM clump. The simplest case is when the DM clump has a constant velocity perpendicular to the plane of the detector. In this work, we focus on DM clumps with a constant velocity during their encounter with the detector, but in an arbitrary direction. Consequently, numerical methods are required to compute the Fourier transforms. To this end, we use the Fast Fourier Transform (FFT) algorithm implemented in the \texttt{scipy.fft} Python module \cite{scipy2024fft}. 

\subsection{Shapiro effect}\label{sec:Shapiro_effect}

The strain produced by the Shapiro time delay effect depends on the gravitational potential generated by the DM clump. We denote this potential as $\phi_{DM}$, and we only consider the situation in which this potential is very weak, $|\phi_{DM} / c^2| \ll 1$. In this context, $\phi_{DM}$ is the classic gravitational potential, and the invariant length of a line element around the DM clump is
\begin{equation}
ds^2 = -\left(1+\frac{2\,\phi_{DM}}{c^2}\right)(c\, dt)^2+\left(1-\frac{2\,\phi_{DM}}{c^2}\right)(dx^2+dy^2+dz^2) \ .
\label{eq:GR_interval} 
\end{equation}
The line element of a light ray traveling from one end mirror to the beam splitter has null length, and hence,
\begin{equation}
dt = \frac{1}{c}\left(1-\frac{2\,\phi_{DM}}{c^2}\right)d\rho,
\label{eq:foton_dt}
\end{equation}
where $d\rho$ is a small segment covered by the light ray in the arm. The amount of time the light ray takes to go from one end mirror to the beam splitter is
\begin{equation}
T = \frac{L}{c} -\frac{1}{c}\int\frac{2\,\phi_{DM}}{c^2}\,d\rho = \frac{L}{c} +\frac{\Delta{L}}{c},
\label{eq:foton_t}
\end{equation}
with
\begin{equation}
\Delta{L} = -\frac{2}{c^2}\displaystyle\int  \phi_{DM}\,d\rho \ ,
\label{eq:Delta_l}
\end{equation}
where the integral is performed along the corresponding arm. The change in the optical path length of the light ray is $\Delta{L}$. Therefore, the strain caused by such a Shapiro effect can be read from Eq.~\eqref{eq:strain}, with $\Delta{L}_{x}$ and $\Delta{L}_{y}$ replaced by the integral of Eq.~\eqref{eq:Delta_l} performed along the corresponding arm. Note here that the DM clump does not need to intersect one of the arms for there to be a Shapiro effect on the interferometer light; the gravitational potential of the clump obviously extends beyond its interior clump geometry to its exterior, thus affecting both arms.

Let us present the strain due to the Shapiro effect for two specific examples. Consider a spherical DM clump that crosses the plane of the detector through the point $\vec p = (x_{0},y_{0},0)$ at $t=0$, moving with a constant velocity $v_{DM}$ perpendicular to the plane. Let us further assume that the velocity of the DM clump is small enough that the gravitational potential can be treated as constant during the time that it takes the light ray to cover a distance of ${\cal{O}}(L)$. Then, if the DM clump never crosses the arms, the radius of the DM clump does not matter, and we can express the strain as
\allowdisplaybreaks[4]
\begin{equation}
\begin{split}
& h_{S}(t)=h_{Sx}(t)+h_{Sy}(t)\ , \\\\
& h_{Sx}(t) = -\frac{2\,M_{DM}\,G}{L\,c^2}\ln\left({\frac{\sqrt{(v_{DM}\,t)^2 + {y_{0}}^2 + (L-x_{0})^2} + (L-x_{0})}{\sqrt{(v_{DM}\,t)^2 + {y_{0}}^2 + {x_{0}}^2}  -x_{0}}}\right) \ , \\\\
& h_{Sy}(t)=\frac{2\,M_{DM}\,G}{L\,c^2}\ln\left({\frac{\sqrt{(v_{DM}\,t)^2 + {x_0}^2 + (L-y_0)^2} + (L-y_0)}{\sqrt{(v_{DM}\,t)^2 + {x_0}^2 + {y_0}^2}  - y_0}}\right) \ ,
\label{eq:shapiro1}
\end{split}
\end{equation}
where $h_{Sx}$ and $h_{Sy}$ can be obtained from the integral of Eq.~\eqref{eq:Delta_l} along the $x$- and $y$-arms, respectively, divided by $L$.

\begin{figure}[t]
\includegraphics[width=.9\linewidth]{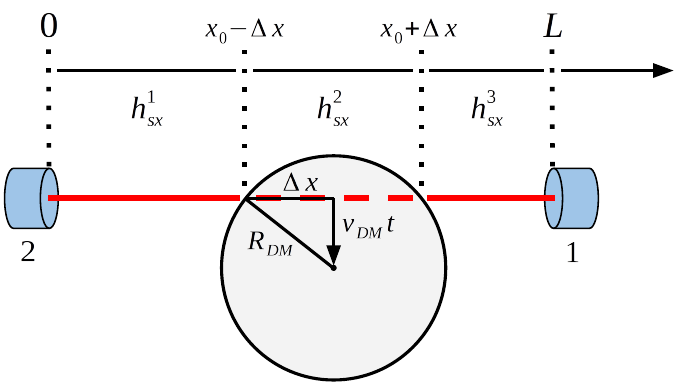}
\caption{Illustration of the crossing of the DM clump through the 
$x$-arm, showing the three regions relevant to computing the integral of Eq.~\eqref{eq:Delta_l}. Each region is labeled with its corresponding integral term.}\label{fig:Shapiro1}
\end{figure}

When the DM clump crosses one of the arms, we must take into account that the light rays will travel through the DM clump itself for some amount of time, and, in that case, the DM clump radius, $R_{DM}$, must be taken into account. Consider then the case in which the DM clump is at ${y}_{0} = 0$ and its radius is such that $0\leq x_0-R_{DM}$ and $x_0+R_{DM}\leq L$. These conditions describe a DM clump that passes through the $x$-arm without colliding with the mirrors. In this case, the integral of Eq.~\eqref{eq:Delta_l} along the $y$-arm is the same as in the case in which the DM clump does not pass through the arms, $\Delta L_{y}$, and thus $h_{S y}$ is unchanged from our previous result in Eq.~\eqref{eq:shapiro1}. The integral of Eq.~\eqref{eq:Delta_l} along the $x$-arm, however, is not the same as before. Let us then further assume that the DM clump has a constant density along its radius, so that we can evaluate the integral analytically. We can further classify this case into two subcases. The first subcase is when the condition $ |t| \geq  R_{DM}/v_{DM} $ is satisfied, so that the DM clump does not touch the $x$-arm, and the strain is given by setting ${y}_{0} = 0$ in Eq.~\eqref{eq:shapiro1}. The second subcase is when the DM clump is crossing the $x$ arm, so that $ |t| \leq  R_{DM}/v_{DM} $, and then $h_{Sx}$ is given by,
\begin{equation}
\begin{split}
& h_{Sx}(t) = h_{Sx}^{1}(t) + h_{Sx}^{2}(t) + h_{Sx}^{3}(t)  \ , \\\\
& h_{Sx}^{1}(t)= - \frac{2\,M_{DM}\,G}{L\,c^2}\ln\left({\frac{\sqrt{(v_{DM}\,t)^2 + \Delta x^2}-\Delta x}{\sqrt{(v_{DM}\,t)^2 + {x_0}^2} - x_0}}\right) \ , \\\\
& h_{Sx}^{2}(t) = \frac{2\,M_{DM}\,G}{L\,c^2}\left[\frac{\Delta x}{R_{DM}}\left(\left(\frac{v_{DM}\,t}{R_{DM}}\right)^2-3\right) + \frac{1}{3} \left(\frac{\Delta x}{R_{DM}}\right)^3\right] \ , \\\\
& h_{Sx}^{3}(t) = -\frac{2\,M_{DM}\,G}{L\,c^2}\ln\left({\frac{\sqrt{(v_{DM}\,t)^2 + (L-x_0)^2} + (L-x_0)}{\sqrt{(v_{DM}\,t)^2 + \Delta x^2} + \Delta x}}\right) \ , \\\\
&\Delta x \equiv \sqrt{R_{DM}^2 - (v_{DM}\,t)^2} \ ,
\label{eq:shapiro2}
\end{split}
\end{equation}
where $h_{Sx}^{1}$, $h_{Sx}^{2}$, and $h_{Sx}^{3}$ are obtained from the integral of Eq.~\eqref{eq:Delta_l} along the $x$ arm in the region bounded by $x = 0$ and $x = x_0 - \Delta x$, in the region inside the DM clump bounded by $x = x_0 - \Delta x$ and $x = x_0 + \Delta x$, and in the region bounded by $x = x_0 + \Delta x$ and $x = L$, respectively.  Figure~\ref{fig:Shapiro1} illustrates the crossing stage of the DM clump through the $x$-arm, showing the three regions described above, each with their corresponding integral terms. 

\begin{figure*}[t]
\includegraphics[height=6.665cm, keepaspectratio]{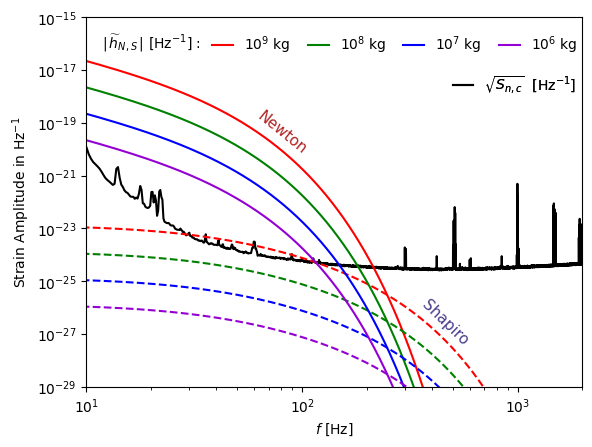}
\includegraphics[height=6.665cm, keepaspectratio]{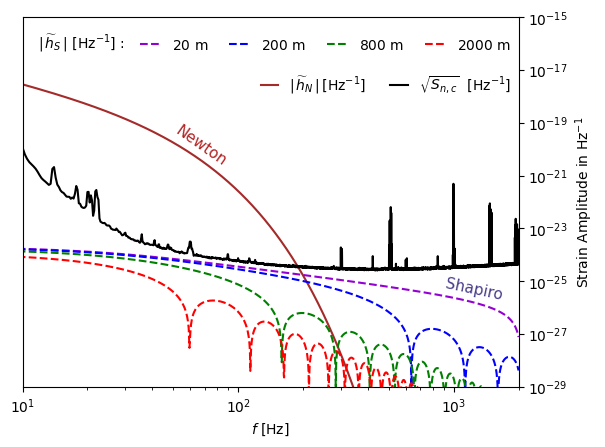}
\caption{Comparison of the Fourier transform of the strain due to the Newtonian and the Shapiro effects. The solid lines represent the Newtonian effect, while the dashed lines represent the Shapiro effect. In the left panel, we consider a DM clump with different values of its mass. This DM clump has a size of $R_{DM} < 0.5 \ \text{km}$ and passes through the detector at $x_0 = 2.0 \ \text{km}$ and $y_0 = 0.5 \ \text{km}$, with a velocity of $v_{DM} = 200 \; \text{km/s}$ perpendicular to the detector's plane. In the right panel, we consider a DM clump with different values of its radius. This DM clump has a mass of $M_{DM} = 10^8 \ \text{kg}$ and passes through the $x$ arm at $x_0 = 2.0 \ \text{km}$ and $y_0 = 0 $, with a velocity of $v_{DM} = 200 \ \text{km/s}$ perpendicular to the detector's plane. Observe in the right panel that a characteristic frequency related to the radius of the DM clump, given by $v_{DM}/R_{DM}$, enters the Fourier transform of the strain in the Shapiro effect case. Observe also that, in both panels, the Newtonian effect is more significant than the Shapiro effect in the relevant frequency range, which is determined by the characteristic spectral noise density shown in black. Both the Fourier transform of the strain and the square root of the characteristic spectral noise density have units of $\text{Hz}^{-1}$.}
\label{fig:newton_shapiro12}
\end{figure*}

\subsection{Comparison of the Newtonian and the Shapiro effect of a crossing DM clump}\label{sec:Comparison of the Newtonian and the Shapiro effect of a crossing DM clump}

We now compare the previous calculations to determine which effect dominates. The Fourier transforms of the strain due to the Newtonian and the Shapiro effects are shown in Fig.~\ref{fig:newton_shapiro12} with solid and dashed lines respectively. Observe that both classes of Fourier transforms decay with frequency. This is just a property of the Fourier transform of any continuous and differentiable function, as we have assumed for $d^2h_{N}/dt^2$ as a function of time, which is further exacerbated by the $1/f^2$ denominator e.g.~in Eq.~\eqref{eq:rule_FT}. Observe also that, as the frequency goes to zero, the Fourier transform of the strain due to the Shapiro effect remains roughly constant, while it grows in the Newtonian effect case. In the latter, this is again because of the $1/f^2$ denominator in Eq.~\eqref{eq:rule_FT}, which is related to the motion imparted by the DM clump on the mirrors. Unlike the Newtonian interaction, the Shapiro effect does not impart a final motion to the mirrors, which implies that the strain does not have significant components at low frequencies.

Figure~\ref{fig:newton_shapiro12} presents the Fourier transform of the strain for two different classes of cases. In the left panel, we consider different masses of a DM clump that never crosses the interferometer arms, and, therefore, the value of the DM radius is irrelevant. Conversely, in the right panel, we consider cases in which the DM clump does cross the $x$ arm, so we study different DM clump radii with a fixed mass of $10^8 \ \text{kg}$. Observe that the Fourier transforms of the strain due to the Shapiro effect in the right panel present oscillatory behavior. This is because, when the DM clump crosses the arm and the radius of the clump becomes significant, a characteristic frequency appears in the Fourier transform. This result can be understood by thinking of the clump-crossing as acting in a fixed, rectangular or ``top-hat'' window with characteristic length of $2 \, R_{DM} / v_{DM}$. In this case, the piecewise (or “top-hat”) nature of the modification to the time-domain strain leads to oscillatory behavior in the Fourier transform, with a characteristic frequency of $v_{DM} / R_{DM}$.

Figure~\ref{fig:newton_shapiro12} indicates that the Fourier transform of the strain for the Newtonian effect is much larger than that for the Shapiro effect, but should we include both in our studies? To find the answer to this question, we also plot in Fig.~\ref{fig:newton_shapiro12} the square root of the characteristic spectral noise density of Advanced LIGO, based on the first three months of O3 in the Hanford detector \cite{ligo2020t2000012,ligo2012p1200087}. The characteristic spectral noise density is defined as the spectral noise density of gravitational-wave detectors divided by the frequency, $S_{n,c} = S_{n}/f$, and, as can be seen from the equations in Sec.~\ref{sec:detectors_noise}, it has units of inverse square frequency, $\text{Hz}^{-2}$. This definition is such that the square of the signal-to-noise ratio (SNR) $\rho_h^2$ of template $\tilde{h}$ can be written as 
\begin{align}
\rho_h^2 = 4 \int \frac{|\tilde{h}|^2}{S_n} df = 4 \int \left( \frac{|\tilde{h}|}{\sqrt{S_{n,c}}} \right)^2 d \ln f\,,
\end{align}
so that the integrand gives us the square of the SNR per logarithmic frequency interval. Observe that the Fourier transform of the strain due to the Newtonian effect is above the characteristic spectral noise density for frequencies below $\sim$ 200 Hz, suggesting that, in principle, signals like this could contribute a non-negligible SNR. On the other hand, the Fourier transform of the strain due to the Shapiro effect is always below the characteristic spectral noise density curve. In particular, the Shapiro effect becomes more significant than the Newtonian one only in the region where the signal is already below the detector noise. Consequently, the Shapiro effect plays a subdominant minor role when constructing a DM clump model.

\subsection{Effective Models}
\label{sec:effective-models}

We can construct approximate or ``effective'' models for the gravitational-wave strain caused by the Newtonian and the Shapiro effects associated with the passage of a DM clump by a ground-based interferometer. In the previous subsections, we considered signals produced by a DM clump that passes closer to the $x$ arm than the $y$ arm. When this is the case, we can approximate the problem by considering only the part of the strain that corresponds to the $x$ arm. Let us further make the approximation that the acceleration of the mirrors due to the passage of the DM clump is zero when the DM clump is farther than approximately a distance $L$, and is constant and nonzero when the DM clump is closer. 

With these approximations in hand, let us first focus on the strain due to the Newtonian DM clump effect. If the DM clump is in the middle of the $x$ arm, the Newtonian second derivative of the strain that corresponds to this arm is $8M_{DM}G/L^3$. If we describe the second derivative of the strain by the piecewise function
\begin{equation}
\displaystyle\frac{d^2h_N}{dt^2} = \left\{\begin{array}{lcc} \frac{8M_{DM}G}{L^3} & if & |t| \leq L/v_{DM}  \\
\\ 0 &  if  &  |t| > L/v_{DM} \end{array}\right. \ ,
\label{eq:rough_newton_1}
\end{equation}
we then obtain
\begin{equation}
\widetilde{h}_{N}(f) = -\frac{16  M_{DM}  G}{L^3} \ \frac{\sin\left(2\pi f L/v_{DM}\right)}{(2\pi f)^3} \ . 
\label{eq:rough_newton_2}
\end{equation}
Observe that our approximations have led to oscillatory behavior in the Fourier transform, which is sourced entirely by the piecewise (or ``top-hat'') nature of the time-domain strain. Had we not assumed that the acceleration of the mirrors due to the DM clump passage was zero when the DM clump is farther than a distance $L$, then the sine-behavior of the Fourier transform would not have arisen. 

We can approximate the strain due to the Shapiro effect similarly. Using our approximations, we consider only the integral of Eq.~\eqref{eq:Delta_l} along the $x$ arm. For a DM clump with $R_{DM} < 0.5 \ \text{km}$ that passes through $x_0 = 2.0 \ \text{km}$ and $y_0 = 0.5 \ \text{km}$ at time $t = 0$, we have
\begin{equation}
\Delta{L}_{x}(t=0)=-\frac{2}{c^2}\displaystyle\int_{0}^{L}\phi_{DM}(t=0)\, dx \approx \frac{8M_{DM}G}{c^2} \,,
\label{eq:rough_shapiro_1}
\end{equation}
and thus, 
\begin{equation}  
\displaystyle h_S(t) = \left\{\begin{array}{lcc} -\frac{8M_{DM}G}{Lc^2} & if & |t| \leq L/v_{DM}  \\\\
 0 &  if  &  |t| > L/v_{DM} \end{array}\right. \ ,
\label{eq:rough_shapiro_2}
\end{equation}
and
\begin{equation}
\widetilde{h}_{S}(f) = -\frac{16M_{DM}G}{Lc^2} \ \frac{\sin\left(2\pi f L/v_{DM}\right)}{2\pi f} \ . 
\label{eq:rough_shapiro_3}
\end{equation}
As before, our piecewise approximation of the strain has led to oscillatory behavior in the Fourier transform. 

\begin{figure}[t]
\includegraphics[width=\linewidth]{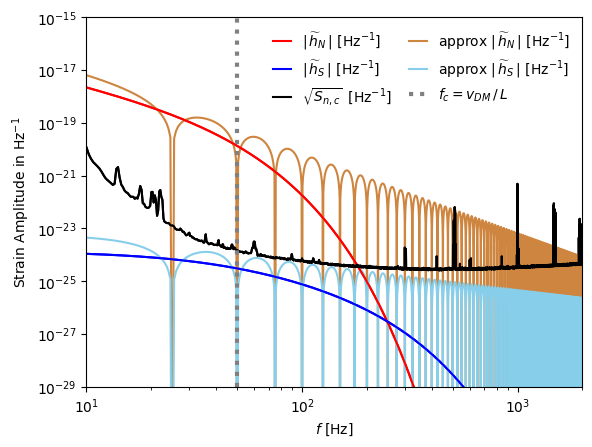}
\caption{Approximate and exact Fourier transforms of the strain for the Newtonian and Shapiro effects for a DM clump with mass $M_{DM} = 10^8 \ \text{kg}$, velocity $v_{DM} = 200 \ \text{km/s}$, and location $x_0 = 2.0 \ \text{km}$ and $y_0 = 0.5 \ \text{km}$. Observe that there is good agreement between the approximate and the exact models for frequencies below $f_c = v_{DM}/L$.
}\label{fig:newton_shapiro3}
\end{figure}

Figure~\ref{fig:newton_shapiro3} compares the approximate Fourier transforms of the strain to the Fourier transform of the strain without approximations. Observe that for frequencies lower than 
$f_c = v_{DM}/L$, there is a fair agreement between the two sets of Fourier transforms at the order of magnitude level, which allows us to make a straightforward comparison between the two effects for our current purposes.  For frequencies lower than $f_c = v_{DM}/L$, we can write \begin{equation}
\frac{\widetilde{h}_{N}}{\widetilde{h}_{S}}= \left(\frac{c}{2\pi f L}\right)^2 \sim \left(\frac{c}{f L}\right)^2\,, 
\label{eq:rough_newton_shapiro}
\end{equation}
which explains why the Newtonian effect is much larger than the Shapiro effect below $\sim 100 \ \text{Hz}$. The factor of $1/L^2$ here appears because the Newtonian interaction scales inversely with separation cubed. The factor of $1/f^2$ enters because the Newtonian acceleration is proportional to the second time derivative of the strain. The combined factor $1/(f L)^2$ is the inverse of a characteristic velocity squared, which for the length of ground-based interferometer and typical frequencies is about $v_{\rm c} \sim f L \sim 10^2 \; {\rm{km/s}} \ll c$. In other words, the Newtonian effect depends on the size of the detector (characterized through $L$), which is a very small length at the typical frequencies at which LVK detectors operate. On the other hand, the Shapiro effect does not depend on the size of the detector. Instead, this effect can be thought of as due to a ``glass ball'' in the path that connects the mirrors of the detector arms; in this analogy, the Shapiro effect produced by the glass ball does not depend on whether the ball is placed in the middle of the arm or closer to one of the mirrors.
\begin{figure*}[t]
\includegraphics[height=0.46\linewidth]{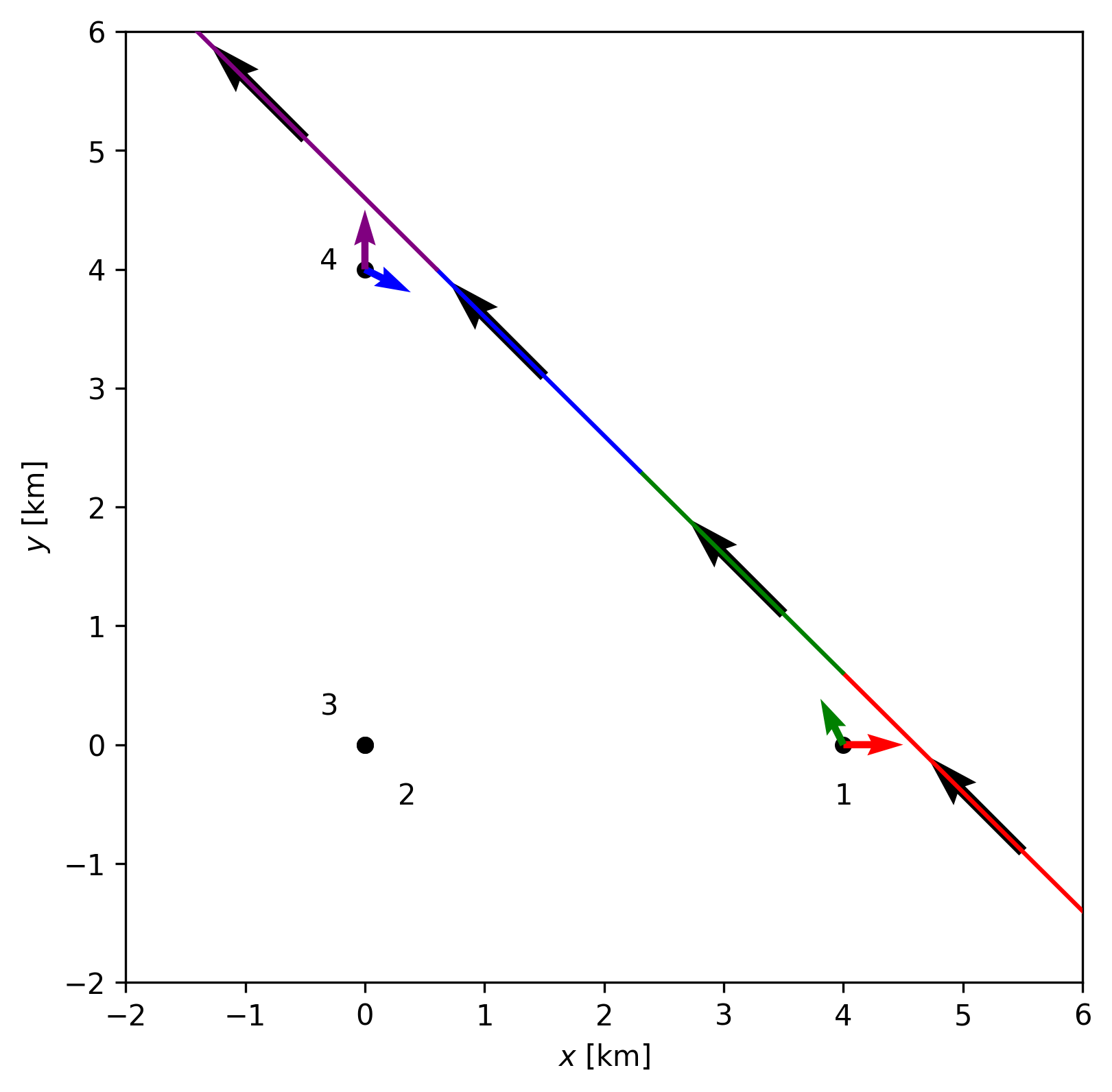}
\includegraphics[height=0.46\linewidth]{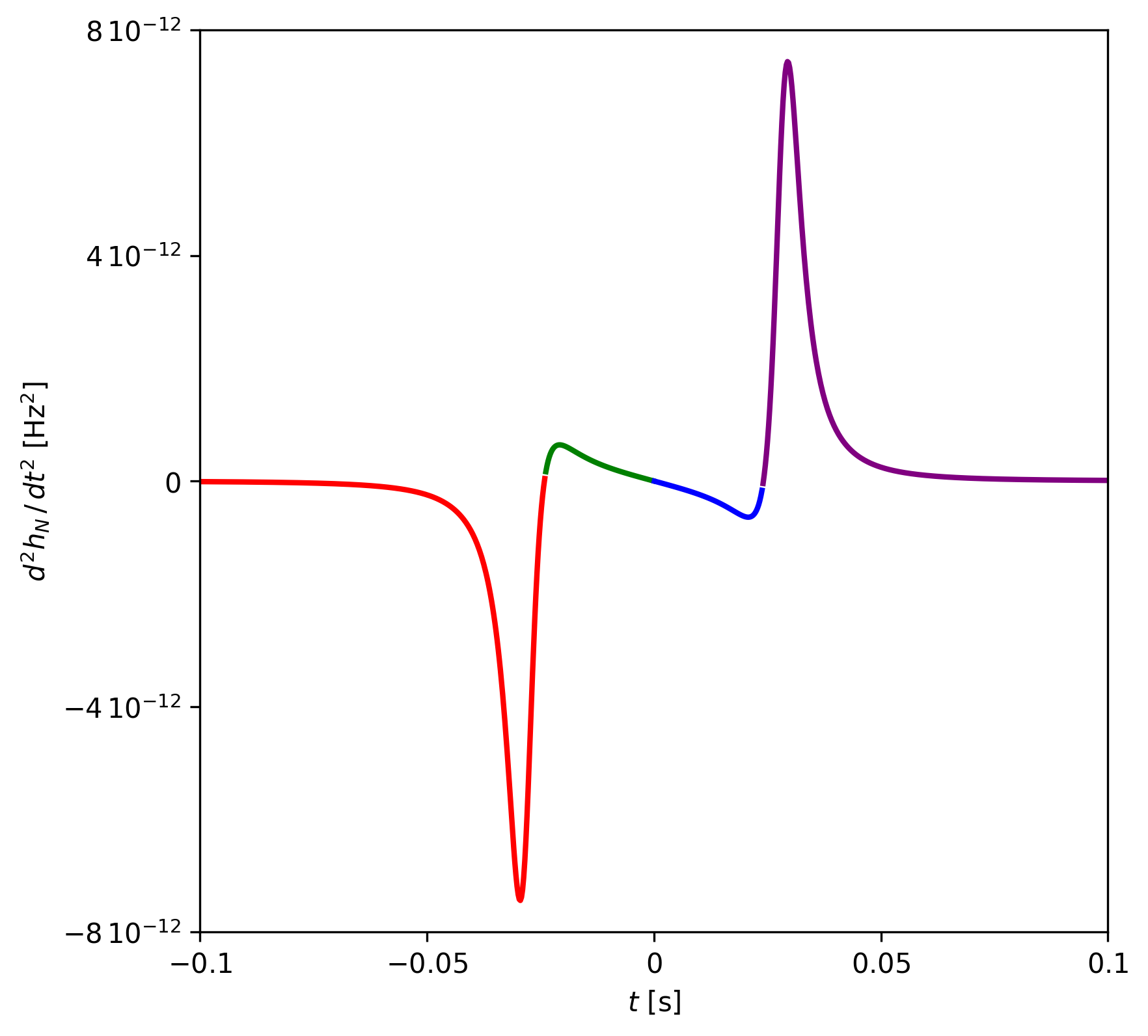}
\caption{Second derivative of the strain (right panel) produced by a DM clump following the trajectory depicted in the left panel. The black points and the numbers indicate the mirrors and their corresponding labels, see text for details on how the strains are produced at each stage of the trajectory.}
\label{fig:trajectory_d2h.png}
\end{figure*}

The above analysis leads us to think that if we increase the size of the detector, there will be some $L$ for which the Shapiro effect may become comparable to the Newtonian effect. Unfortunately, increasing the size of the detector also changes the frequency of the gravitational waves that can be detected. For a gravitational-wave detector of the size of advanced LIGO ($L = 4 \; {\rm{km}}$) and frequencies of $100 \; {\rm{Hz}}$, we have ${\widetilde{h}_{N}}/{\widetilde{h}_{S}} \sim 10^4$. For a next-generation, ground-based instrument, like Cosmic Explorer ($L = 40 \; {\rm{km}}$), we have ${\widetilde{h}_{N}}/{\widetilde{h}_{S}} \sim 10^{2}$ again at frequencies of $100 \; {\rm{Hz}}$. For space-based gravitational-wave detectors, like the LISA mission ($L = 10^6 \; {\rm{km}}$), ${\widetilde{h}_{N}}/{\widetilde{h}_{S}} \sim 10^{1}$ at frequencies of $10^{-2} \; {\rm{Hz}}$. We cannot increase the frequency for the LISA detector, because its sensitivity curve rises sharply above $10^{-2} \; {\rm{Hz}}$ due to the response function of the instrument~\cite{Robson:2018ifk}. Therefore, in general one can expect that the Newtonian effect always dominates.

As we can see, the approximations made in this section allow us to carry out a straightforward comparison between the two effects and to better understand the physics underlying them. However, for the remainder of this work, we use the full strain model described by the equations presented in Secs.~\ref{sec:Newtonian_effect} and~\ref{sec:Shapiro_effect}.

\section{Dark Matter Clump Model}\label{sec:Dark Matter Clump Model}

In this section, we construct a DM clump model, where we neglect the Shapiro effect, since the Newtonian effect dominates. The DM clump model we use is described by Eq.~\eqref{eq:Newton2}, in which we assume the DM clump has spherical symmetry around its center, and it does not collide with the mirrors. This latter assumption sets an upper bound on the possible values the DM clump radius can take: when implementing the inference model we have to take into account that Eq.~\eqref{eq:Newton2} has a fictitious divergence at the mirrors; this divergence is fictitious because it would physically occur only if the DM clump had zero radius.  To avoid this, we set in the model a minimum distance to the mirrors that ranges from $10$m to $100$m in the different setups, and we report the minimum distance reached by the DM clump to any mirror along its trajectory.  For the solutions to be consistent with the model, the DM clump radius cannot be larger than this minimum distance.  Moreover, to have solutions that are physically reasonable, we also study the gravitational acceleration produced on the surface of these DM clumps with their maximum allowed radius, yielding accelerations of order ${\cal O}(10^{-6})$ g or less. Finally, an upper limit on the DM clump radius of about \( 4 \, \mathrm{km} \) arises naturally when we account for the fact that a large DM clump enveloping the detector would produce a negligible signal.

Under these assumptions, the DM clump is described only by its mass, $M_{DM}$, and its trajectory in time (i.e.,~its worldline), $\vec r_{DM}(t)$. Considering that the DM clump moves with a constant velocity 
$\vec v_{DM} (v_{DM},\theta,\phi)$, its trajectory is given by Eq.~\eqref{eq:DM_position1} with
\begin{align}
\vec{r}_{DMx}(t) &= {x}_{0}-v_{DM}\sin\theta\cos\phi \ (t-t_c) \ , \nonumber \\
\vec{r}_{DMy}(t) &= {y}_{0}-v_{DM}\sin\theta\sin\phi \ (t-t_c) \ , \nonumber \\  
\vec{r}_{DMz}(t) &= -v_{DM}\cos\theta \ (t-t_c) \ , 
\label{eq:DM_position1}
\end{align}
where $t_c$ is the crossing time of the DM clump at the plane of the detector, and $x_{0}$ and $y_{0}$ are its corresponding crossing coordinates. Therefore, the DM clump model has in total seven parameters that define the second derivative of the strain over time and consequently its Fourier transform,
\begin{align}
&\frac{d^2h_{N}}{dt^2} = \frac{d^2h_{N}}{dt^2}(t;\{M_{DM},x_0,y_0,v_{DM},\theta,\phi,t_c\})  \ , \nonumber \\
 &\widetilde{h}_{N}(f) = -\frac{1}{(2 \pi f)^2}
\widetilde{\frac{d^2h_{N}}{dt^2}}(f;\{M_{DM},x_0,y_0,v_{DM},\theta,\phi,t_c\}) \ , \nonumber \\  
&\widetilde{h}_{N}(f) = \widetilde{h}_{N}(f;\{M_{DM},x_0,y_0,v_{DM},\theta,\phi,t_c\}) \ .  
\label{eq:DM_d2h}
\end{align}
We refer collectively to these seven parameters as $\Theta = \{M_{DM},x_0,y_0,v_{DM},\theta,\phi,t_c\}$.

The right panel of Fig.~\ref{fig:trajectory_d2h.png} shows the second derivative of the DM clump model strain for a DM clump moving through the trajectory shown in the left panel of the same figure. The parameters of this realization of the DM clump model are 
$ M_{DM} = 10^8\,$kg, $x_0 = 2.3\,$km, $y_0 = 2.3\,$km, $v_{DM} = 10^2\,$km/s, $\theta = \pi/2$, $\phi = -\pi/4$, and $t_c = 0$. In the figure, different parts of the DM clump trajectory are colored, with each color representing different intervals of time, approximately limited by the roots of the second derivative of the strain. The DM clump starts approaching mirror $1$ from the region $x>4 \,$km, pushing mirror $1$ outward and generating a large negative second derivative of the strain as it passes close and in the direction of motion of the mirror. Then, when the DM clump reaches $x = 4 \,$km, the force applied on mirror 1 is perpendicular to its direction of motion, and therefore its acceleration --and hence the second derivative of the strain-- is reduced to zero. As the DM clump reaches the region $x < 4 \,$km, where it pushes mirror $1$ to the left and upward, it produces a positive second derivative of the strain, again with small relative intensity because of the angle between the force felt by the mirror and its direction of allowed motion.  When the DM clump reaches the point $y=x$, the second derivative of the strain is null because the force on all mirrors cancel their contribution to the second derivative of the strain (see Eq.~\eqref{eq:Newton1}). Finally, by symmetry, the same story is repeated but with opposite sign as the DM clump flies away from the symmetric $y=x$ point. 

\begin{figure}[t]
\includegraphics[width=\linewidth]{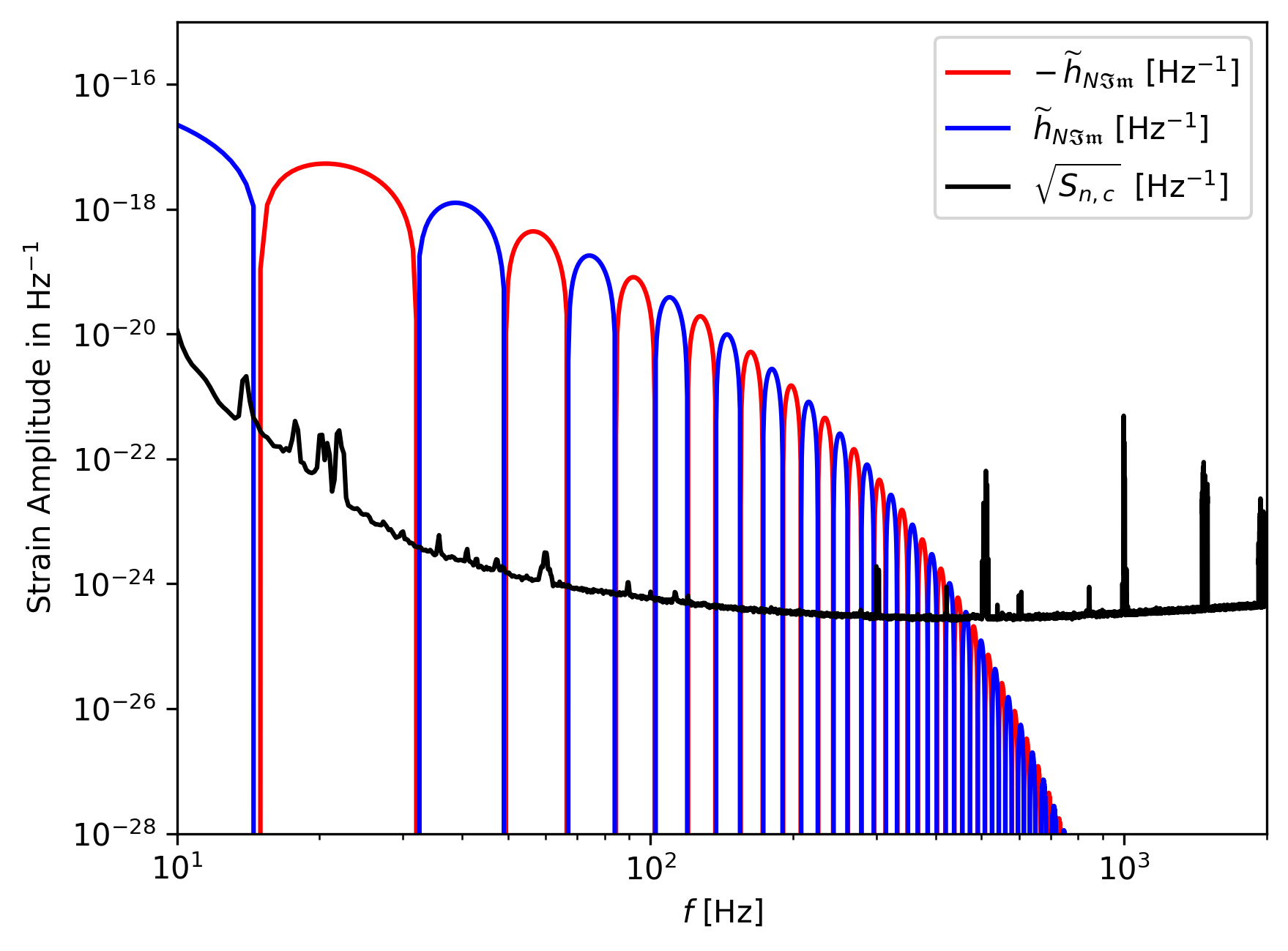}
\caption{Imaginary part of the Fourier transform of the strain produced by the DM clump realization shown in Fig.~\ref{fig:trajectory_d2h.png}. 
Negative values, which cannot be displayed in a log scale, have been sign-flipped and are shown in red; positive values are plotted in blue without modification.} 
\label{fig:fh_trajectory_strain.png}
\end{figure}

Figure~\ref{fig:fh_trajectory_strain.png} shows the imaginary part of the Fourier transform of the strain produced by the DM clump realization depicted in Fig.~\ref{fig:trajectory_d2h.png}. Since the second derivative of the strain for this DM clump is antisymmetric in time, its Fourier transform---and therefore the Fourier transform of the strain---is purely imaginary. Due to the impossibility of displaying negative values in a log scale, we have flipped the sign of the negative parts of the Fourier transform and colored them red, whereas the positive parts remain unchanged and are plotted in blue. Figure ~\ref{fig:trajectory_d2h.png} reveals that the structure formed by one of the major peaks together with its consecutive minor peak is approximately \( 0.05 \) seconds long. This implies that the Fourier transform of the strain exhibits positive maxima and negative minima approximately every \( 20 \; \mathrm{Hz} \). 
 
\section{Inferring a posterior distribution}\label{sec:improved_methods}

Given the data of a glitch, we would like to test the hypothesis that a DM clump passing through the detector is responsible for the glitch. To carry out this test, we will perform Bayesian parameter estimation to infer the posterior distribution of the DM clump model parameters, given some data set. In this section, we begin with a brief introduction of the statistical properties of the noise of the LVK detectors for completeness (following mostly~\cite{Maggiore2007}). Both the statistical properties of the noise and the DM clump model are crucial for constructing the posterior distribution of the model parameters and to test the above hypothesis. Finally, we demonstrate the construction of the posterior distribution using data of a given glitch, the statistical properties of the noise, and the DM clump model.

\subsection{The detector noise}\label{sec:detectors_noise}

In the absence of a signal, the detector is expected to exhibit stationary, Gaussian, stochastic noise, denoted as \( n(t) \), which is a real function of time. The stationary characteristic of the noise implies that the different components of the noise corresponding to different frequencies are uncorrelated \cite{Maggiore2007},
\begin{equation}
\langle \, \widetilde{n}(f) \ \widetilde{n}^*(f') \, \rangle  = \frac{1}{2} \, S_{n}(f) \, \delta(f-f') \ .
\label{eq:n(f)_2_and_Sn}
\end{equation}
In this equation, the average is taken over multiple possible realizations of the experiment, each with infinite time duration. The quantity \( S_{n}(f) \) is the noise power spectral density, mentioned in Sec.~\ref{sec:Comparison of the Newtonian and the Shapiro effect of a crossing DM clump}. This noise power spectral density can also be related to the Fourier transform of the correlation function \( C(\tau) = \langle n(t)\, n(t+\tau)\rangle \), where, as in the above equation, the average is taken over multiple possible realizations of the experiment \cite{Maggiore2007}.

In Eq.~\eqref{eq:n(f)_2_and_Sn}, the Dirac delta function arises because the derivation of the equation assumes an infinite number of realizations, each one infinitely long in time. However, in reality, there is only one realization, and it is not infinitely long. In practice, one divides this single realization into many segments, each with the same duration $T$, and treats each segment as an independent realization\footnote{Here, we have to ensure that the correlation between the different segments is negligible, which can be achieved by separating each segment from the others by some multiple of the autocorrelation length of the system.}. In this scenario, $\widetilde{n}(f)$ is defined as 
\begin{equation}
\widetilde{n}(f) = \int_{-T/2}^{T/2} dt \, n(t) \, e^{-i 2 \pi f t} \ ,
\label{eq:n_definition}
\end{equation}
and Eq.~\eqref{eq:n(f)_2_and_Sn} becomes
\begin{equation}
\langle \, \widetilde{n}(f) \ \widetilde{n}^*(f') \, \rangle = \frac{1}{2} \, S_{n}(f) \, \frac{\sin(\pi T (f-f'))}{\pi (f-f')} \ .
\label{eq:n(f)_2_and_Sn_T}
\end{equation}
Having this, we can compute the amplitude fluctuation of the noise at the specific frequency $f$,
\begin{align}
\langle \, |\widetilde{n}(f)|^2 \, \rangle  &= \frac{1}{2} \, S_{n}(f) \, T \label{eq:eq:n(f)_2_and_Sn_T_f} 
=  2 \, \langle \, \widetilde{n}_{\mathfrak{Re}}(f)^2 \, \rangle  = 2 \, \langle \, \widetilde{n}_{\mathfrak{Im}} (f)^2 
 \, \rangle \ ,
\end{align}
where subscripts $\mathfrak{Re}$ and $\mathfrak{Im}$ refer to the real and imaginary parts, respectively. Since the different components of the noise corresponding to different frequencies are randomly shifted in time, the symmetric and antisymmetric parts in a noise realization are statistically equivalent. This equivalence explains the equality, $\langle \widetilde{n}_{\mathfrak{Re}}(f)^2 \rangle  = \langle \widetilde{n}_{\mathfrak{Im}}(f)^2 \rangle$. 

Throughout this work, we use data segments that are two-seconds long ($T = 2$ seconds) with a sample rate of $4096$ Hz. This means the Fourier transform of each segment is computed using a definite integral, as shown in Eq.~\eqref{eq:n_definition}, with integration limits of $-T/2$ and $T/2$, respectively. Consequently, we obtain information about the Fourier transform of each segment with a frequency discretization of $0.5$ Hz, up to approximately $2000$ Hz, which corresponds to the Nyquist frequency.

The stationary Gaussian stochastic noise 
$n(t)$ can sometimes be accompanied by a signal 
$s(t)$, resulting in the total strain data $d(t) = n(t) + s(t)$. 
The presence of a signal can be discovered by studying the behavior of the arrays $\{D_{\mathfrak{Re}j}\}$ and 
$\{D_{\mathfrak{Im}j}\}$, defined via
\begin{align}
D_{\mathfrak{Re}j} = \frac{\widetilde{d}_{\mathfrak{Re}}(f_j)}{\sqrt{\frac{1}{4}S_{n}(f_j)T}} \ , \qquad D_{\mathfrak{Im}j} = \frac{\widetilde{d}_{\mathfrak{Im}}(f_j)}{\sqrt{\frac{1}{4}S_{n}(f_j)T}}\,.
\label{eq:D_randD_I}
\end{align}
where
\begin{align}
& f_{j} = j \times \Delta f \ , \ \ \ \  \Delta f = \frac{1}{T} \ , 
\label{eq:fjjDF}
\end{align}
with $j$ taking positive integer values. One can think of these quantities as square-root of the integral that defines $\rho_d^2$, the square of the SNR of the data, 
\begin{align}
\rho_d^2 &= 4 \int \frac{|\tilde{d}|^2}{S_n(f)} df = 4 \int \frac{[ \, (\tilde{d_{\mathfrak{Re}}})^2+(\tilde{d_{\mathfrak{Im}}})^2 \, ]}{S_n(f)} df\,.
\nonumber \\
&= 4 \sum_j \frac{[ \, \tilde{d_{\mathfrak{Re}}}(f_j)^2+\tilde{d_{\mathfrak{Im}}}(f_j)^2 \, ]}{S_n(f_j)} \Delta f\,,
\nonumber \\
&= \sum_j \left(D_{\mathfrak{Re} j}^2 + D_{\mathfrak{Im} j}^2\right)\,, 
\label{Dj}
\end{align}
where in the second to last line we converted the integral into a Riemann sum.
In the absence of a signal, $\{D_{\mathfrak{Re}j}\}$ and $\{D_{\mathfrak{Im}j}\}$ are expected to be uncorrelated and to follow a normal Gaussian distribution for each value of $j$. In the presence of a signal, of course, this is no longer true. 

\begin{figure}[t]
\includegraphics[width=1.0\linewidth]{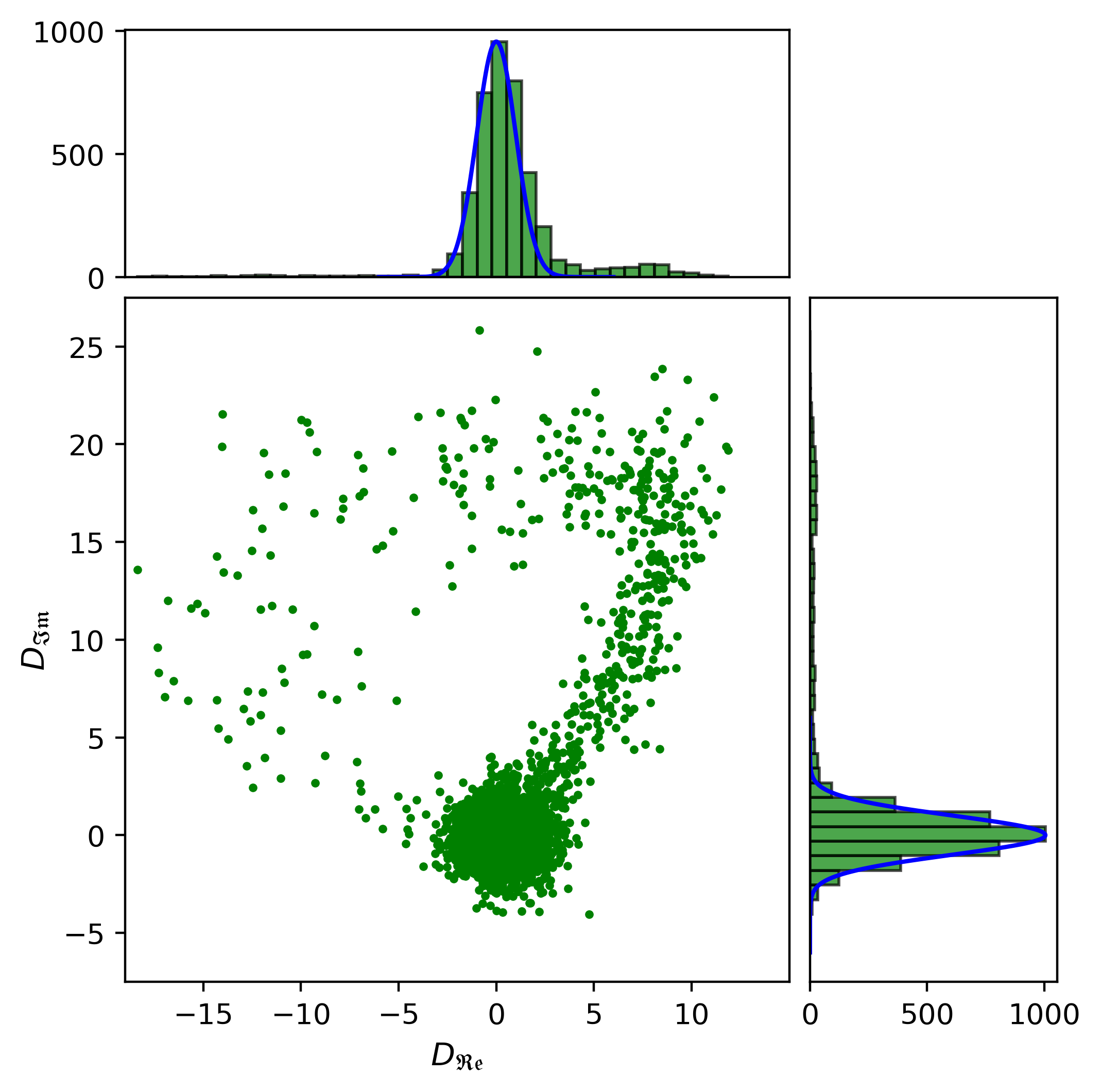}
\caption{Tracks of $D_{\mathfrak{Re}}$ vs $D_{\mathfrak{Im}}$, computed for a two-second-long segment of data containing the glitch that occurred in the Hanford detector at $T_g = 1174109651.84082$ seconds. Observe that almost all points fall around the circle centered at the origin, and follow the blue normal Gaussians, behaving as stationary, Gaussian, stochastic noise states. However, there are other points that fall outside the origin, indicating the presence of the glitch.}
\label{fig:normal_noise.png}
\end{figure}

Figure~\ref{fig:normal_noise.png} shows $\{D_{\mathfrak{Re}j}\}$ vs $\{D_{\mathfrak{Im}j}\}$, computed for a two-second-long segment of data centered around a real glitch that occurred in the Hanford detector at $T_{g} = 1174109651.84082 \; {\rm{secs}}\,$,
in GPS time. The $S_{n}(f)$ used here was computed by averaging thirty segments of data, each two seconds long, with these segments spanning from $T_{g} - 61$ seconds to $T_{g} - 1$ second. Since the duration of a glitch is generally much smaller than a second, the data spanning $T_{g} - 61$ seconds to $T_{g} - 1$ second is expected to contain only stationary, Gaussian, stochastic noise. Each green point in Fig.~\ref{fig:normal_noise.png} corresponds to a certain integer value of $j$. Observe that almost all points fall around the circle centered at the origin and follow the blue normal Gaussian distributions, exhibiting behavior corresponding to stationary, Gaussian, stochastic noise. However, there are some points that fall outside that circle, following an arch and indicating the presence of a signal, which in this case, is a glitch. 

Let us now study the hypothesis that this glitch originates from the passage of a DM clump. To test this, we combine the coefficients \(\{D_{\mathfrak{Re}j}\}\) and \(\{D_{\mathfrak{Im}j}\}\) with the DM clump model to construct the arrays \(\{\Delta_{\mathfrak{Re}j}\}\) and \(\{\Delta_{\mathfrak{Im}j}\}\):
\begin{equation}
\begin{split}
& \Delta_{\mathfrak{Re}j} = \frac{\widetilde{d}_{\mathfrak{Re}}(f_j)-\widetilde{h}_{N\mathfrak{Re}}(\Theta,f_j)}{\sqrt{\frac{1}{4}S_{n}(f_j)T}} \ , \\\\
& \Delta_{\mathfrak{Im}j} = \frac{\widetilde{d}_{\mathfrak{Im}}(f_j)-\widetilde{h}_{N\mathfrak{Im}}(\Theta,f_j)}{\sqrt{\frac{1}{4}S_{n}(f_j)T}}  \ .
\label{eq:D_randD_I_1}
\end{split}
\end{equation}
These arrays are essentially the square-root of the integrand of the log-likelihood, and thus, they measure the plausibility of the DM clump hypothesis; their characteristic behavior must be consistent with stationary, Gaussian, stochastic noise when the DM clump model is evaluated at the maximum of its posterior distribution. This posterior distribution is constructed based on the glitch data, the statistical properties of the noise, and the DM clump model, as we will discuss in the next subsection.

\subsection{Posterior Distribution}\label{sec:Likelihood}

Given the noise power spectral density \( S_n(f) \), the probability of obtaining a particular realization of stationary, Gaussian, stochastic noise, \( \widetilde{n}(f) \), is given by \cite{Maggiore2007}
\begin{equation}
P[\widetilde{n}(f)] = C \, \exp{ \int  - \, \frac{|\widetilde{n}(f)|^2 }{\frac{1}{2} \, S_{n}(f)} \, df } \ ,
\end{equation}
where \( C \) is a normalization constant. Under the assumption that the data is composed of stationary, Gaussian, stochastic noise and a glitch that is explained by the DM clump model, $\widetilde{d}(f) = \widetilde{n}(f) + \widetilde{h}_{N}(\Theta, f)$, we can obtain the probability of observing our data given a model with parameters $\Theta$, which one usually calls the likelihood, namely
\begin{equation}
\mathcal{L}(d|\Theta) = C \, \exp{ \int - \, \frac{|\widetilde{d}(f)-\widetilde{h}_{N}(\Theta,f)|^2 }{\frac{1}{2} \, S_{n}(f)} \, df } \ .
\label{eq:eq_likelihood}
\end{equation}

In the vernacular of gravitational wave literature, it is very common to express quantities in terms of the inner product \( (A|B) \), defined as
\begin{align}
(A|B) &= 2 \int \frac{\widetilde{A}(f)\widetilde{B}^*(f) + \widetilde{A}^*(f)\widetilde{B}(f)}{S_{n}(f)} \, df 
\nonumber \\
&= \mathfrak{Re} \int \frac{\widetilde{A}(f)\widetilde{B}^*(f)}{\frac{1}{4} S_{n}(f)} \, df \ .
\label{eq:inner}
\end{align}
In terms of this inner product, the likelihood can be written as
\begin{align}
\mathcal{L}(d|\Theta) &= C \, \exp{\Big[ \, -\frac{1}{2} \big( \, d-h_{N}(\Theta) \, \big| \, d-h_{N}(\Theta) \, \big) \, \Big]} \ , 
\nonumber \\
&\propto \exp{\Big[ \, \big( \, d \, \big| \, h_{N}(\Theta) \, \big) - \frac{1}{2} \big( \, h_{N}(\Theta) \, \big| \, h_{N}(\Theta) \, \big) \, \Big]} \ ,
\label{eq:eq_likelihood1}
\end{align}
where, in the last step, we only write down the part of the likelihood that depends on the model.

Finally, by combining the likelihood with the prior \( P(\Theta) \), we obtain the posterior distribution for the parameters of the DM clump model
\begin{equation}\label{eq:posterior}
\mathcal{P}(\Theta|d) =  \frac{\mathcal{L}(d|\Theta) P(\Theta)}{\int \mathcal{L}(d|\Theta) P(\Theta) \, d\Theta} \ .
\end{equation}
The priors correspond to uniform distributions along the following ranges:  $\ln \, ( \, \text{M} \times \text{kg}^{-1} \, ) \,\in ( \, \ln (10^4) \, , \,  \ln (10^9 ) \,) $; $x_0,\, y_0 \in(-4\,\text{km},8\,\text{km})$;  $\ln \, ( \, \text{v} \times (\text{m/s})^{-1} \, ) \,\in ( \, \ln (10^4) \, , \,  \ln (10^7 ) \, )$; $t_c \in (-0.04, 0.04) \, \text{s}$;  $\cos(\theta) \in (0,1)$; and $\phi$ in its full range.  In a few cases, when the posterior accumulates at the edge of the prior, we extend the range and produce more samples until the convergence is within the new limits. In practice, we always sample in the log of the likelihood.

\section{Model exploration and validation}
\label{sec:model-exploration}

In this section, we present {\tt BayesShip} \cite{Perkins:2021mhb}, the software we use throughout this work to infer posterior distributions. Furthermore, we introduce the metrics we use to determine whether a particular glitch could have been produced by the passage of a DM clump. Finally, we evaluate the performance of the entire procedure (posterior inference using {\tt BayesShip}, plus the computation of metrics) using two examples: one where we inject a DM clump signal into detector noise, and another where we inject a Cos-Gaussian signal assuming the same noise model. As we will show, our procedure is capable of ruling out a DM clump as the cause of a glitch, when the data is not produced by the passage of a DM clump. As a complement to this section, in Appendix~\ref{app:Degeneracies}, we discuss the degeneracies of the DM clump model. These degeneracies play a crucial role in the posterior inference process.

\subsection{Posterior sampling and {\tt BayesShip}}\label{sec:Likelihood sampling and BayesShip}

The posterior in Eq.~\eqref{eq:posterior} exists in a seven-dimensional parameter space, defined by the parameter vector $\Theta$, which characterizes our DM model.
As discussed in detail in Appendix~\ref{app:Degeneracies}, the posterior has strong degeneracies among various parameters, resulting in multi-modality in the distribution.
Sampling from this kind of distribution requires sophisticated algorithms, typically falling into two classes: nested sampling, which estimates the evidence directly, and Markov-chain Monte Carlo (MCMC), which stochastically drifts through the posterior surface along the parameter space via a Markovian process. 
As our primary focus is to draw samples of $\Theta$ from the posterior, and not to estimate the evidence, we work with a MCMC algorithm, as it can be more efficient when properly tuned.

The efficiency of MCMC sampling lies almost exclusively in the accuracy of the proposal distributions used to move through the parameter space.
The implementation used in this work utilizes the software built in~\cite{Perkins:2021mhb}, which includes two techniques to improve efficiency.
First, we use several types of proposals to increase the MCMC efficiency, beyond the standard symmetric Gaussian proposals. Namely, we use the technique of differential evolution to propose steps in parameter space and a Gaussian mixture model trained on past samples using {\tt Armadillo}~\cite{Sanderson2016,2018arXiv181108768S} to propose new steps in parameter space. 
Differential evolution works by proposing future points along vectors defined by connecting randomly selected past points in the past Markov chain. 
This allows the sampler to explore correlations and degeneracies in the parameter space efficiently, while ensuring that detailed balance is maintained asymptotically.

Secondly, we employ parallel tempering~\cite{PhysRevLett.57.2607,B509983H} to address the difficulty of sampling from a multi-modal distribution.
This method involves running multiple different samplers in parallel, with all but one chain sampling from a modified likelihood, defined by $\mathcal{L}(h|\Theta)^{1/\displaystyle \tau}$ for some $\tau \in[1,\infty)$.
These chains are said to be ``tempered'' and allow the easier sampling of various parts of parameter space. 
While the samples from these tempered chains cannot be used directly in any final analyses, because they are not samples from the true posterior, they can be used as proposals for the Markov chain exploring the posterior distribution of interest, the untempered distribution.
This is accomplished by periodically allowing different chains to ``swap'' positions in parameter space, while being careful to enforce detailed balance.

\subsection{Metrics to identify a possible DM clump signal in a glitch}\label{sec:Metrics}

In order to explore a DM clump model given a glitch, we proceed as follows. Given the glitch data, we infer the parameters of the DM clump model of Sec.~\ref{sec:Dark Matter Clump Model} by sampling the posterior distribution of Sec.~\ref{sec:Likelihood} with {\texttt{BayesShip}}, through a parallel-tempered MCMC algorithm. Once the inference is complete, we evaluate the DM clump model at the maximum \textit{a posteriori} (MAP) point of parameter space and compute the MAP residual SNR
\begin{align}
Q &= \left(d - h_N(\Theta_{MAP}) \left.\right| d - h_N(\Theta_{MAP})\right)\,,
\nonumber \\
& = 
4 \int_{f_{min}}^{f_{max}} \frac{\left|\widetilde{d}(f) -\widetilde{h}_{N}(f;\Theta_{MAP})\right|^2}{S_{n}(f)} df \,,
\label{eq:Q-int}
\end{align}
where $f_{min}$ and $f_{max}$ are some lower and upper limits of integration, respectively. Practically, we discretize on frequencies and therefore this integral is computed via summation, which can be written as
\begin{align}
Q &= \sum_{j=j_{min}}^{j_{max}} \left[\frac{\left( \widetilde{d}_{\mathfrak{Re}}(f_j)-\widetilde{h}_{N\mathfrak{Re}}(f_j) \right)^2}{\frac{1}{4}S_{n}(f_j)T} + \frac{\left( \widetilde{d}_{\mathfrak{Im}}(f_j)-\widetilde{h}_{N\mathfrak{Im}}(f_j) \right)^2}{\frac{1}{4}S_{n}(f_j)T}\right] \,,
\nonumber \\
Q &= \sum_{j=j_{min}}^{j_{max}}  \left(\Delta_{\mathfrak{Re} j}^2 + \Delta_{\mathfrak{Im} j}^2 \right)\,,
\label{eq:Q}
\end{align}
where $j_{min}$ and $j_{max}$ are the bin indices corresponding to $f_{min}$ and $f_{max}$. $\Delta_{\mathfrak{Re} j}$ and  $\Delta_{\mathfrak{Im} j}$ can be read from the above equations, but were also defined in Eq.~\eqref{eq:D_randD_I_1}. Note that the metric $Q$ is both the residual SNR at the MAP point of parameter space, and it is related to the logarithm of the likelihood evaluated at this point.

While the residual SNR is a meaningful start to assessing model performance, additional metrics of model accuracy can help to refine the sensitivity of our search. Fully Bayesian methods, such as calculating the Bayes' factors and posterior odds ratios, require the full specification of competing models. As a well-motivated glitch model driven by physics (and not by signal phenomenology) is not available, we choose to instead evaluate a test statistic on the residual, which does not require a full set of \emph{signal} models. This approach does assume knowledge of the underlying noise, but this aspect of the data is better understood than any signal that might be present. Because of the form of the likelihood and assumptions about the noise model of the detectors, the argument of the discrete sum in $Q$ at each frequency ($\Delta_{\mathfrak{Re} j}$ and  $\Delta_{\mathfrak{Im} j}$) is expected to follow a normal Gaussian distribution. Moreover, $Q$ itself is expected to follow a $\chi^2$ distribution, with $k = 2 \times (j_{max} - j_{min})$ degrees of freedom. If this occurs, we cannot rule out the DM clump hypothesis, and the glitch could have originated from a DM clump.

\begin{figure*}[t]
\includegraphics[width=0.496\linewidth]{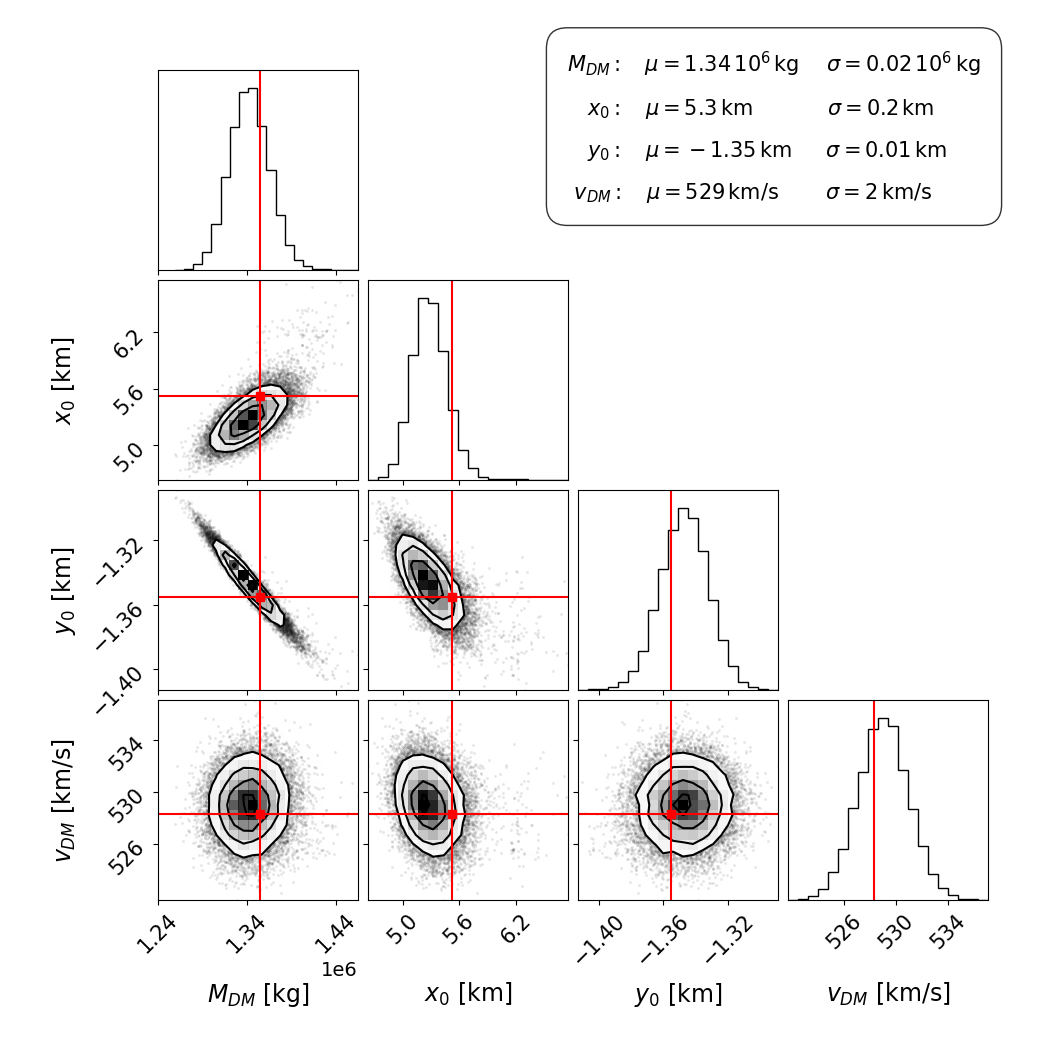}
\includegraphics[width=0.496\linewidth]{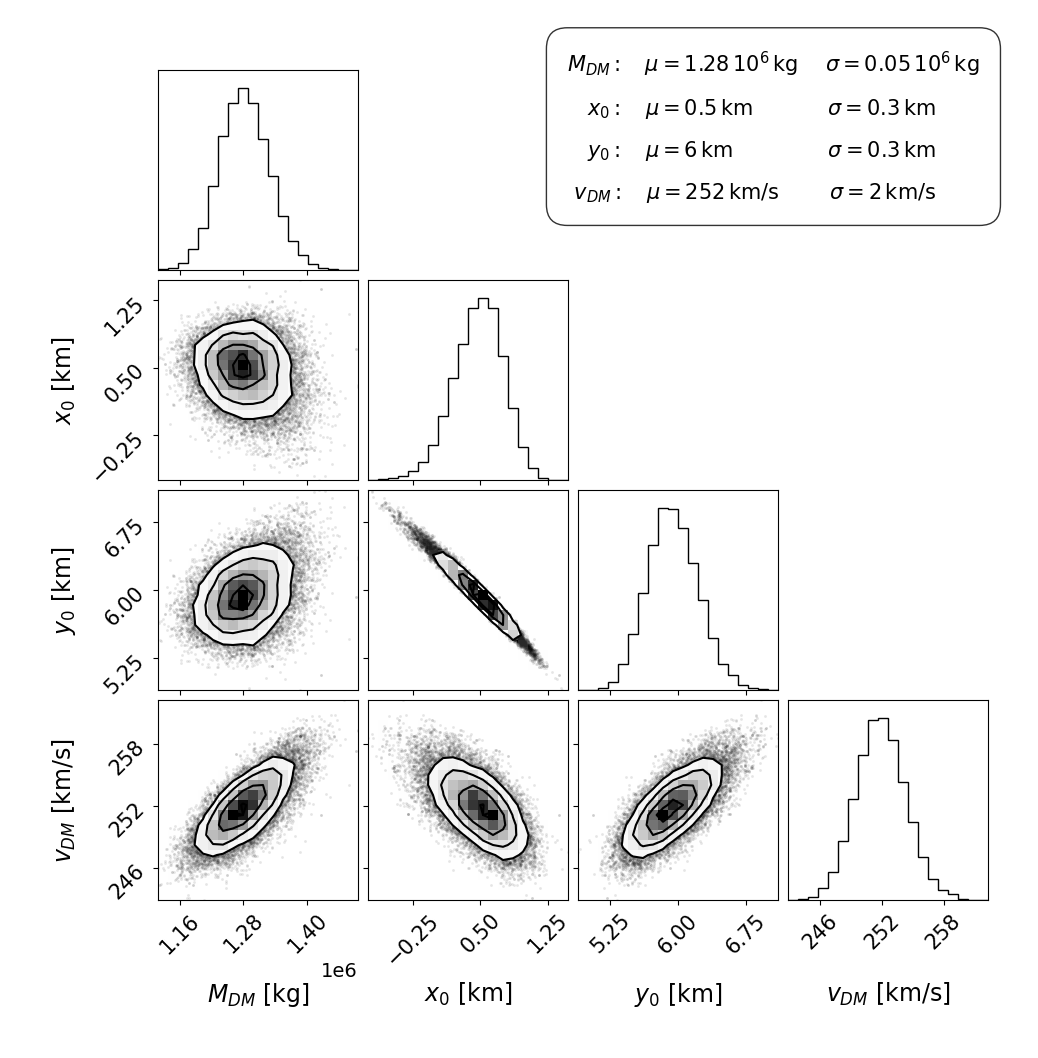}
\caption{Marginal posterior distributions obtained through Bayesian inference with {\tt{BayesShip}} using the DM clump model on a DM clump injected signal (left) and a Cos-Gaussian injected signal (right), with mean parameters and their 1$\sigma$ uncertainties shown in the legends.  In the left panel, the red lines indicate the true values of the parameters of the injected DM clump, showing a good agreement between these true values and the posterior distribution obtained. The DM clump injected signal yields \( \mathcal{S} = 0.8 \) when analyzed in its selected frequency range. Therefore, a second inference over the full frequency range was performed, and we present the marginal posterior distribution resulting from this second inference. In contrast, the Cos-Gaussian injected signal yields \( \mathcal{S} = 56.8 \) in its selected frequency range, and thus we present the marginal posterior obtained from the inference in its selected frequency range.} \label{fig:corner_clump_cos.pdf}
\end{figure*}

We will use the $Q$ quantity to assess how well the DM clump model fits the data by computing the metric $\mathcal{S}$, defined as
\begin{equation}
\mathcal{S} = \frac{|Q - k|}{\sqrt{2k}} \ .
\end{equation}
Since $k$ is the mean and $\sqrt{2k}$ is the standard deviation of the corresponding $\chi^2$ distribution, the metric $\mathcal{S}$ quantifies the deviation of $Q$ from the mean in units of standard deviations. If the DM clump model is a poor fit of the data, $\mathcal{S}$ should be a ``large'' number, and in such cases we can discard the DM clump hypothesis. Conversely, when ${\cal{S}}$ is ``small'' we cannot discard the hypothesis that the glitch was caused by a DM clump. In this paper, we adopt a threshold value of ${\cal{S}}$ above which we shall assume the glitch was not produced by a DM clump, i.e.~when ${\cal{S}} > {\cal{S}}_{\rm threshold} = 5$ the data does not support the hypothesis that the glitch was produced by a DM clump, and thus, we shall discard it. Doing so is similar to saying that we discard the glitch because its signal is $5\sigma$ away from what would be expected, if it would have been produced by a DM clump.

Another metric we can use to assess the similarity of a glitch to a given DM clump is the arccosine of the fitting factor ($FF$), defined as
\begin{equation}
\mathcal{A} = \arccos\left(FF\right) = \arccos\left(\frac{(h_{N}|d)}{\sqrt{(h_{N}|h_{N})(d|d)}}\right) \,,
\label{eq:alpha}
\end{equation}  
where $h_{N}$ is evaluated at the MAP point. The quantity $\mathcal{A}$ defines an angle that ranges in $[ \, 0 \, , \, \pi/2 \, ]$. In the hypothetical case where the glitch corresponds to a DM clump, we expect $\mathcal{A}$ to be small, with $\mathcal{A} \ll \pi/2$. For example, for a model with $FF=0.999$, we have that $\mathcal{A} \sim 0.044$, while if the model has $FF = 0.99$, we then have $\mathcal{A} \sim 0.14$.

We observe that in the definition of the fitting factor (and hence $\mathcal{A}$) in Eq.~\eqref{eq:alpha}, if the model is much smaller than the noise above certain frequency, then the remaining of the integrals corresponds to $h_{N}$ being very small and $d$ being stochastic noise.  This is the case even if the model matches the data.  Therefore, in this range of frequencies the $FF$ decreases because the denominator becomes larger than the numerator.  However, this decrease is not an indicator of a disagreement between model and data; rather, it is just a feature of the $FF$ in regions where the model prediction is much below the noise.  We analyze this behavior in Appendix~\ref{app:FF-behavior}.  In order to improve $\mathcal{A}$ as an assessment of the matching between model and data, we define the inner products in Eq.~\eqref{eq:alpha} restricted to be within the range of frequencies $f_{min}$ and $f_{max}$ in which the signal is above the noise, as defined below.  On the other hand, the metric $\mathcal{S}$ represents the significance of a $\chi^2$ distribution, which, by definition, takes into account the noise bins and the expectation of how much they add up to, as a function of the number of degrees of freedom\footnote{It is true that adding too many spurious noise bins to a $\chi^2$ at some point can artificially improve the matching to a data set. We analyze this behavior below.}.

Let us conclude this subsection by describing our procedure to assign a metric to each glitch, while reducing the load of the numerical methods.  Since ${\cal S}$ can discard a glitch as a possible DM clump (but they cannot confirm it), we then first compute this metric on a reduced, selected, frequency range, defined by $f_{min}$ and $f_{max}$, inside which the glitch has a larger excess above the noise and the metrics are expected to be more sensitive.  This allows us to discard about 2/3 of the glitches, and then perform the full calculation using all the available frequencies (10-1600 Hz) on the remaining 1/3 of the glitches. To this end, we should define the recipe for the more sensitive, reduced, frequency range, and the limits we set to say that a glitch can be discarded as a DM clump.

Given the full span of frequencies, we select $ f_{\text{min}} $ and $ f_{\text{max}} $ to be the minimum and maximum frequencies of a set on which any of the coefficients $D_{\mathfrak{Re}}$ or $D_{\mathfrak{Im}}$ all lie above the \( 2\sigma \)-level of the noise in the ranges $( f_{\text{min}}, f_{\text{min}} + Df )$ and $( f_{\text{max}} - Df, f_{\text{max}} )$, respectively, with  $Df = 10 \, \text{Hz}$. From Eq.~\eqref{eq:fjjDF}, we see that $f_{\text{min}} = j_{\text{min}} \times \Delta f$ and $f_{\text{max}} = j_{\text{max}} \times \Delta f$. We infer the posterior distribution of the DM clump parameters, using this selected frequency range for all the glitches in the original Koi-Fish dataset. Then, based on the MAP parameters obtained from this inference, we compute the metric \( \mathcal{S} \) restricted to the same frequency range, and discard all cases with \( \mathcal{S} > \mathcal{S}_{\mathrm{threshold}} = 5 \). As we will see, this allows us to discard 51 out of 84 glitches. For the remaining glitches, we perform a second posterior inference using the full frequency range (\( 10\text{-}1600 \, \mathrm{Hz} \)), and using the MAP parameters obtained from this second inference, we compute the metric \( \mathcal{S} \) over this full frequency range. As mentioned before, we compute and present \( \mathcal{A} \) in the selected frequency range, using the MAP parameters from the second inference (when a second inference is performed), or from the first inference otherwise.

When performing the analysis in the full frequency range, we find small variations in the metric ${\cal S}$ of ${\cal O}(1)$ in comparison to their value in the reduced frequency range.   In the results presented below, we indicate the metric of each glitch according to whether they were discarded or not using their corresponding selected frequency range analysis.  For those discarded in the first pass, we use the metrics of the first pass. For those that were not discarded in the first pass, we use the ${\cal S}$-metric that yields the largest value between the first pass and the second pass with the full frequency range.  The reason for this is that ${\cal S}$ can only be used to discard a glitch if its result is not good in any predetermined frequency range, and hence, we use its worst value. This procedure might result in a slight mismatch when mixing the two glitch populations. However, this is not a major concern because glitches with ${\cal S} > {\cal{S}}_{\rm threshold} = 5$ are already excluded from our analysis, and this trade-off significantly reduces the computational load.

\subsection{Testing the metrics through injection of DM clump and Cos-Gaussian signals}\label{sec:DM Clump and Cos-Gaussian}

Let us now assess the reliability of the metrics $\mathcal{S}$ and $\mathcal{A}$ to test the hypothesis that the glitch data is produced by a DM clump model. If these metrics are effective, we would expect $\mathcal{S}$ and $\mathcal{A}$ to exhibit significantly different behaviors for data sets containing a DM clump signal compared to those containing other types of glitch-like, non-DM-related signals. In particular, one would expect $\mathcal{S} = {\cal{O}}(1)$ and $\mathcal{A} \ll \pi/2$, if the data set is composed of a DM clump signal. Conversely, one would expect $\mathcal{S} \gg 1$ and $\mathcal{A} = {\cal{O}}(\pi/2)$, if the data cannot be explained by the DM clump model.

\begin{figure*}[t]
\includegraphics[width=8.75cm, keepaspectratio]{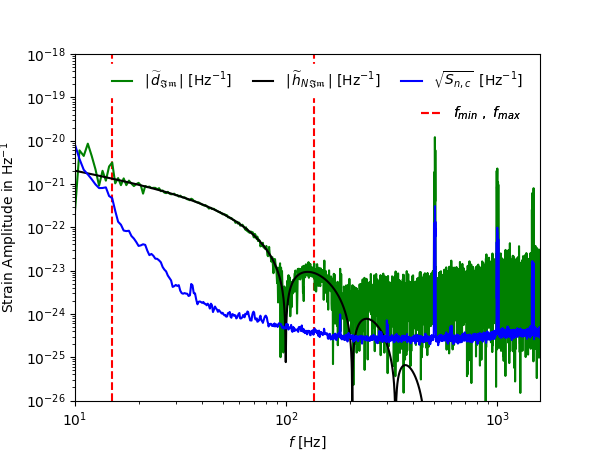}
\includegraphics[width=8.75cm, keepaspectratio]{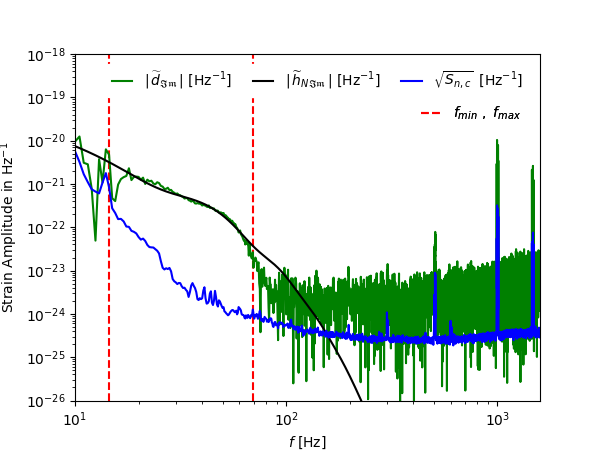}
\includegraphics[width=8.75cm, keepaspectratio]{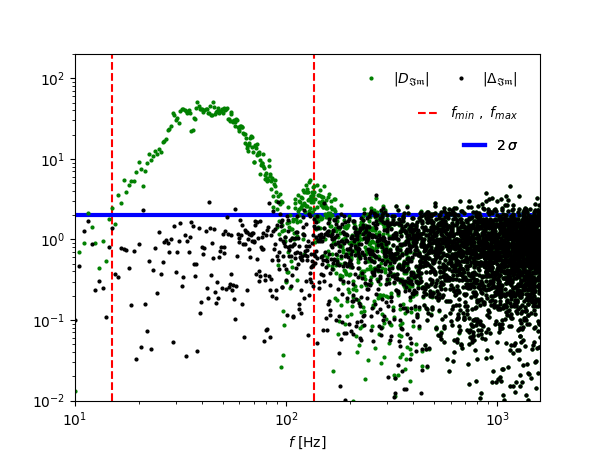}
\includegraphics[width=8.75cm, keepaspectratio]{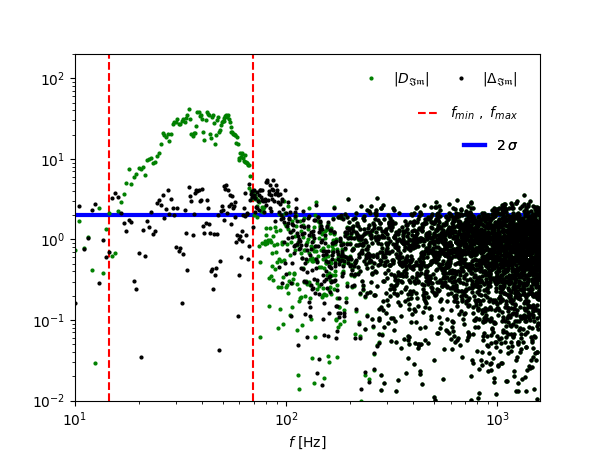}
\caption{
Left: DM-clump injected signal; Right: Cos-Gaussian injected signal.  Top: Absolute value of the imaginary part of the Fourier transform of the data and MAP; Bottom: Coefficients $|D_{\mathfrak{Im}j}|$ and $|\Delta_{\mathfrak{Im}j}|$ evaluated at the MAP (see Eqs.~\eqref{eq:D_randD_I} and \eqref{eq:D_randD_I_1}). The DM-clump injected signal (left) yields ${\cal S}=0.8$ when analyzed in the selected frequency range, and therefore we extend its analysis to the full frequency range yielding ${\cal S}=0.15$. Whereas the Cos-Gaussian injected signal (right) yields ${\cal S}=56.8$ in its selected frequency range ($f_{min}$-$f_{max}$) and therefore it does not need to be analyzed in the full frequency range.   Observe in the bottom panel the residual plot and how the black points in the DM-clump injected (left) are considerably below the 2$\sigma$ noise level (blue line), whereas for the Cos-Gaussian injected they emerge above the $2\sigma$ noise level, even for the MAP of the inference process.  We find that even though both signals have a good convergence in the DM clump model, these residual plots and their corresponding metrics are good indicators to discard a possible DM clump.  Observe that $D$ and $\Delta$ coefficients are dimensionless (see Eq.~\ref{Dj}), hence the vertical axis in the lower panel has no units.}
\label{fig:Residual_I_clump_cos.png}
\end{figure*}

In order to conduct this test, we create two different sets of synthetic data injections: one containing a DM clump signal and the other containing a Cos-Gaussian signal. The latter is a linear combination of cosine functions multiplied by exponentially decaying functions in the time domain, namely
\begin{equation}
h_{C-G}(t) = \sum_{j=1}^{2} A_{j} \, \cos \, \left(2 \, \pi \, f_{j} \, t + \phi_{j}\right) \, e^{-\left(t/T_{Cj}\right)^2}  \ .
\end{equation}
The Cos-Gaussian injection has parameters 
\allowdisplaybreaks[4]
$(A_{1}, A_{2}) = (3.075, 3.113) \times 10^{-17}$, 
$(f_{1}, f_{2}) = (17.864, 17.718)$ Hz, 
$(\phi_{1}, \phi_{2}) = (0.712, 3.854)$ radians,
and $(T_{C1}, T_{C2}) = (0.0149, 0.01484)$ seconds. The DM clump injection has parameters 
$(M_{DM}, v_{DM}) = (1.35 \, 10^6 \mbox{ kg},528\mbox{ km/s})$, 
$(x_0,\,y_0) = (5.5, -1.3)$ km, 
and $(\theta,\phi)=(1.3, 6.2)$ radians. Both signals are injected into synthetic Advanced LIGO detector noise, generated using the power spectral density computed between $( T_{g} - 61 )$ seconds and $( T_{g} - 1 )$ seconds, where $T_g$ refers to the time of a real glitch. That is, we use noise from real data around a real glitch, but we don't include the signal from the real glitch in the analysis, and instead inject either a DM clump or a Cos-Gaussian artificial signal.  Using {\tt BayesShip}, we then find their corresponding posteriors, and we compute their corresponding metrics $\mathcal{S}$ and $\mathcal{A}$. 

Figure~\ref{fig:corner_clump_cos.pdf} shows the inferred marginal posterior distributions for the parameters \( M_{DM} \), \( x_{0} \), \( y_{0} \), and \( v_{DM} \), corresponding to the DM clump injected signal on the left panel and the Cos-Gaussian injected signal on the right panel (see Appendix ~\ref{app:Completed_corner_plots} for the full corner plots). In the left panel, the red lines represent the true (injected) values of the parameters used to create the DM-clump injected signal, showing agreement between these values and the posterior distributions found with {\tt BayesShip}. The left panel then clearly shows that the DM-clump model can recover a DM-clump signal, but the right panel shows that the former can even recover a Cos-Gaussian signal. Therefore, an investigation of only the marginalized posteriors of a given model are not enough to determine whether the signal was generated by that model.

The metrics to assess the agreement of the data with the model, however, yield different diagnosis depending on whether the signal is a DM clump or a Cos-Gaussian.   When the data is an injected DM clump signal, we find that $\mathcal{S} = 0.8$ and $\mathcal{A} = 0.02$, while when the data is an injected Cos-Gaussian signal, we find that $\mathcal{S} = 56.8$ and $\mathcal{A} = 0.1$.

The differences in the values of $\mathcal{S}$ and $\mathcal{A}$ can be easily explained by examining Fig.~\ref{fig:Residual_I_clump_cos.png}.  In the top-row plots, we show the absolute value of the imaginary part\footnote{we find very similar results for the real part of the Fourier transform of the data, so we only show the imaginary part here.} of the Fourier transform of the data and of the MAP DM clump model, for both the DM clump injected signal (left panel) and the Cos-Gaussian injected signal (right panel). Very few discrepancies can be observed between the absolute value of the imaginary part of the Fourier transforms in the DM clump injected signal case, whereas discrepancies are larger in the Cos-Gaussian injected signal case, especially near the red vertical lines, which show $f_{\text{min}}$ and $f_{\text{max}}$ (see Sec.~\ref{sec:Metrics}).  These discrepancies are magnified in the bottom row of the figure, which shows the absolute value of the coefficients \(\{D_{\mathfrak{Im}j}\}\) and \(\{\Delta_{\mathfrak{Im}j}\}\), evaluated at the MAP DM clump model (see Sec.~\ref{sec:detectors_noise}). Observe the larger number of black points above the blue horizontal line ($2\sigma$-level of the noise) in the case of the Cos-Gaussian injected signal, especially between the red dashed lines. This indicates that the Cos-Gaussian data set does not reduce to stationary, Gaussian, stochastic noise after the DM clump model subtraction. 

This synthetic experiment is intended to demonstrate the usefulness of the metrics $\mathcal{S}$ and $\mathcal{A}$ for determining whether a certain data set containing a glitch could be produced by a DM clump or not. However, while we have \( \mathcal{S} \gg 1 \), as expected, for the Cos-Gaussian injected signal, we have \( \mathcal{A} \ll \pi/2 \) in that case, even though, we have $\mathcal{A}$ significantly lower for the DM clump injected signal. Therefore, the results obtained in this synthetic experiment suggest that the metric \( \mathcal{A} \) is not as sensitive as the metric \( \mathcal{S} \) for detecting non-Gaussian deviations of the data from the model. We have studied the behavior of these metrics in other similar synthetic experiments, as well as through toy models and across the set of real glitches presented in this work, reinforcing this conclusion. Therefore, from this point forward, we discard the DM clump hypothesis based solely on the analysis of the \( \mathcal{S} \) metric, and report the values of \( \mathcal{A} \) only as complementary information.

\section{Bayesian Parameter Estimation Results on a set of real glitches}
\label{sec:Results on a sample of glitches}

In this section, we apply the Bayesian inference procedure (posterior inference using {\tt{BayesShip}}, followed by the computation of the $\mathcal{S}$ and $\mathcal{A}$ metrics) described in the previous section to a real dataset consisting of 84 Koi-Fish glitches from the LIGO Hanford detector \cite{Glanzer_2023,gravityspy}. The motivation for studying this glitch category is discussed in Sec.~\ref{sec:Glitches at LIGO}.

\begin{figure}[t]
\includegraphics[width=1.0\linewidth]{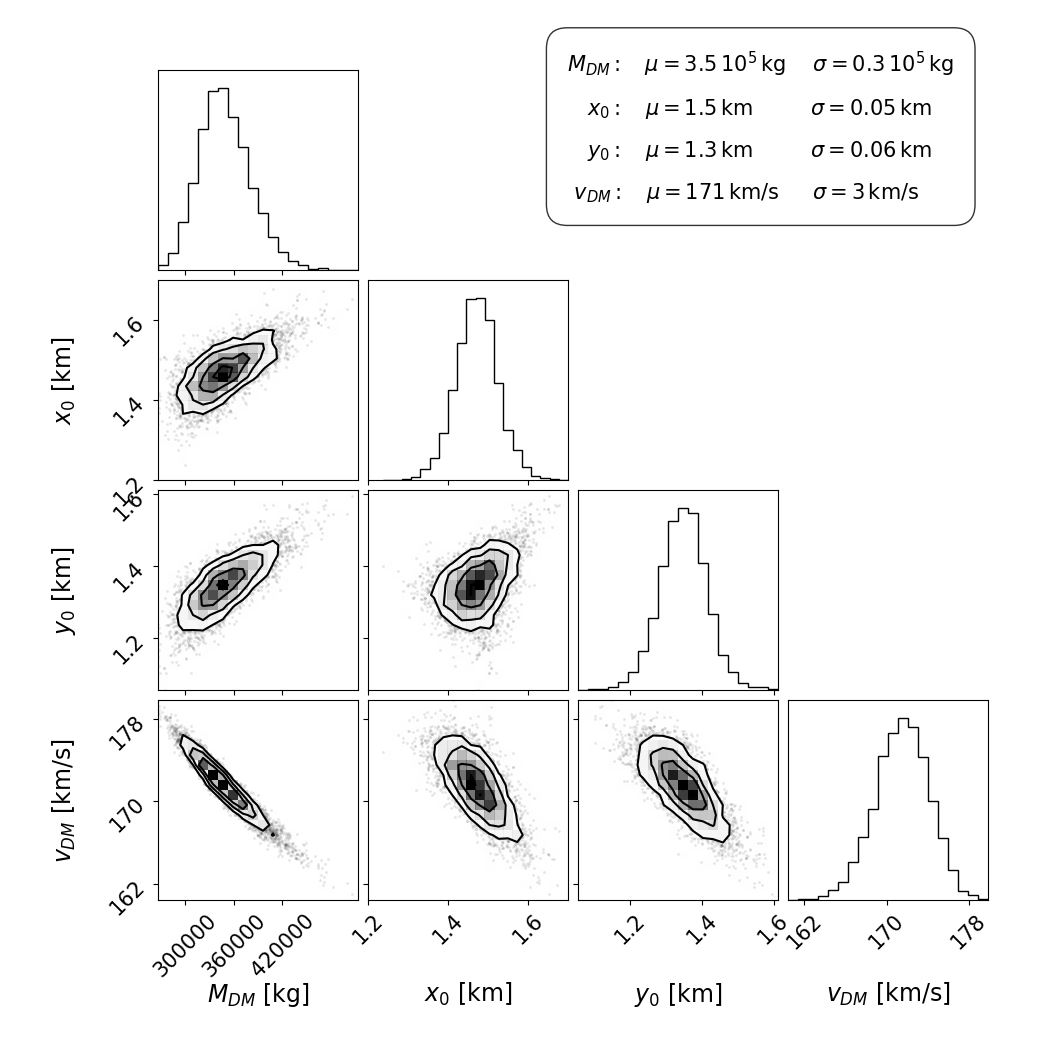}
\caption{Marginal posterior distribution obtained after performing Bayesian inference using the DM clump model on real data that contains the Koi-Fish glitch that occurred at Hanford at $T_g$ (see text). The legend indicates the mean parameter values and their uncertainties that result from the inference study.  The posterior chains have good convergence, and the marginalized posterior distributions are approximately Gaussian.}\label{fig:corner_1183157056.15234.png}
\end{figure}

\begin{figure*}[t]
\includegraphics[width=0.497\linewidth]{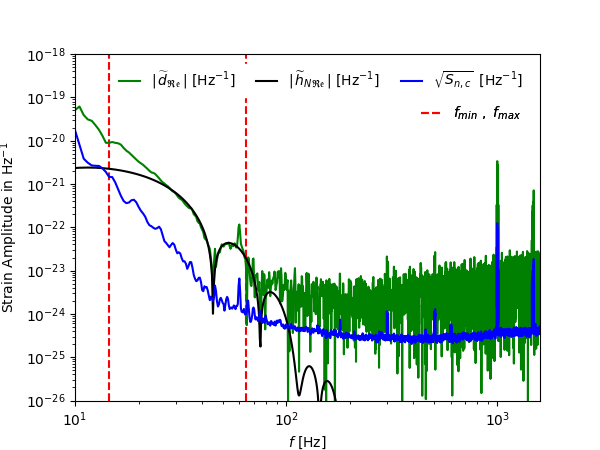}
\includegraphics[width=0.497\linewidth]{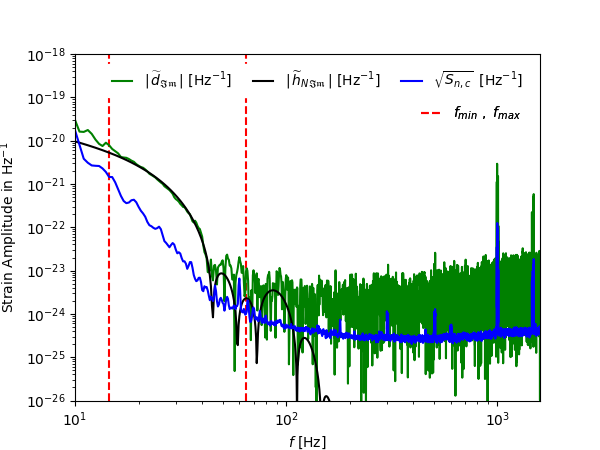}
\includegraphics[width=0.497\linewidth]{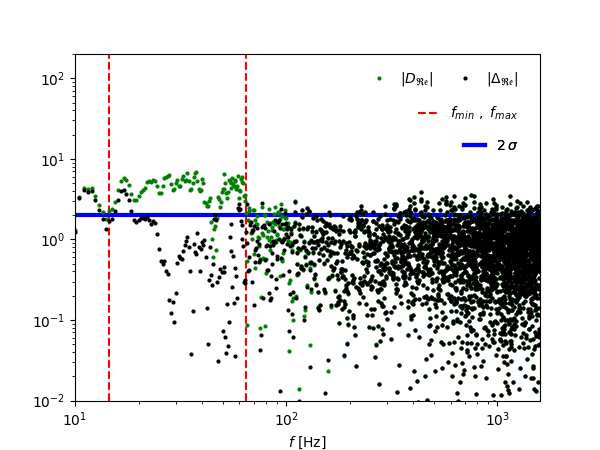}
\includegraphics[width=0.497\linewidth]{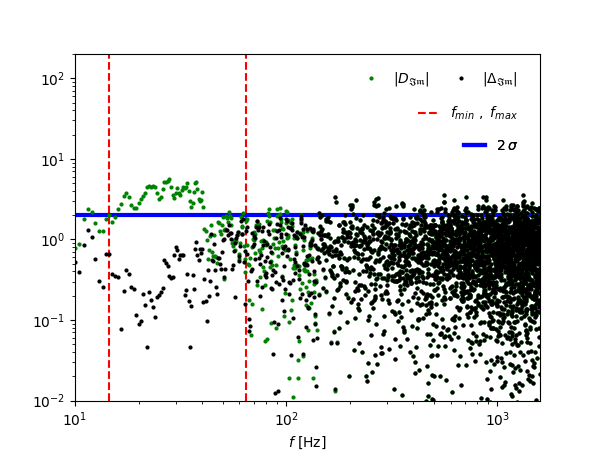}
\caption{
Koi-Fish glitch at $T_g$ (see text) analyzed with the DM-clump model.  Top: Absolute value of the Fourier transform of the data and the model at the MAP point.  Bottom: Coefficients $D$ and $\Delta$ evaluated at the MAP point (see Eqs.~\eqref{eq:D_randD_I} and \eqref{eq:D_randD_I_1}).  Left and Right: real and imaginary parts, respectively.   As it can be seen from the plots, the DM-clump MAP has good agreement with the data, yielding good metrics.  In this case we have ${\cal S} = 2.8$ and ${\cal A}=0.28$, which cannot discard it as a DM clump (and neither confirm it). Observe that $D$ and $\Delta$ coefficients are dimensionless (see Eq.~\ref{Dj}), hence the vertical axis in the lower panel has no units.}
\label{fig:Residual_1183157056.15234.png}
\end{figure*}

We start this section by describing in detail the results of the Bayesian inference procedure applied to one of the Koi-Fish glitches in our sample, in order to show the details of the calculations.  We then complete the section with a global description of the results of the Bayesian inference procedure applied to the entire sample. 
This global description allows us to  explore interesting implications if we assume the final set of glitches that are consistent with being DM clump. 
As we show below, this assessment is not conducted solely through the study of the metrics $\mathcal{S}$ and $\mathcal{A}$, but also through the examination of the trajectories, crossing points and velocities the Bayesian inference procedure reveals for the hypothetically inferred DM clumps.

\begin{figure*}[t]
\includegraphics[width=0.495\linewidth]{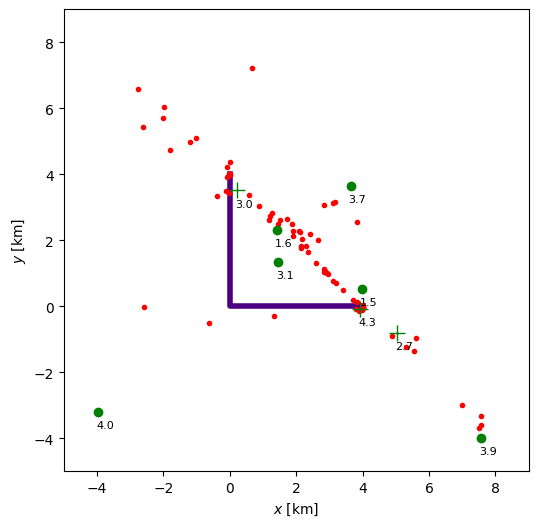}
\includegraphics[width=0.495\linewidth]{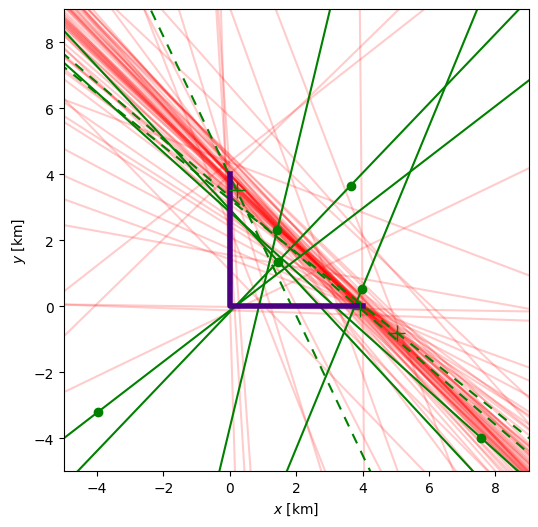}
\caption{Left: Crossing MAP positions of hypothetical DM clumps inferred from the Koi-Fish glitch sample. Right: Projections of the MAP trajectories over the detector's plane. We color in red those DM-clump MAP models with ${\cal S}>5$ and in green those with ${\cal S}<5$.   Among the latter, we use cross and dashed lines to indicate those passing closer than 200 m to a mirror. The numbers on top of the green points and crosses in the left panel correspond to their ${\cal S}$ value.}\label{fig:xy.png}
\end{figure*}

\subsection{Inference procedure and analysis of a single glitch}\label{sec:One Case in Detail}

We pick an example Koi-Fish glitch that occurred in the Hanford detector at $T_g = 1183157056.15234$ seconds in GPS time as our data and carry out Bayesian parameter estimation with our DM clump model. This particular glitch yields ${\cal S}<5$ in the selected frequency range; henceforth, we use the procedure with the full frequency range. Figure~\ref{fig:corner_1183157056.15234.png} shows the inferred posterior distributions for the parameters \( M_{DM} \), \( x_{0} \), \( y_{0} \), and \( v_{DM} \) that results from our Bayesian analysis in the full frequency range (see Appendix ~\ref{app:Completed_corner_plots} for the full corner plot).
Evaluating the DM clump model at the MAP point resulting from this full frequency range  inference, we obtain $\mathcal{S} = 2.8$ using the full frequency range\footnote{This glitch yields a slightly more sensitive ${\cal S}$ using the selected frequency range, which according to the procedure described in Sec.~\ref{sec:Metrics}, is the one used in Table \ref{tab:fullwidth}.} and $\mathcal{A} = 0.28$ using the selected frequency range. Figure~\ref{fig:Residual_1183157056.15234.png} shows in the upper panels the amplitude of the real and imaginary parts of
the Fourier transform of the data, the MAP DM clump model, and the square root of the characteristic spectral noise density computed for
this event.  The lower panels show the $D$ and $\Delta$ coefficients from Eqs.~\eqref{eq:D_randD_I} and \eqref{eq:D_randD_I_1}. In both figures, we can see good agreement between the data and the DM-clump MAP model, explaining the low values obtained for $\mathcal{S}$ and $\mathcal{A}$.

Given the above statistical results, we cannot reject the hypothesis that this glitch was produced by the passage of a DM clump. If this hypothesis is correct, the DM clump would have had a mass of approximately $3.5 \times 10^5 \ \mathrm{kg}$, and it would have passed through the point $(x_{0}, y_{0}) \approx (1.5 \ , \ 1.3) \ \mathrm{km}$ with a velocity of approximately $170 \ \mathrm{km/s}$. Furthermore, as explained in Sec.~\ref{sec:Dark Matter Clump Model}, the DM clump cannot have a radius much larger than the size of the detector. Taking this into account, and assuming a uniform DM clump density, we find a DM clump density of about $10^{-7}\, \text{g/cm}^3$. This same calculation can be repeated with other radii, bearing in mind that the Schwarzchild radius for this mass is $\sim 10^{-21}$ m.

Before proceeding, a word of caution is in order. The above analysis suggests that the DM clump model is a good fit to the data, but it does not tell us whether the data prefers the DM clump model over another model that represents an instrumental glitch. In order to address the second question, one would have to carry a model-selection analysis to calculate the Bayes factor of the two models given the data. Unfortunately, instrumental glitch models are still being developed, and the search for additional models and glitch explanations has not yet been exhausted. Therefore, a model-selection study of the DM clump model versus all other possible instrumental glitch models is beyond the scope of this paper. Given this, our analysis is not to be interpreted as suggesting that this given glitch was caused by the passage of a DM clump, i.e.~showing that the data cannot reject this hypothesis is \textit{not} the same as showing that the data supports this hypothesis over all others.

\subsection{Bayesian Inferences on the set of glitches}\label{sec:improved_methods_implementation}

We have performed the same Bayesian inference procedure described in the previous subsection to our entire sample of Koi-Fish glitches from Hanford detector.  We find that 33 out of the 84 original Koi-Fish glitches pass the threshold test ${\cal S}<{\cal{S}}_{\rm threshold} = 5$ in the reduced frequency range, and 9 out of these 33 also pass the test in the full frequency range. The DM-clump MAP for each glitch, as well their ${\cal S}$ and ${\cal A}$ metrics that assess their corresponding agreement to the model, can be found in the Appendix \ref{MAPtable}. In this and the following sections, as well as in Appendix~\ref{MAPtable}, we present the DM clump MAP parameters and the \( \mathcal{S} \) and \( \mathcal{A} \) metrics as follows: if a second inference in the full frequency range was required, we report the MAP resulting from that inference; otherwise, we report the MAP resulting from the inference in the selected frequency range. The metrics \( \mathcal{S} \) and \( \mathcal{A} \) are defined according to the procedure described in Sect.~\ref{sec:Metrics}. Fig.~\ref{fig:xy.png} shows the crossing point of the detector plane (left) and their trajectory projected to this plane (right) for each DM-clump MAP model for all the glitches studied.  We have colored with green those glitches with ${\cal S}<5$, and we have added their metric ${\cal S}$ values in the left panel. In the following paragraphs, we discuss the properties of the DM-clump MAP models, but first let us discuss a potential systematic uncertainty in our procedure.

As we can see in Fig.~\ref{fig:xy.png}, there is a preference of trajectories that pass near the mirrors, and most of the time, with a direction along the diagonal between the outer mirrors.  From Table \ref{tab:fullwidth} in Appendix \ref{MAPtable}, we can also see a preference for trajectories with a polar angle $\theta \sim \pi/2$ (low-level flights).  The reason behind this is connected to the discussion in Sec.~\ref{sec:Dark Matter Clump Model}, where we described how a DM clump passing close to a mirror can easily produce a sudden and short strain with a pattern similar to that of a Koi-Fish glitch.  This drives us to the question of whether there is a selection bias when choosing to analyze glitches in the Koi-Fish category that may be biasing our conclusions. The answer to this question is not straightforward, and thus, it should  be addressed in a different dedicated article, since it is beyond the scope of this work.  One approach would be to repeat the work in this article but with a randomized sample of glitches, which would wash out any potential bias. For the purpose of this work, we will continue to analyze the results we obtained, bearing in mind that there may be a potential bias originating in the glitch-category selection.

\begin{figure}[t]
\includegraphics[width=0.355\textheight]{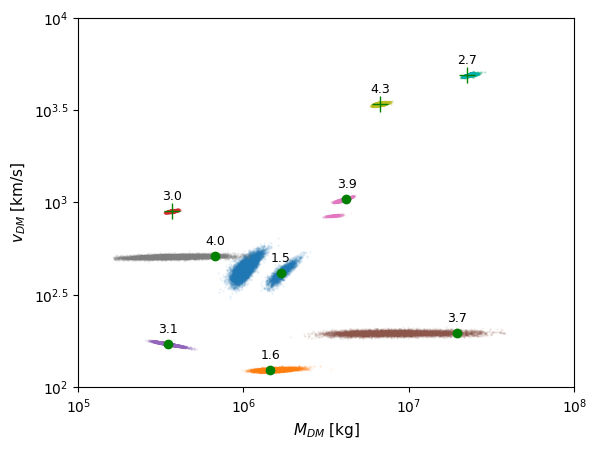}
\caption{Marginalized posterior distributions over mass and velocity for the inferred DM clump in the 9 glitches that cannot be discarded as such.}
\label{fig:M-v-scatter_hist.png}
\end{figure}

This said, one can recognize from Fig.~\ref{fig:xy.png} and Table \ref{tab:fullwidth} that there are 6 out of 9 green DM-clump MAP models where the DM clumps always remain more than 200 meters away from the mirrors.  We use green crosses and dashed lines for those models passing closer than 200 meters to a mirror and green point and solid line for the others. Interestingly, if any of these DM-clump MAP models would actually correspond to a DM clump, then the latter's passage close to a person or object would exert an acceleration on it. Fig.~\ref{fig:M-v-scatter_hist.png} shows the marginalized posterior distribution of the 9 DM candidate glitches in the DM mass-DM velocity plane. Observe that the MAP values for the DM mass and velocity range from $(5 \times 10^5,5 \times 10^7)$ kg and $(10^{-4},10^{-2}) \, c$. If the DM clump radius is smaller than 10 meters and if we place a test object at that distance, then we find that the typical accelerations are less than ${\cal O}(10^{-6} \mbox{m/s}^2)$. If the object size is about 1 meter, then the integrated accelerations, or the overall changes in velocity, are ${\cal O}(10^{-12} \mbox{m/s})$. We see that although the DM clump masses seem large, their effects on material bodies would be tiny.  

\begin{figure}[t]
\includegraphics[width=0.35\textheight]{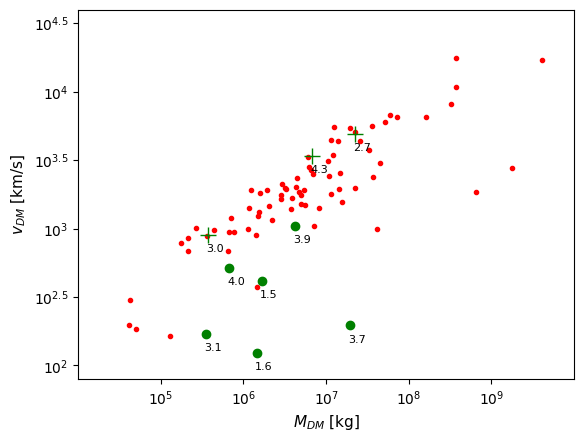}
\caption{Scatter plot of MAP masses and velocities for the DM clump models inferred from the Koi-fish glitch sample. Numbers near the green marks correspond to their ${\cal S}$ value. See text for the discussion on the observed correlation.}
\label{fig:M-v-scatter.png}
\end{figure}

There are many interesting patterns in the DM-clump MAP models of Fig.~\ref{fig:xy.png} and Table \ref{tab:fullwidth}.  The first pattern that arises, as mentioned above, is the accumulation of DM-clump MAP models in azimuthal angle $\phi$ along the diagonal ($\phi=3\pi/4$ and $-\pi/4$) and in polar angle $\theta$ corresponding to low-level flights ($\theta=\pi/2$).  We also find that the closer $\theta$ is to $\pi/2$, the closer the DM clump passes to a mirror ($r_{min}$ decreases).  We can also see that the larger the maximum activated frequency $f_{max}$, the larger the velocity of the DM clump.  This is expected, since a faster DM clump produces sharper variations in the strain.  Along the same lines of reasoning, given a sensitivity span for the SNR in the detector, we expect that detected objects that move faster should have larger masses, as one can see in Fig.~\ref{fig:M-v-scatter.png}.  In this figure, one can see how the green DM-clump MAP models with small ${\cal S}$ are slightly farther away from the main pattern.   We have studied a simple K-means algorithm to assess whether there is a natural number of clusters in the six-dimensional DM-clump MAP models space ($M,\,x,\,y,\,v,\,\theta$ and $\phi$\footnote{We have also explored the four-dimensional space obtained by removing $x$ and $y$, and some other variations without success.}); however, we did not find a marked cluster division.  

\begin{figure}[t]
\includegraphics[width=0.31\textheight]{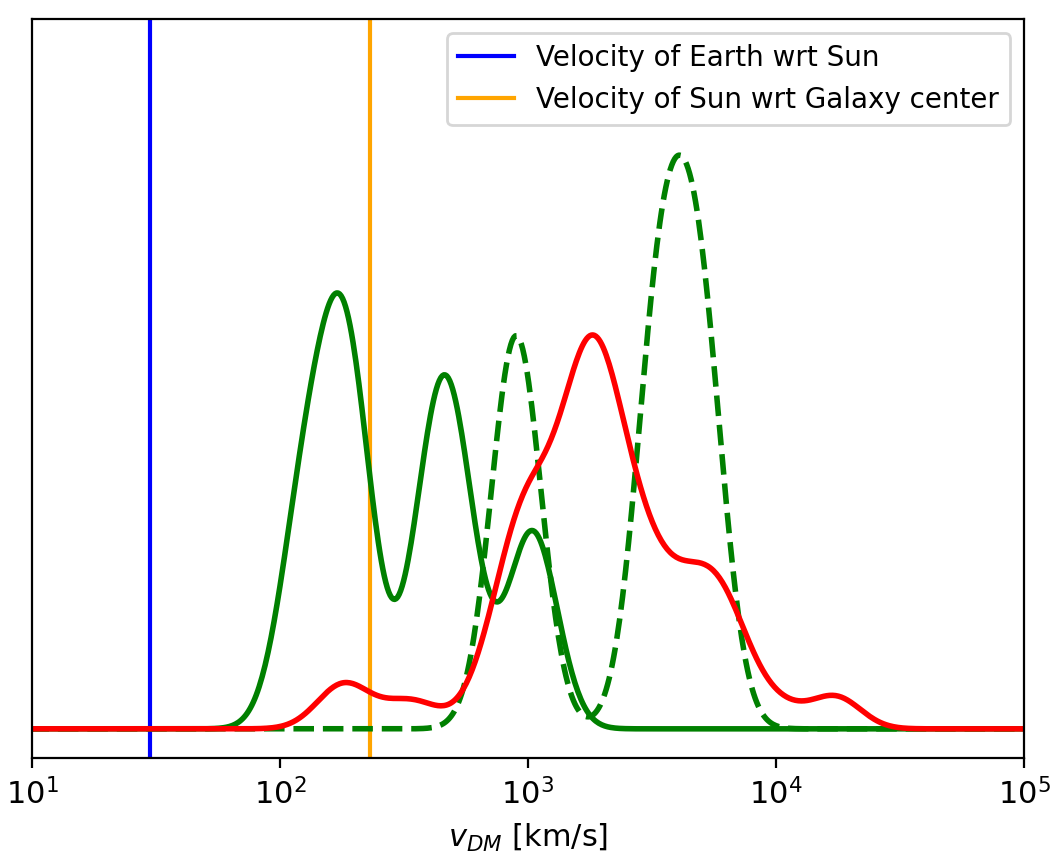}
\caption{MAP velocity distribution for the inferred DM clumps. Vertical lines indicate the orbital velocity of the Earth around the Sun, and the velocity of the Sun around the galactic center, which could coincide to the Sun velocity along the DM halo of the Galaxy. The solid green line corresponds to ${\cal S}<5$ and passing not closer than 200 meters to any mirror, whereas dashed green DM-clump MAP models pass closer than 200 meters to a mirror.  The red line corresponds to glitches with ${\cal S} \geq 5$.   See discussion in text about the slight preference of green DM-clump MAP models to lower velocities.}
\label{fig:v-distribution}
\end{figure}

An interesting line of study is to consider the velocity distribution of the DM-clump MAP models, since we know that DM is distributed in the Galaxy halo and its velocity could have a relation to it.  With this in mind, we show, in Fig.~\ref{fig:v-distribution}, the velocity distribution of the DM-clump MAP models.  We plot in green and red the distributions corresponding to ${\cal S}<5$ and ${\cal S}>5$, respectively.  Moreover, we distinguish in solid green and dashed green the DM-clump MAP models that do not pass close to any mirror and those that do, respectively.  In the same plot, we indicate the Earth velocity with respect to the Sun, and the Sun velocity with respect to the galactic center.  Observe how the green DM-clump MAP models have a slight preference for velocities similar to the Sun's with respect to the galactic center.  This result is far from being a statement, but it is worth noting regardless, as it would be interesting to evaluate whether similar behavior could occur in a new dataset.

In this direction, we should also mention that the posterior velocity distribution in Fig.~\ref{fig:M-v-scatter_hist.png} is likelihood-dominated. We have performed an importance sampling computation of the mean velocities using a Maxwell-Boltzmann prior (instead of a log-uniform prior) with mean value \(v_{\rm mean} = 220\)~km/s and standard deviation \(\sigma_v = 156\)~km/s, both values based on~\cite{Kavanagh:2017cru}. We found that the change of the mean velocities relative to the combined standard deviation 
\begin{equation}
z_{v} = \frac{|v_{U}-v_{MB}|}{\sqrt{\sigma_{U}^2+\sigma_{MB}^2}}
\end{equation}
is \(\lesssim 3\) for the three green DM clumps with higher velocities and \(\lesssim 0.25\) for the remaining green DM clumps. Here, \(v_{U}\) and \(\sigma_U\) denote the mean velocity and its uncertainty from the original posterior samples, while \(v_{MB}\) and \(\sigma_{MB}\) denote the mean velocity and its uncertainty from the importance-sampled (Maxwell-Boltzmann) distribution, respectively.

\subsection{Direct DM density constraints due to DM clumps}\label{sec:Constraints on the DM density due to DM clumps}

Let us close this section by using the previous results to set a rough, yet direct, constraint on the DM density in our position in the halo within the assumptions used in this work. The main assumptions are that the DM is in the form of clumps whose diameter  is not much larger
than the dimensions of the LIGO detectors (see Sec.~\ref{sec:Dark Matter Clump Model}), and that produce strains that are classified as the selected Koi-Fish glitches. Furthermore, we will assume that the detection efficiency of the gravitational wave detector is 1 within the spacetime volumes directly probed by the detector. In this regard, it is worth noting that although the assumptions are too specific, the advantage of this method over other direct detection experiments is that the DM cross-section is known, since it is a gravitational interaction (the only interaction we know for sure that DM feels).  The rationale for how to obtain a rough estimate/constraint for the DM density is as follows.

Let us first assume that the 9 glitches that cannot be discarded from the ${\cal{S}}$-metric test are indeed caused by the passage of DM clumps and ask how many in total there should be in the entire list of O2 observations and what are their properties.  
The total number of glitches in our sample is $84$, and they were randomly selected from a list of $564$ glitches. This list corresponds to all glitches recorded in the Hanford detector during O2 that were classified as Koi-Fish glitches with $95\%$ confidence or higher and had SNRs between 100 and 200 \cite{gravityspy}.  Within our sample, there are 9 glitches with $\mathcal{S} \leq \mathcal{S}_{\rm threshold} = 5$ (we let $\mathcal{S}_{threshold}$ vary below). We can project that in the full list there should be, on average, approximately $N_{DM} \approx 564 \times(9/84) \approx 60$ Koi-Fish glitches with $\mathcal{S} \leq 5$. These $N_{DM}$ glitches would be those that we would not be able to discard as DM clumps. Let us assume that the mass of these hypothetical DM clumps is the average mass of the 9 inferred DM clumps, approximately $M_{DM} \approx 0.5 \times 10^7 \, \text{kg}$.
As the Hanford detector moves across the galactic halo with approximately the same velocity as the Sun $v_S$, the detector covers a volume \( V_{H} \) in a time $T_{O2}$ of approximately
\begin{equation}
V_{H} = \sigma_{H} \times v_{S} \times T_{O2} \,,   
\end{equation}
where \( \sigma_{H} \) is the effective cross-section of the Hanford detector, and \( T_{O2} \approx  2 \times 10^7 \ \text{s} \) is the O2 duration. In Appendix \ref{app:Cross-Section of LIGO} we derive the cross section for a detector as Hanford in a given reasonable scenario, which yields $\sigma_H \approx 100~\text{km}^2$.
Furthermore, assuming DM clumps are not much larger than the detector, we neglect any fraction of their mass lying outside the volume \( V_H \).
 
Using the above approximations, we can set an approximate upper limit for the DM density within our assumptions. The essential argument here is that if there are $N_{\rm DM}$ glitches produced by DM crossings, then the density of DM clumps is, at most, the total mass of those clumps divided by the volume sampled by the detector (as it moved around the Sun with Earth), namely 
\begin{equation}
\rho_{\text{DM clumps}} \lesssim \frac{N_{DM} \times M_{DM}}{V_{H}} \approx 0.5 \times 10^{-15} \,\text{g/cm}^3 ,    
\label{dmbound}
\end{equation}
when ${\cal S}_{threshold}=5$. The above is an upper bound because, if the DM density were higher, we would have detected more glitches that we would not have been able to reject as being caused by the passage of DM clumps.  By inspecting Table \ref{tab:fullwidth}, we see that the nine glitches that were not discarded would correspond to masses of about $10^5$ to $10^7$ kg.

This bound can be compared to the usual indirect estimate of the local Galactic density near the Sun, $\rho_{\rm DM}\approx 10^{-24}\,{\rm g\,cm^{-3}}$, and to the more direct limits from planetary orbital data \cite{Frere:2007pi, Pitjev_2013}, $\rho_{\rm DM}\lesssim 10^{-19}\,{\rm g\,cm^{-3}}$ at Earth’s orbit. However, the latter are derived under the assumption of a Sun‑centered, spherically-symmetric, quasi‑static distribution. By contrast, the result in Eq.~\eqref{dmbound} targets a different signal class: compact, transiting clumps that produce short‑duration, near‑field tidal responses when passing near a terrestrial interferometer. Our inference assumes only that the clump’s characteristic size is smaller than the interferometer baseline (so the point‑perturber tidal model applies), while allowing a broad range of spatial and kinematic distributions. In particular, clumps may trace the Galactic halo or appear as sparse, anisotropic, or transient substructure—Earth‑bound material, streams, or ring‑like features. Conversely, our constraint is, by construction, insensitive on interferometer scales to any spatially smooth component. The two sets of limits are therefore complementary rather than competing. Moreover, the constraint in Eq.~\eqref{dmbound} is, to our knowledge, the first direct bound on compact, transiting DM clumps derived from ground‑based gravitational‑wave data, and it is expected to tighten with increased exposure: at fixed sensitivity it scales inversely with the product of the number of operating observatories and the accumulated observing time. Hence, if no glitches are due to DM clumps (see below), the present framework provides a straightforward way to project how the bound will improve as the network and observing runs grow.

\begin{figure}[t]
\includegraphics[width=0.35\textheight]{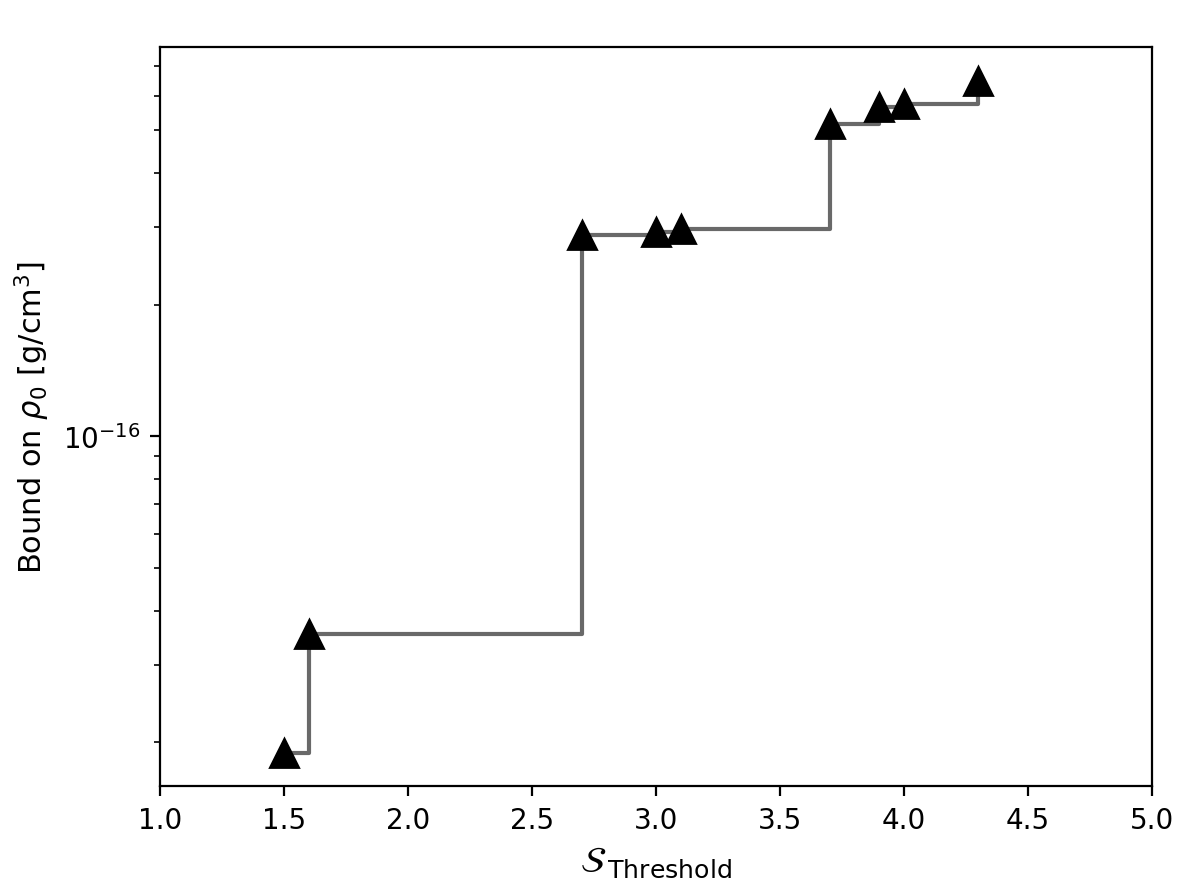}
\caption{Upper limits on $\rho_{\text{DM clumps}}$ as a function of the threshold used to discard the DM clump hypothesis on a glitch. }
\label{fig:DM_density_constraints.png}
\end{figure}

Let us now vary ${\cal S}_{threshold}$ to study how the constraint in Eq.~\eqref{dmbound} varies with this quantity.  Figure~\ref{fig:DM_density_constraints.png} shows the upper limit on the DM clump density as a function of ${\cal S}_{threshold}$. Observe also that, in Eq.~\eqref{dmbound}, we approximated the number of glitches that cannot be discarded as DM clumps, $N_{DM}$, as a simple linear extrapolation of the non-discarded glitches to the whole Koi-Fish dataset. This approximation loses precision as the number of non-discarded glitches goes to zero.  In particular, Eq.~\eqref{dmbound} is not valid for $N_{DM} = 0$ and a different treatment is needed then.

Let us then consider the case that none of the glitches in O2 were produced by the passage of a DM clump. In that case, the number of detected DM clumps (i.e.~zero) is a draw from a Poisson distribution (ignoring any spatial correlations in the DM clump distribution along the path of Earth) with parameter $\mu$, namely
\begin{equation}
     0 \sim \mbox{Poisson}(\mu) .
\label{observation}
\end{equation}
Therefore, to determine the posterior probability distribution on $\mu$ given the observation in Eq.~\eqref{observation}, we use Bayes' theorem and find
\begin{equation}
    p(\mu | 0) = \frac{ p(0|\mu) p_{prior}(\mu)}{\int p(0|\mu)\,p_{prior}(\mu) d\mu}\,,
    \label{bayes}
\end{equation}
where $p_{prior}$ is the prior on $\mu$, $p(\mu | 0)$ is the posterior on $\mu$ given the observation of zero clumps, and $p(0|\mu)$ is the likelihood. Assuming no prior knowledge on $\mu$, we set $p_{prior}(\mu) =$ constant, and the prior cancels in Eq.~\eqref{bayes}. The likelihood is simply the probability density of zero event drawn from a Poisson distribution with mean $\mu$. Using this equation, we find the 95\% credible interval for $\mu < \mu_0$ by solving
\begin{equation}
0.95 = \int_0^{\mu_0} e^{-\mu}  \; d\mu\,,
\label{eq:bound-no-DM}
\end{equation}
which yields $\mu < \mu_0 \approx 3$.  Given, therefore, the aforementioned hypothesis and assuming DM clumps of mass $M_{DM}$, we can make a rough estimate of the upper bound in the 95\% credible interval for the DM density by using $N_{DM} =3$ in Eq.~\eqref{dmbound} to obtain

\begin{equation}
    \rho_{\text{DM clumps}} \lesssim 2.5 \times 10^{-17} \mbox{ g/cm}^3\,.
    \label{bound with no observation}
\end{equation}

We see then that, even when we assume that none of the glitches detected were caused by the passage of a DM clump, we can still place an upper bound on the DM clump density. This bound is stronger than what we find when we assume all 9 candidate glitches were caused by the passage of a DM clump (i.e.~Eq.~\eqref{bound with no observation} is more stringent than Eq.~\eqref{dmbound}). This is because the statement that none of the 564 glitches detected were caused a DM clump implies a stronger consequence on the plausible DM density in the path of Earth than assuming that some of the 564 glitches were caused by DM. Of course, if were able to confidently state that at least 1 of the glitches observed was indeed caused by the passage of a DM clump, then we would be able to place a \textit{lower} bound, instead of an upper bound, as we would have a confident detection.

Since this limit is derived from O2 data at LIGO Hanford alone, we can make a rough extrapolation to the full network by folding in additional sites and exposure. For a single‑site and  threshold of SNR \(>10\), the surveyed cross‑section scales with the square of the detection horizon; with SNR \(\propto b^{-3}\) for a point‑perturber tidal signal at closest approach \(b\), this gives \(\sigma_H\propto {\rm SNR}^{-2/3}\). In practice, adding LIGO‑Livingston, Virgo, and KAGRA increases the effective geometric exposure by roughly the number of operating sites, since a clump passing near any one detector would be captured, and the accumulated observing time has grown by about an order of magnitude relative to Hanford O2. Using these rough scalings, Eq.~\eqref{bound with no observation} extrapolates to \(\rho_{\rm DM}\lesssim 10^{-18}\,{\rm g\,cm^{-3}}\). While this remains less stringent than the ephemeris limits \(\rho_{\rm DM}\lesssim 10^{-19}\,{\rm g\,cm^{-3}}\) \cite{Frere:2007pi, Pitjev_2013}, our constraint applies to Galactic‑halo and other non‑heliocentric distributions, including sparse or transient substructure, and it scales approximately inversely with the product of the number of detectors and the accrued observing time—both of which are steadily increasing. A more precise forecast would need to account for duty cycles, network‑sensitivity variations, and selection efficiencies at each site; we leave that bookkeeping to future work.

\section{Outlook and Conclusions}\label{sec:conclusions}

We have studied the theoretical possibility that the passage of a small DM clump through one of the arms of a gravitational-wave interferometer could produce a measurable glitch in the strain. We began by creating a model for the strain that would be caused by a passing DM clump, focusing on the Newtonian acceleration caused by the clump on the mirrors and the Shapiro time delay experienced by photons that cross the clump. We found that the former effect is dominant for ground- and space-based gravitational wave interferometers. We then compared this DM clump model to 84 Koi-Fish glitches through a Bayesian parameter estimation analysis, where the likelihood was explored through parallel-tempered MCMC techniques. We find that for all but 9 of the glitches analyzed we can reject the hypothesis that the glitches were produced by a crossing DM clump. For the remaining 9 glitches, this hypothesis cannot be rejected or confirmed, given the lack of other competitive models to compare the DM clump model. If these glitches were produced by DM clumps, our MCMC parameter estimation study suggests that their most-likely average density would have been about $(10^{-7},10^{-9}) {\rm{g}}/{\rm{cm}}^3$. 

One may wonder whether these densities are too large that such DM clumps would have been observed by other means on Earth. The DM clumps we consider are all restricted to be smaller than the size of the gravitational-wave interferometers, so that the clump can cross one of the arms without touching the mirrors. Given this, the DM clumps would have to be smaller than roughly $4$ km for the LIGO instruments, and their total mass would range between $(10^7,10^5)$ kg. The corresponding densities are only about a factor of $10^3$ smaller than the density of water vapor (and $10^7$ times smaller than the density of particles in a sand storm). DM clumps, however, cannot be seen by the naked eye because they do not interact electromagnetically. For the same reason, DM clumps would not emit any kind of electromagnetic radiation, even if they are traveling at $\sim 1\%$ the speed of light (as suggested by the maximum likelihood points of the 9 candidate glitches), or even if they are slowed down as they cross the interferometer arms.

Although we cannot confirm the hypothesis that any of the glitches were produced by the crossing of DM clumps, we can use our analysis to place an upper limit on the local mass density in compact clumps of a given size in Earth’s neighborhood. If DM clumps with radii smaller than 4\,km existed and crossed a detector arm, then their density would have to be $\lesssim 10^{-15}\,\mathrm{g\,cm^{-3}}$ (Eq.~\ref{dmbound}) for such events to remain consistent with the observed glitches under the $\chi^2$ discriminator defined here, and $\lesssim 10^{-17}\,\mathrm{g\,cm^{-3}}$ (Eq.~\ref{bound with no observation}) in order not to have produced any detectable candidates in our search. This provides a direct bound which—although less stringent than the constraint inferred from planetary orbital data, $\sim 10^{-19}\,\mathrm{g\,cm^{-3}}$—applies to more general, non‑heliocentric (and potentially transient) DM distributions. Moreover, by accumulating interferometer data and focusing on clumps that would yield $\mathrm{SNR}>10$, the limits reported here are expected to improve by about an order of magnitude or more. Further work will refine these projections by incorporating duty cycles, network sensitivity, and selection efficiencies.

Our work opens the door to several possible avenues for future research. The models we created here are but the first initial steps toward the creation of an accurate DM clump strain. Such models could be improved by accounting for non-spherical DM clump geometries and non-constant DM energy densities. Our data analysis could also be improved if a non-DM, physical glitch model were developed. If so, one could then compare the evidences of these two models through a Bayes factor, thus allowing the data to decide which hypothesis is better supported. 

\acknowledgements
This work was performed under the auspices of the U.S. Department of Energy by Lawrence Livermore National Laboratory under Contract DE-AC52-07NA27344. The document number is \IMRELEASENO{}.  N.~Y.~ acknowledges support from the Simons Foundation through Award No. 896696, the NSF through Grant No.~PHY-2207650, and NASA through Grant No.~80NSSC22K0806.

\appendix

\section{Degeneracies of the DM clump model}
\label{app:Degeneracies}
\begin{figure}[t]
\includegraphics[width=0.6\linewidth]{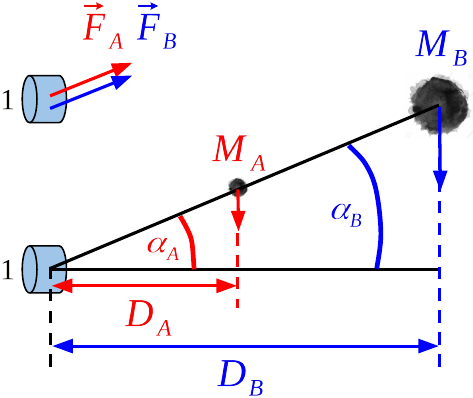}
\caption{Two DM clumps passing near mirror 1, such that gravitational interactions between them and the other mirrors can be neglected. DM clump B, with a mass of $M_B = \lambda^2 M_A$, a velocity of $v_B = \lambda v_A$, and an impact parameter of $D_B = \lambda D_A$ from mirror 1, generates an almost identical second derivative of strain as DM clump A.}\label{fig:simetry0.pdf}
\end{figure}

To illustrate the degeneracy present in the Newtonian model, we consider the situation shown in Fig.~\ref{fig:simetry0.pdf}. This figure shows two DM clumps moving perpendicularly to the plane of the detector, such that they cross the plane through the points $\vec{x}_A = (L+D_A,0,0)$ and $\vec{x}_B = (L+D_B,0,0)$. The respective masses and velocities of the DM clumps are $M_A$ and $v_A$, and $M_B$ and $v_B$. DM clump A generates a second derivative of the strain $d^2h_{NA}/dt^2$, while DM clump B generates $d^2h_{NB}/dt^2$. In the regime where $D_A$ and $D_B$ are much smaller than the characteristic size of the detector, and when these, the masses, and the velocities are related to each other via $D_B = \lambda D_A$, $M_B = \lambda^2 M_A$, and $v_B = \lambda v_A$, both DM clumps generate almost the same second derivative of the strain over time,
\begin{equation}
\frac{d^2h_{NA}}{dt^2}(t) \approx \frac{d^2h_{NB}}{dt^2}(t).
\label{eq:simetry}
\end{equation}

To prove this result, we can consider the gravitational interaction forces between the DM clumps and mirror 1, $\vec{F}_A$ and $\vec{F}_B$, with $\alpha_A$ and $\alpha_B$ being the angles formed between these forces and the plane of the detector. Under the conditions described above, these forces remain equal over time, since the relations $|\vec{F}_A| = |\vec{F}_B|$ and $\alpha_A = \alpha_B$ are satisfied at all times. On the other hand, the gravitational interaction forces between the DM clumps and the other mirrors can be neglected, because the clumps pass much farther from those mirrors than from mirror 1 ($D_A \ll L$ and $D_B \ll L$). Therefore, Eq.~\eqref{eq:simetry} is satisfied.

\begin{figure*}[t]
\includegraphics[width=0.49\linewidth]{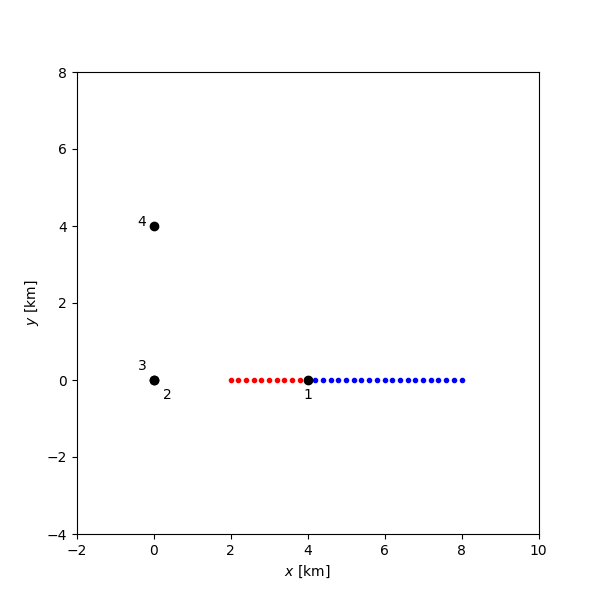}
\includegraphics[width=0.49\linewidth]{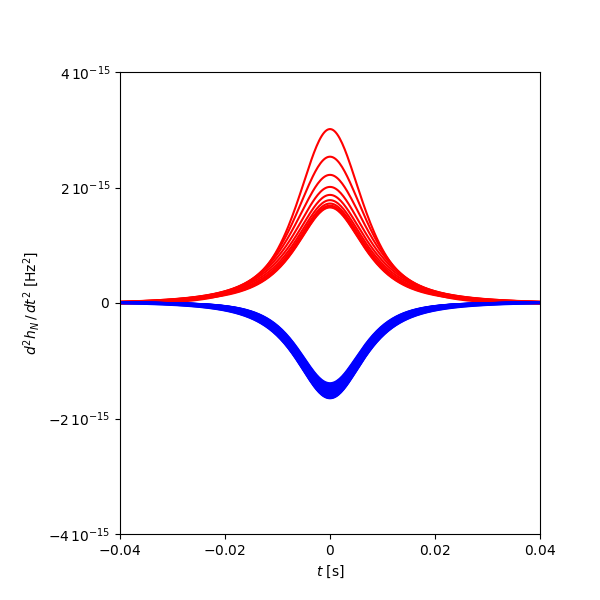}
\caption{Second derivative of the strain (right plot) for DM clumps crossing through the colored points displayed in the left plot; the colors in both plots correspond to each other. The DM clumps have a perpendicular velocity with respect to the plane of the detector, and the minimal distances to mirror 1 they reach, their masses, and velocities are related as described in the text. The DM clumps that pass a distance equal to $1 \ \text{km}$ from mirror 1 have $M = 10^5 \, \text{kg}$ and $v = 10^2 \, \text{km/s}$. We can see that the second derivative of the strain produced by the DM clumps crossing through the blue points are almost equal to each other, whereas the second derivative of the strain produced by the DM clumps crossing through the red points present some dissimilarities due to  the effect of the other mirror.}\label{fig:symmetry.png}
\end{figure*}

Figure~\ref{fig:symmetry.png} shows this degeneracy by presenting the second derivative of the strain produced by different DM clumps moving perpendicular to the plane of the detector. The smallest distances these DM clumps get to mirror 1, as well as their masses and velocities, are related through the aforementioned relations. As one can see from the figure, the second derivative of the strain produced by the DM clumps crossing through the blue points exhibits striking similarities, even though some of them pass at distances from mirror 1 comparable to the characteristic size of the detector. Conversely, some dissimilarities are observed in the second derivative of the strain produced by the DM clumps crossing through the red points. This discrepancy arises due to the proximity that these DM clumps reach relative to the other mirror through their trajectories.

The above examples illustrate a scenario where DM clumps can yield nearly identical signals upon altering three specific parameters according to a prescribed relation. However, in that scenario, the signals could remain relatively unchanged upon altering the direction of their velocities through a rotation around the $x$-axis, introducing the possibility of incorporating an additional parameter into the existing set of three parameters. This result leads us to consider the possibility of incorporating more parameters in the list of parameters that can be changed without producing significant changes in the signals. In fact, we could consider incorporating all the parameters of the model into this list, producing trajectories in the parameter space that connect different points where the model, evaluated at these points, produces almost the same signal. 

\begin{figure*}[t]
\includegraphics[width=0.49\linewidth]{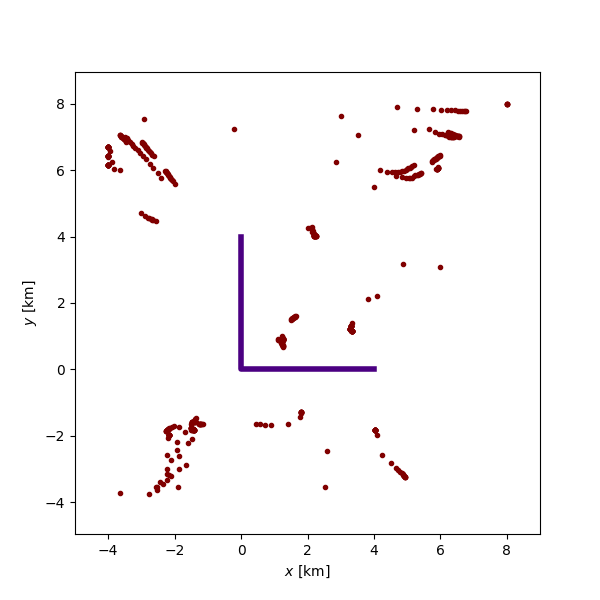}
\includegraphics[width=0.49\linewidth]{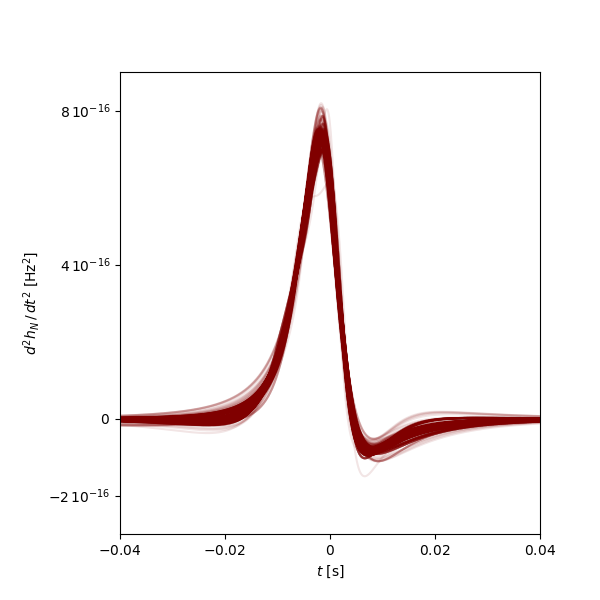}
\caption{Second derivative of the strain (right plot) produced by the DM clumps that cross the plane of the detector through the brown points shown in the left plot. Here, the detector is represented by the violet arms. Each DM clump has a different mass, a different velocity, and a different incidence angle, highlighting the challenge in accurately determining the actual parameters of a DM clump that produces a certain signal.}\label{fig:symmetry1.png}
\end{figure*}

Figure~\ref{fig:symmetry1.png} shows how the model evaluated at different points distributed across various regions of the parameter space can produce nearly the same signal. The left panel of the figure displays the spatial distribution of these points with respect to the detector, while the right panel depicts the second derivative of the strain corresponding to the model evaluated at these points. For clarity, we only show the spatial distribution of these points with respect to the detector. However, it can be seen that these points encompass different masses, velocities, and incident angles, highlighting the challenge of accurately determining the actual parameters of a DM clump that produces a certain signal.

\section{Behavior of FF}
\label{app:FF-behavior}
\begin{figure*}[th]
\centering
\includegraphics[width=0.497\linewidth]{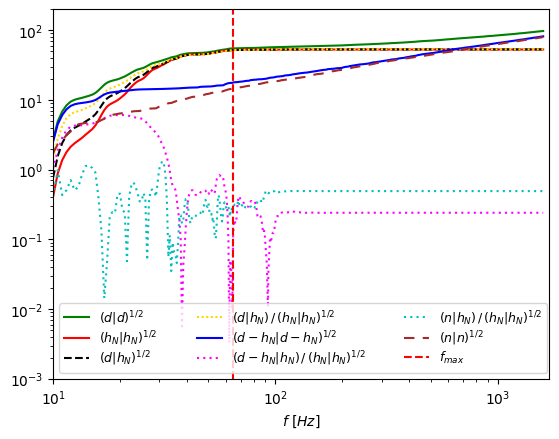}
\caption{Dependence of the inner products (as described in the legend) on the upper bound of the frequency range, defined from \( 10 \, \mathrm{Hz} \) to \( f \), shown on the x-axis.}
\label{fig:inner_1183157056.15234.png}
\end{figure*}

As discussed in Sec.~\ref{sec:Metrics}, the fitting factor ($FF$) is given by 
\[
FF = \frac{(h_{N} \mid d)}{\sqrt{(h_{N} \mid h_{N})(d \mid d)}}\, .
\]
However, the inner products used to compute the $FF$ for glitches depend strongly on the frequency range over which the integral defining the inner product is evaluated (see Eq.~\eqref{eq:inner}). For example, taking as reference the glitch analyzed in Sec.~\ref{sec:One Case in Detail}, Fig.~\ref{fig:inner_1183157056.15234.png} illustrates the dependence of several inner products on the upper limit of the frequency range used in the integral. These inner products are constructed using the corresponding strain data \( d \), the noise power spectral density \( S_{n} \), and the inferred MAP DM clump model \( h_{N} \) for that event. The frequency range always starts at \( 10 \, \mathrm{Hz} \) and extends up to the frequency \( f \), indicated on the x-axis. Additionally, to provide a reference for interpreting the behavior of some of these inner products, we also include inner products involving a synthetic Gaussian noise realization \( n \).

In Fig.~\ref{fig:inner_1183157056.15234.png}, we can observe that there is a frequency near \( f_{\mathrm{max}} \) at which all inner products of the form \( (\,\cdot\,| h_{N}) \) stop increasing with the upper frequency limit of integration $f$. This behavior occurs because \( h_{N} \) falls below the noise level at frequencies beyond \( f_{\mathrm{max}} \). On the other hand, the inner products \( (d|d) \), \( (d - h_{N} | d - h_{N}) \), and \( (n|n) \) grow indefinitely due to the cumulative effect of persistent noise. Another effect of the persistent noise is revealed in the similar behavior of \( (d - h_{N}  |  d - h_{N}) \) and \( (n  |  n) \) at high frequencies, where \( h_{N} \) has fallen below the noise level and \( d \) effectively becomes pure Gaussian noise. Furthermore, we observe that in certain regions of the frequency space, the quantities \( (d | d) \), \( (h_{N} | h_{N}) \), \( (d | h_{N}) \), and \( (d | h_{N}) / (h_{N} | h_{N})^{1/2} \) become similar. This occurs because \( d \) and \( h_{N} \) are very similar to each other in those regions of the frequency domain. Finally, we observe that the quantities \( (d - h_{N} | d - h_{N}) / (h_{N} | h_{N})^{1/2} \) and \( (n | n) / (h_{N} | h_{N})^{1/2} \) are of order unity or less than unity, as expected for data containing only noise.
\begin{figure*}[th]
\includegraphics[width=0.497\linewidth]{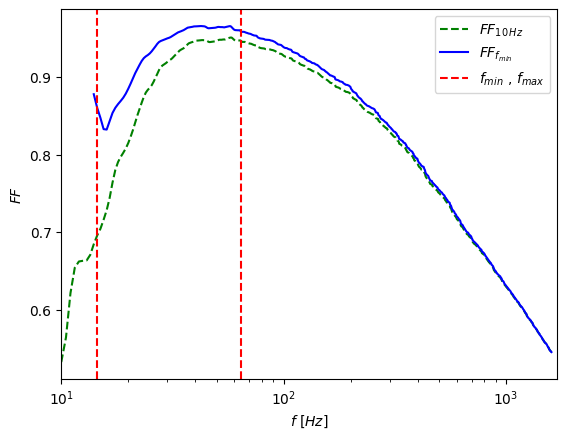}
\caption{Dependence of the fitting factor \( FF \) on the upper bound of the frequency range (indicated on the x-axis) used to compute the inner products. The dashed green curve corresponds to a range starting from \( 10 \, \mathrm{Hz} \), while the blue curve corresponds to a range starting from \( f_{\min} \).}\label{fig:FF_1183157056.15234.png}
\end{figure*}

Since \( (h_{N} | h_{N}) \) and \( (d | h_{N}) \) stop growing at frequencies higher than \( f_{\mathrm{max}} \), whereas \( (d | d) \) continues to grow indefinitely, the fitting factor \( FF \) will reach a maximum at some frequency and then decrease indefinitely beyond that point. For example, Fig.~\ref{fig:FF_1183157056.15234.png} shows the fitting factor as a function of the upper frequency limit of integration used in the inner product computations, with this upper limit indicated on the x-axis. Two curves are shown: one corresponding to a frequency range starting from \( 10 \, \mathrm{Hz} \) (dashed green curve), and the other corresponding to a frequency range starting from \( f_{\mathrm{min}} \). Observe that both curves are similar, indicating that the starting limit of integration is not as important. As discussed in Sec.~\ref{sec:Metrics}, in this work we define the \( FF \) as the value computed using the frequency range from \( f_{\min} \) to \( f_{\max} \). In this example case, we obtain \( FF \approx 0.96 \) and \( \mathcal{A} \approx 0.28 \).

\section{Corner plots}
\label{app:Completed_corner_plots}

\begin{figure*}[th]
\includegraphics[width=1.0\linewidth]{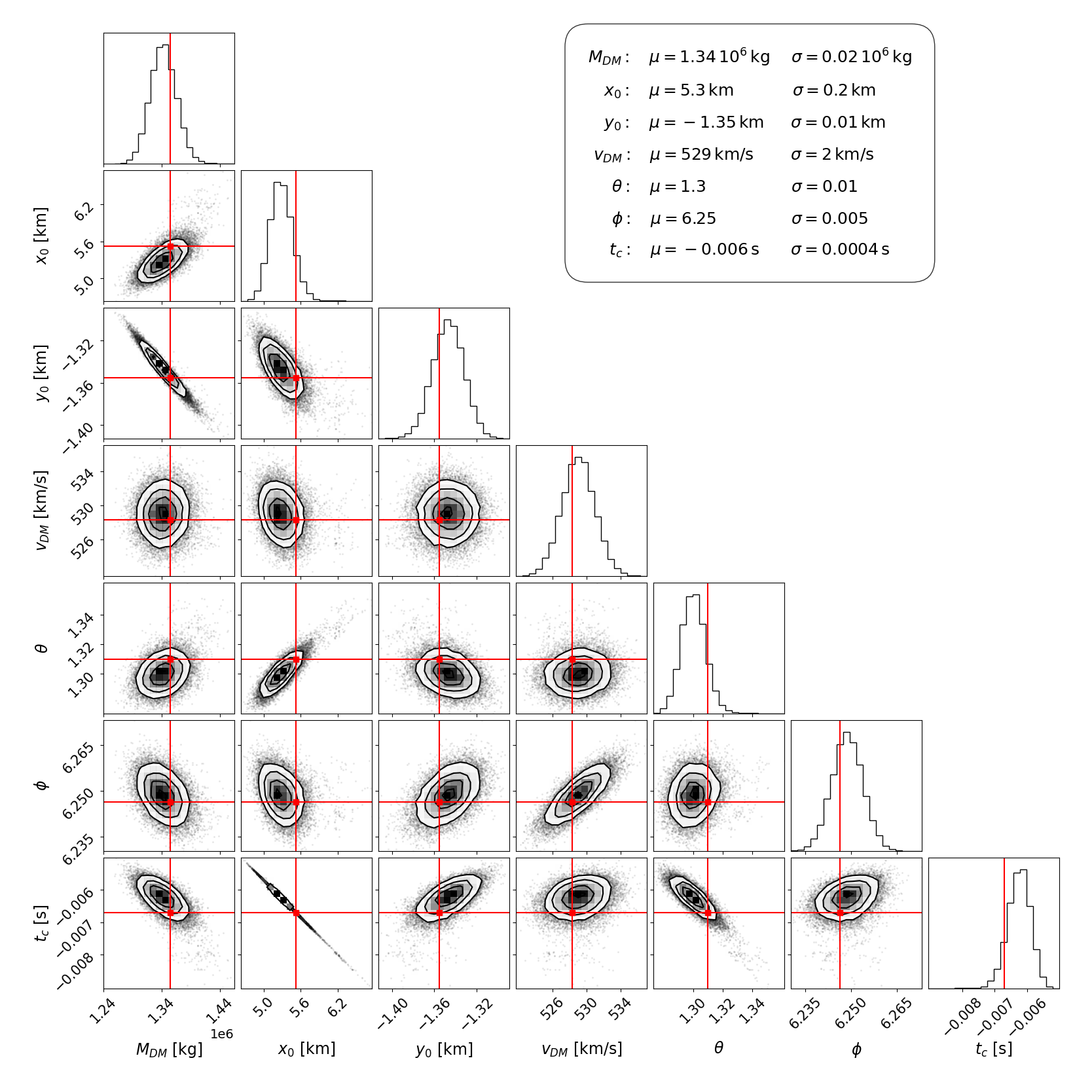}
\caption{Full posterior distribution corresponding to the DM clump injected signal of Sec.~\ref{sec:DM Clump and Cos-Gaussian}.}
\label{cp1}
\end{figure*}

\begin{figure*}[th]
\includegraphics[width=1.0\linewidth]{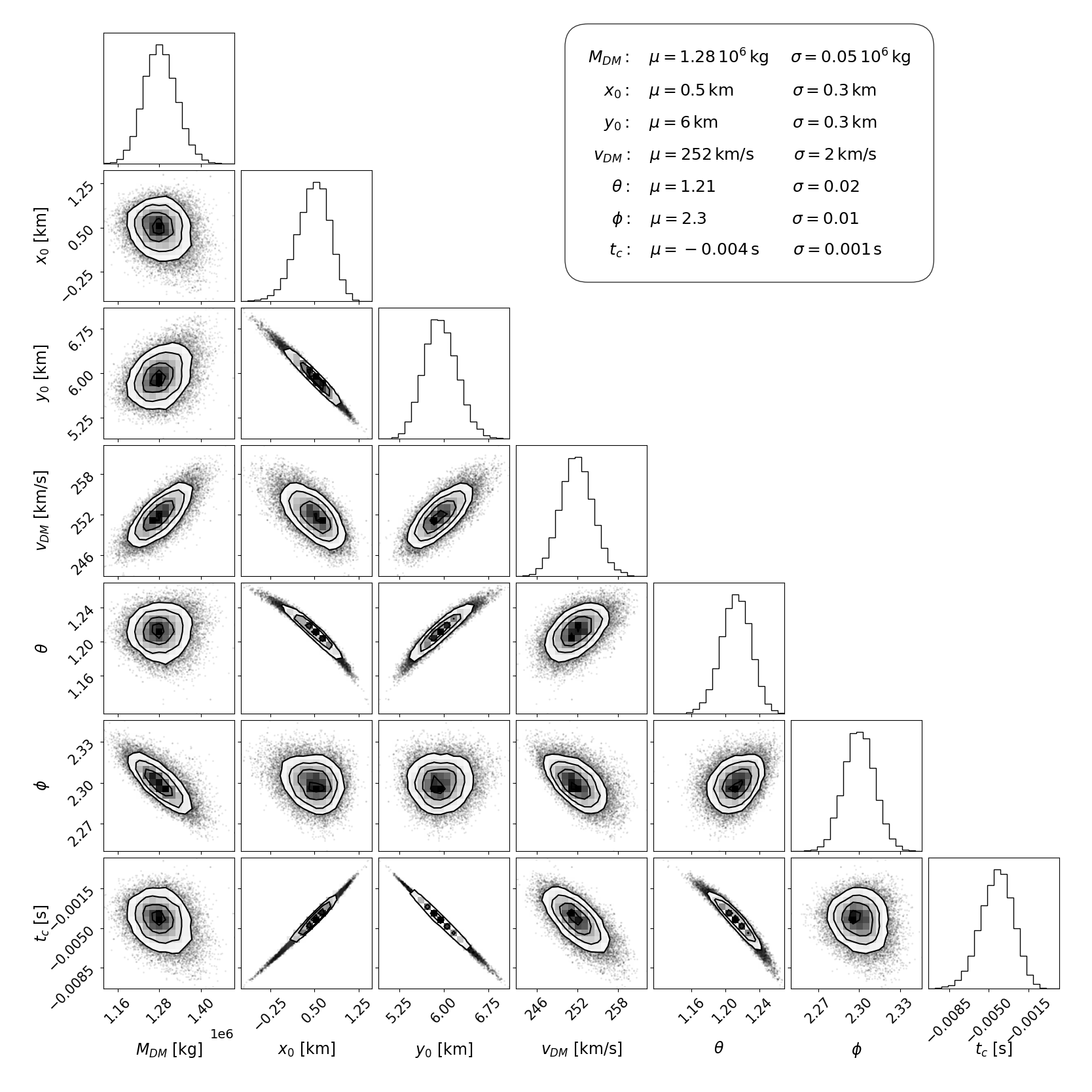}
\caption{Full posterior distribution corresponding to the Cos-Gaussian injected signal of Sec.~\ref{sec:DM Clump and Cos-Gaussian}.}
\label{cp2}
\end{figure*}

\begin{figure*}[th]
\includegraphics[width=1.0\linewidth]{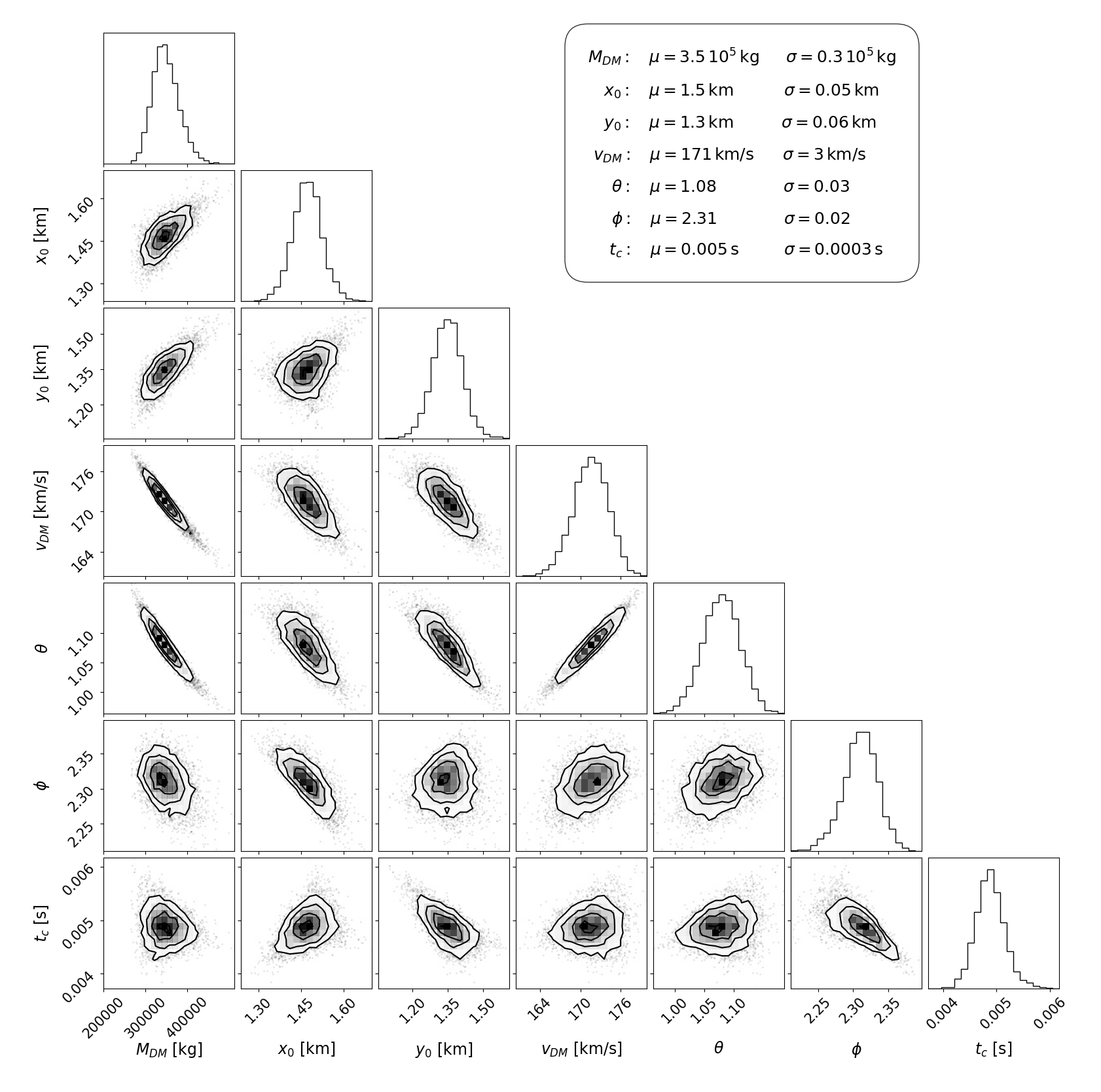}
\caption{Full posterior distribution corresponding to the glitch studied in   Sec.~\ref{sec:One Case in Detail}.}
\label{cp3}
\end{figure*}

Inferred marginal posterior distributions for all the parameters of the DM clump model corresponding to the DM clump (Fig.~\ref{cp1}) and Cos-Gaussian (Fig.~\ref{cp2}) injections of Sec.~\ref{sec:DM Clump and Cos-Gaussian}, as well as the glitch (Fig.~\ref{cp3}) studied in Sec.~\ref{sec:One Case in Detail}. These posteriors were inferred using {\tt BayesShip}. As in Sec.~\ref{sec:DM Clump and Cos-Gaussian}, the red lines in the corner plot of the DM clump injected signal indicate the true (injected) parameter values used to generate the DM clump. Observe the good agreement between these injected values and the inferred posterior distribution.

By examining the three corner plots, we observe that they are not sufficient to rule out any of these three types of signals as being caused by the passage of a DM clump, since they do not show appreciable differences. Therefore, we need the \( \mathcal{S} \) metric to discriminate between them.

\section{LIGO Cross-Section for DM Clumps}
\label{app:Cross-Section of LIGO}

We are interested in computing an approximated cross-section for DM Clumps activating SNR above a given threshold at LIGO.  To obtain a reference value for this cross-section, we propose a scenario as described below.  More detailed calculations, or different specific cases, are straightforward extensions of the calculations below.

\begin{figure*}[htbp]
    \centering
        \centering
        \includegraphics[width=0.49\textwidth]{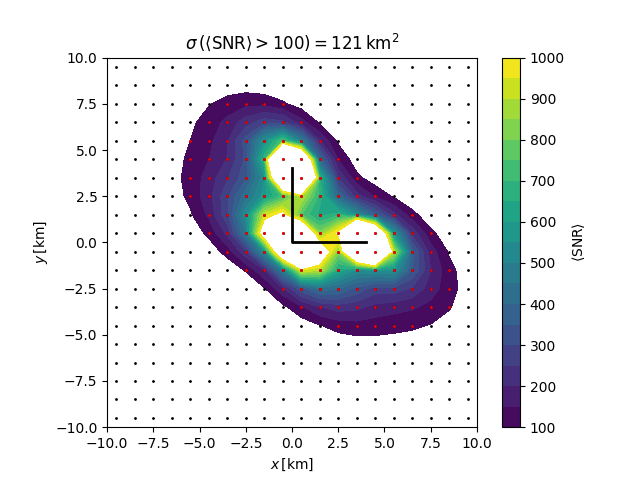} 
    \hfill
        \centering
        \includegraphics[width=0.49\textwidth]{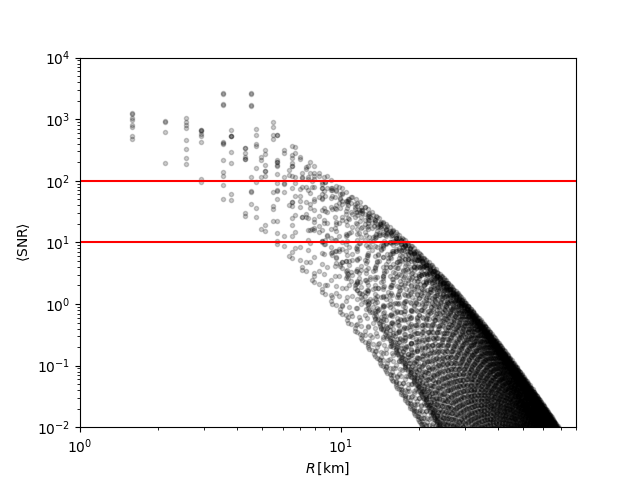} 
    \caption{{\it Left:} One-day hourly averaged SNR values over the detector plane for each \(1\,\text{km}^{2}\) cell for $10^6$ kg DM clumps comoving with the Galactic DM halo. The LIGO interferometer is indicated by the black lines at the center. Red (black) points correspond to cells with an average SNR above (below) 100.  The total area of cells whose $\langle$SNR$\rangle > 100$ yields $121\,\text{km}^{2}$, which we define as the cross-section for a reference scenario (see text). {\it Right:} Average SNR as a function of the cell distance to the detector beam splitter for all the cells in the grid. The spread in SNR values at a given distance corresponds to cells located at the same radial distance but with different orientations relative to the detector arms.} 
    \label{xsection}
\end{figure*}

The SNR depends on the DM Clump mass, velocity and trajectory. For the sake of concreteness we use a mass of $10^6$ kg and proceed as follows for velocity and trajectory.  We compute the relative velocity to the LIGO detector that would have a DM Clump which is still in the DM Galactic halo, at a given time~\cite{Kavanagh:2017cru}.  We compute the SNR that this DM Clump would produce in the LIGO detector if it crosses the detector plane at a given position.  We create an imaginary grid with cells of 1 km$^2$ in the detector plane and perform this computation for each one of the grid's cell.  We perform the described calculation for each hour of a given day, that is 24 times for each grid cell.  We average the SNR in each cell and determine whether this mean SNR is larger than a given threshold, SNR$_{th}$.  We define the approximated cross-section as the total area of all the cells in the grid in which the mean SNR is larger than the threshold SNR$_{th}$.
In Fig.~\ref{xsection} we present the results of the above calculation for SNR$_{th}$ = 100, which yields $\sigma( \langle SNR \rangle > 100)  \approx 120$ km$^2$.

Although along this work we use glitches with SNR $> 100$, it is interesting to notice the dependence of the cross section with SNR$_{th}$, since the volume explored by the detector is directly proportional to this cross section.  For instance, $\sigma( \langle SNR \rangle > 10)  \approx 400$ km$^2$ which represents an important increase. However, in this case one should also study the performance of the parameter inference within the DM clump model to assess the overall increase.

\section{MAP on DM-clump model parameters}
\label{MAPtable}

Table \ref{tab:fullwidth} shows some features of the outcome of the inference process applied to the 84 glitches in the studied dataset, sorted by the ${\cal S}$ metric. Observe that ${\cal A}$ is correlated with ${\cal S}$, but this correlation is lost for small values of ${\cal S}$.  The reason for this is that the ${\cal S}$ metric can differentiate whether small deviations are coming from noise fluctuations or not.  Hence, the ${\cal{S}}$ metric is more sensitive to the detection of small disagreements between a model and the data. 

\begin{table*}[!htbp]
\centering
\fontsize{8pt}{8pt}
\selectfont
\setlength\tabcolsep{3pt}
\begin{tabular}{lllllllllllllll}
\hline
T$_{GPS}$~[s] & $f_{min}$~[Hz] & $f_{max}$~[Hz] & ${\cal S}$ & $\mathcal{A}$ & $M$~[kg] & $v$~[km/s] & $x_0$~[km] & $y_0$~[km] & $\theta$ & $\phi$ & $t_{c}$~[s] & $r_{min}$~[km] \\
\hline
1182507861.02734 & 12.0 & 146.0 & 1.5 & 0.09 & 1679593.0 & 414.0 & 3.982 & 0.533 & 0.18 & 1.182 & 0.0001 & 0.526\\
1183070482.59766 & 10.0 & 59.0 & 1.6 & 0.23 & 1450085.0 & 124.0 & 1.428 & 2.312 & 0.897 & 1.337 & -0.0034 & 1.822\\
1176210190.10498 & 44.5 & 684.5 & 2.7 & 0.16 & 22473669.0 & 4894.0 & 5.03 & -0.794 & 1.437 & 2.465 & 0.0005 & 0.176\\
1166341030.73535 & 35.0 & 296.5 & 3.0 & 0.12 & 367748.0 & 896.0 & 0.209 & 3.516 & 1.432 & 5.154 & 0.0005 & 0.075\\
1183157056.15234 & 14.5 & 64.5 & 3.1 & 0.28 & 349141.0 & 171.0 & 1.467 & 1.356 & 1.075 & 2.316 & 0.0048 & 1.571\\
1183060560.41797 & 13.5 & 51.5 & 3.7 & 0.09 & 19598500.0 & 196.0 & 3.638 & 3.662 & 1.081 & 3.953 & 0.0167 & 2.431\\
1177840413.0332 & 20.0 & 182.0 & 3.9 & 0.09 & 4196291.0 & 1044.0 & 7.574 & -3.988 & 1.269 & 2.406 & 0.0058 & 1.68\\
1173156104.74609 & 15.0 & 151.5 & 4.0 & 0.09 & 671755.0 & 512.0 & -3.968 & -3.199 & 1.45 & 0.659 & 0.0086 & 0.62\\
1176191047.99951 & 29.0 & 520.5 & 4.3 & 0.18 & 6721442.0 & 3395.0 & 3.932 & -0.066 & 1.435 & 2.429 & 0.0009 & 0.094\\
1186032085.92676 & 32.0 & 256.5 & 5.6 & 0.13 & 1516051.0 & 1235.0 & 1.459 & 2.498 & 1.334 & 5.486 & 0.0007 & 0.491\\
1167798730.26611 & 26.0 & 419.5 & 5.6 & 0.18 & 1221419.0 & 1916.0 & 2.586 & 1.321 & 1.57 & 5.479 & -0.0014 & 0.005\\
1167882035.76465 & 25.5 & 312.5 & 5.6 & 0.17 & 5391661.0 & 1899.0 & 1.507 & 2.618 & 1.12 & 5.457 & -0.0006 & 0.903\\
1174632952.45068 & 25.5 & 690.0 & 5.8 & 0.14 & 25598257.0 & 4357.0 & 3.211 & 0.697 & 1.303 & 5.613 & 0.0004 & 0.284\\
1171459625.33496 & 29.0 & 246.0 & 6.0 & 0.15 & 709178.0 & 1195.0 & 7.007 & -2.977 & 1.509 & 2.376 & 0.0056 & 0.27\\
1165146772.71924 & 25.0 & 483.5 & 6.1 & 0.13 & 6564559.0 & 2684.0 & 0.881 & 3.037 & 1.376 & 5.438 & 0.0001 & 0.253\\
1170698943.40381 & 29.5 & 519.5 & 6.3 & 0.14 & 6128950.0 & 2805.0 & 7.577 & -3.312 & 1.54 & 2.385 & 0.0022 & 0.157\\
1175515520.50879 & 21.0 & 326.0 & 6.4 & 0.12 & 4976831.0 & 1768.0 & 2.842 & 1.046 & 0.98 & 6.104 & -0.0001 & 1.106\\
1177890163.95117 & 27.5 & 98.0 & 6.8 & 0.05 & 1474445.0 & 377.0 & 0.657 & 7.235 & 1.338 & 5.058 & 0.0133 & 1.833\\
1182862581.83838 & 33.0 & 942.0 & 6.9 & 0.17 & 159870469.0 & 6554.0 & -2.007 & 5.711 & 1.443 & 5.476 & -0.0 & 0.429\\
1169886089.75928 & 31.0 & 669.5 & 7.0 & 0.2 & 4029211094.0 & 17143.0 & -2.606 & 5.444 & 1.057 & 5.495 & 0.0002 & 1.632\\
1176788918.29395 & 18.0 & 374.0 & 7.0 & 0.13 & 14322304.0 & 1939.0 & 2.103 & 2.245 & 1.008 & 2.327 & -0.0002 & 1.486\\
1182681768.83496 & 23.0 & 266.5 & 7.0 & 0.13 & 5483049.0 & 1498.0 & 1.73 & 2.648 & 1.001 & 5.468 & -0.0002 & 1.217\\
1174734951.90723 & 25.0 & 319.0 & 7.0 & 0.13 & 2852132.0 & 1772.0 & 2.966 & 0.968 & 1.256 & 2.432 & 0.001 & 0.443\\
1176431859.02832 & 46.5 & 360.0 & 7.3 & 0.12 & 4694738.0 & 1864.0 & -1.98 & 6.054 & 1.396 & 5.413 & 0.0018 & 0.53\\
1172600858.11426 & 21.5 & 257.0 & 7.4 & 0.15 & 8207601.0 & 1424.0 & 2.672 & 2.003 & 0.898 & 2.371 & -0.0006 & 1.55\\
1176100533.08496 & 23.5 & 190.0 & 7.6 & 0.13 & 364880.0 & 882.0 & 4.893 & -0.901 & 1.475 & 2.385 & 0.0032 & 0.129\\
1181206295.57715 & 26.5 & 436.5 & 7.9 & 0.15 & 4448346.0 & 2353.0 & 0.586 & 3.372 & 1.426 & 5.46 & 0.0007 & 0.124\\
1170153278.22754 & 25.5 & 272.5 & 8.1 & 0.15 & 2881845.0 & 1647.0 & 7.496 & -3.67 & 1.438 & 2.381 & 0.0043 & 0.713\\
1171022340.57715 & 22.5 & 422.0 & 8.3 & 0.13 & 1572891.0 & 1817.0 & 4.004 & -0.01 & 1.501 & 2.385 & 0.0007 & 0.005\\
1182359936.18408 & 24.5 & 455.5 & 8.4 & 0.13 & 6860250.0 & 2516.0 & 2.872 & 1.045 & 1.365 & 2.397 & 0.0003 & 0.314\\
1167808222.25684 & 29.0 & 364.5 & 8.4 & 0.17 & 2895662.0 & 2105.0 & 5.321 & -1.221 & 1.422 & 2.435 & -0.0001 & 0.275\\
1165344218.48926 & 26.0 & 384.5 & 8.6 & 0.11 & 1548547.0 & 1314.0 & 2.938 & 1.013 & 1.42 & 2.385 & -0.0012 & 0.22\\
1185794664.26074 & 25.5 & 397.0 & 9.1 & 0.14 & 3276444.0 & 1959.0 & 7.576 & -3.592 & 1.526 & 2.368 & 0.004 & 0.237\\
1170708743.44043 & 20.0 & 224.0 & 9.2 & 0.14 & 1147344.0 & 985.0 & 1.284 & 2.818 & 1.188 & 5.481 & -0.0003 & 0.659\\
1174646797.39795 & 22.0 & 497.0 & 9.4 & 0.12 & 44512496.0 & 3007.0 & 1.883 & 2.512 & 0.972 & 5.469 & -0.0001 & 1.384\\
1168054016.3706 & 22.5 & 417.0 & 9.6 & 0.16 & 14677645.0 & 2526.0 & 2.283 & 1.837 & 1.118 & 5.528 & -0.0 & 1.11\\
1172715628.83203 & 10.5 & 105.0 & 9.8 & 0.11 & 127578.0 & 163.0 & 1.338 & -0.294 & 1.365 & 2.336 & 0.0055 & 0.796\\
1167924967.85254 & 30.0 & 279.5 & 9.9 & 0.17 & 2043342.0 & 1472.0 & 2.838 & 1.131 & 1.225 & 2.423 & -0.0007 & 0.555\\
1174993903.74902 & 23.5 & 466.5 & 10.2 & 0.14 & 38223515149.0 & 4056.0 & 3.171 & 3.163 & 0.023 & 2.778 & 0.0002 & 3.27\\
1182110368.07568 & 25.5 & 551.5 & 10.4 & 0.16 & 11966591.0 & 3455.0 & 5.544 & -1.362 & 1.501 & 2.4 & 0.0009 & 0.149\\
1178050067.20801 & 18.5 & 212.0 & 10.6 & 0.13 & 664711.0 & 944.0 & 2.873 & 1.076 & 1.344 & 2.379 & -0.0002 & 0.35\\
1175396864.18262 & 37.0 & 399.5 & 11.6 & 0.12 & 4305668.0 & 2016.0 & 3.114 & 0.767 & 1.304 & 2.44 & -0.0006 & 0.31\\
1174429654.08789 & 15.0 & 139.5 & 12.0 & 0.12 & 41355.0 & 196.0 & 0.001 & 4.389 & 1.534 & 3.809 & 0.0022 & 0.305\\
1177672444.92822 & 36.0 & 631.0 & 12.6 & 0.15 & 36117798.0 & 5598.0 & 3.994 & -0.079 & 1.349 & 5.674 & 0.0002 & 0.069\\
1174955080.72461 & 10.5 & 156.5 & 14.3 & 0.08 & 49746.0 & 184.0 & -0.638 & -0.487 & 1.334 & 3.593 & -0.005 & 0.244\\
1174109651.84082 & 18.5 & 272.0 & 16.6 & 0.1 & 3782031.0 & 1391.0 & 1.908 & 2.125 & 1.134 & 2.291 & 0.0003 & 1.146\\
1167862578.08008 & 15.0 & 193.0 & 20.6 & 0.17 & 41278306.0 & 997.0 & 2.845 & 3.065 & 0.43 & 5.525 & -0.0014 & 2.774\\
1173683970.98145 & 15.0 & 441.0 & 21.7 & 0.14 & 439771.0 & 971.0 & -0.067 & 4.224 & 1.57 & 5.01 & 0.0012 & 0.001\\
1178288223.76367 & 20.5 & 294.0 & 21.7 & 0.18 & 657921.0 & 691.0 & -2.769 & 6.586 & 1.425 & 5.481 & 0.0075 & 0.582\\
1183157294.72754 & 20.0 & 358.5 & 23.2 & 0.13 & 1175522.0 & 1417.0 & 3.709 & 0.201 & 1.34 & 5.714 & -0.0003 & 0.082\\
1182329551.97119 & 24.5 & 456.0 & 24.1 & 0.17 & 11588782.0 & 4466.0 & 4.023 & -0.033 & 1.288 & 5.934 & -0.0 & 0.025\\
1171783807.51758 & 18.5 & 276.0 & 24.2 & 0.13 & 213792.0 & 847.0 & 3.791 & -0.072 & 1.543 & 6.268 & -0.0018 & 0.075\\
1186713727.15723 & 38.0 & 296.0 & 24.5 & 0.12 & 776962.0 & 951.0 & 1.227 & 2.729 & 1.413 & 5.482 & -0.0006 & 0.278\\
1177854631.26123 & 25.0 & 389.0 & 26.7 & 0.14 & 1905369.0 & 1900.0 & -0.009 & 4.013 & 1.513 & 5.477 & 0.0016 & 0.003\\
1182484842.58008 & 19.5 & 191.0 & 28.4 & 0.12 & 1424877.0 & 892.0 & 1.892 & 2.274 & 1.144 & 5.528 & -0.0028 & 1.061\\
1183294027.69434 & 15.5 & 317.5 & 30.3 & 0.09 & 4912758.0 & 1522.0 & 2.362 & 1.654 & 1.236 & 2.36 & -0.0007 & 0.764\\
1165850072.31885 & 24.0 & 435.0 & 39.0 & 0.23 & 5950263.0 & 3318.0 & -0.021 & 4.003 & 1.317 & 5.183 & -0.0002 & 0.018\\
1168289373.06494 & 22.5 & 650.5 & 41.8 & 0.21 & 60008044.0 & 6773.0 & 3.986 & -0.102 & 1.272 & 5.718 & -0.0002 & 0.095\\
1173228411.30957 & 19.0 & 410.0 & 53.1 & 0.14 & 3138831.0 & 1968.0 & 3.827 & 0.133 & 1.496 & 2.374 & 0.0017 & 0.029\\
1177915073.64819 & 13.0 & 1024.5 & 53.9 & 0.24 & 327320330.0 & 8175.0 & -0.03 & 3.43 & 1.294 & 5.516 & 0.0006 & 0.443\\
1175657147.96191 & 19.0 & 276.0 & 60.5 & 0.24 & 643929653.0 & 1848.0 & 3.117 & 3.138 & 0.168 & 2.293 & -0.0008 & 3.202\\
1167799474.10205 & 15.0 & 685.0 & 68.1 & 0.25 & 22040761.0 & 5080.0 & 4.01 & -0.043 & 1.241 & 5.792 & -0.0006 & 0.035\\
1167908031.27515 & 10.5 & 837.5 & 70.6 & 0.28 & 52212156.0 & 5964.0 & 5.6 & -0.959 & 1.567 & 2.465 & 0.0004 & 0.254\\
1183233224.15137 & 22.0 & 333.5 & 72.2 & 0.22 & 3870714.0 & 1672.0 & 2.132 & 1.826 & 1.251 & 2.348 & 0.0006 & 0.821\\
1168502591.35742 & 15.0 & 405.0 & 79.8 & 0.27 & 15348113.0 & 1558.0 & 4.0 & 0.033 & 1.346 & 1.58 & -0.002 & 0.007\\
1168325749.5498 & 24.0 & 651.0 & 81.4 & 0.28 & 22057524.0 & 1981.0 & 2.41 & 2.189 & 0.937 & 2.359 & -0.0001 & 1.639\\
1166476456.51807 & 25.5 & 457.0 & 91.0 & 0.2 & 10940160.0 & 2407.0 & 2.131 & 1.758 & 1.292 & 2.364 & -0.0006 & 0.708\\
1185995935.97412 & 14.5 & 756.0 & 117.0 & 0.24 & 32874060.0 & 3751.0 & -1.21 & 4.969 & 1.475 & 5.5 & 0.0013 & 0.222\\
1184590507.66601 & 15.0 & 557.0 & 118.3 & 0.26 & 7210764.0 & 1049.0 & -0.001 & 4.001 & 0.095 & 3.918 & -0.0001 & 0.001\\
1177927452.96191 & 17.0 & 435.5 & 118.9 & 0.22 & 11344824.0 & 1785.0 & 2.164 & 2.035 & 1.081 & 2.343 & 0.0005 & 1.293\\
1185772471.04395 & 15.0 & 444.0 & 129.8 & 0.23 & 2201893.0 & 1145.0 & 1.168 & 2.634 & 1.273 & 2.208 & -0.0003 & 0.541\\
1173030250.41211 & 15.0 & 347.5 & 132.3 & 0.23 & 1777397326.0 & 2760.0 & 3.846 & 2.568 & 0.189 & 1.936 & -0.0002 & 2.531\\
1169679101.79101 & 16.5 & 399.0 & 134.8 & 0.33 & 174991.0 & 785.0 & -1.008 & 5.101 & 1.57 & 2.288 & -0.0055 & 0.036\\
1167881008.30127 & 29.0 & 588.5 & 151.9 & 0.37 & 372288114.0 & 17742.0 & -0.377 & 3.351 & 1.567 & 5.084 & -0.0006 & 0.587\\
1169000234.35303 & 19.0 & 423.0 & 172.9 & 0.35 & 12346924.0 & 5561.0 & -0.097 & 3.929 & 1.567 & 4.773 & -0.0013 & 0.101\\
1172841128.26538 & 8.5 & 1095.5 & 260.7 & 0.3 & 368579956.0 & 10763.0 & -1.8 & 4.729 & 1.569 & 2.5 & -0.0002 & 0.319\\
1173727086.04785 & 29.5 & 251.0 & 272.8 & 0.38 & 42384.0 & 301.0 & 4.002 & -0.006 & 1.552 & 5.488 & -0.0007 & 0.002\\
1178238927.93018 & 9.0 & 768.5 & 279.1 & 0.33 & 19330389.0 & 5400.0 & -2.571 & -0.03 & 1.568 & 3.107 & -0.0001 & 0.119\\
1170580345.62744 & 22.0 & 578.5 & 320.0 & 0.41 & 13931706.0 & 4384.0 & 4.007 & -0.028 & 1.187 & 5.866 & 0.0002 & 0.024\\
1185561257.70605 & 15.0 & 391.0 & 333.3 & 0.35 & 211888.0 & 681.0 & 0.003 & 3.986 & 1.469 & 4.782 & 0.0014 & 0.003\\
1169769827.71826 & 23.0 & 480.0 & 364.5 & 0.37 & 10481211.0 & 3099.0 & 3.42 & 0.504 & 1.286 & 5.634 & 0.0009 & 0.222\\
1182148609.39697 & 15.0 & 827.5 & 423.3 & 0.54 & 71711015.0 & 6549.0 & 3.887 & -0.141 & 1.44 & 2.362 & 0.002 & 0.18\\
1186683197.42871 & 21.0 & 282.0 & 901.2 & 0.7 & 268809.0 & 1005.0 & -0.098 & 3.495 & 1.489 & 1.601 & -0.001 & 0.12\\
1185582032.97168 & 14.5 & 529.5 & 1033.1 & 0.49 & 37095624.0 & 2386.0 & 2.088 & 2.299 & 1.015 & 5.509 & -0.0015 & 1.436\\
\end{tabular}
\caption{Relevant features, ${\cal S}$ metric, and MAP points for DM-lump model parameters for the 84 selected Koi-Fish glitches analyzed in this work.}
\label{tab:fullwidth}
\end{table*}

\bibliography{biblio.bib}
\end{document}